\documentclass[aps,prc,twocolumn,showpacs,preprintnumbers,nofootinbib,float,superscriptaddress,longbibliography]{revtex4-1}
\usepackage{graphicx}
\usepackage{xcolor}
\usepackage[dvipsnames]{xcolor}
\usepackage{breakurl}
\usepackage[colorlinks=true, pdfstartview=FitV, linkcolor=RedOrange, citecolor=ForestGreen, urlcolor=blue]{hyperref}
\usepackage{amsmath}
\usepackage{amssymb}
\usepackage{bm}
\usepackage{physics}
\usepackage{placeins}
\usepackage{adjustbox}
\usepackage[normalem]{ulem}

\DeclareMathOperator{\im}{Im}

\usepackage[maxfloats=256]{morefloats}
\begin{document}
\title{The production of lepton pairs in heavy-ion collisions at SPS and RHIC energies}

\author{Xiang-Yu Wu} \email{xiangyu.wu2@mail.mcgill.ca}
\affiliation{Department of Physics, McGill University, 3600 University Street, Montreal, QC H3A 2T8, Canada}
\author{Han Gao} \email{han.gao3@mail.mcgill.ca}
\affiliation{Department of Physics, McGill University, 3600 University Street, Montreal, QC H3A 2T8, Canada}
\author{Charles Gale} \email{charles.gale@mcgill.ca}
\affiliation{Department of Physics, McGill University, 3600 University Street, Montreal, QC H3A 2T8, Canada}
\author{Sangyong Jeon} \email{sangyong.jeon@mcgill.ca}
\affiliation{Department of Physics, McGill University, 3600 University Street, Montreal, QC H3A 2T8, Canada}
\author{Ralf Rapp} \email{rapp@comp.tamu.edu}
\affiliation{Cyclotron Institute and Department of Physics and Astronomy, Texas A\&M University, College Station, TX 77843-3366, USA}

\begin{abstract}
We perform a quantitative study of dilepton production and elliptic flow in In+In collisions at $\sqrt{s_{\rm NN}}=17.3$~GeV and isobar collisions at $\sqrt{s_{\rm NN}}=200$~GeV, concentrating on the low and intermediate invariant mass regions. 
The medium evolution is simulated event-by-event with the multistage iEBE-MUSIC framework, 
and the electromagnetic emission  combines for the first time  quark-gluon plasma (QGP) emission  calculated at next-to-leading order (NLO) in the strong coupling with in-medium radiation calculated with hadronic many-body theory. Freeze-out and surviving primordial $\rho$ decays are also included.
The calculations are compared with the high-precision dimuon spectra measured by the NA60 Collaboration,  and with STAR preliminary measurements of dielectron spectra and elliptic flow. We obtain effective temperatures extracted from the transverse-mass spectra, and we calculate the elliptic flow of the different dilepton sources.
Our results suggest that dilepton measurements in the low-mass and high $p_T$ region can provide insight into higher-order corrections to the QGP emission rate. 
\end{abstract}
\maketitle

\section{Introduction}
\label{sec:intro}

The quark-gluon plasma (QGP) is an exotic state of strongly interacting matter, where the relevant degrees of freedom are quarks and gluons. The transition at high temperatures from an ensemble of composite hadrons to a QGP has been predicted by lattice QCD \cite{Bazavov:2009zn}, and its hallmark signatures have been observed in the high-energy collisions of atomic nuclei \cite{Busza:2018rrf}. 
Lepton pairs (dileptons) are penetrating probes of the hot and dense QCD matter created in relativistic heavy-ion collisions~\cite{Gale:2025ome,Gale:2003iz,Geurts:2022xmk,Rapp:1999ej,Bailhache:2025kwa,Harris:2023tti}. Once produced, they undergo negligible final-state interactions and escape the medium undistorted. Their invariant mass spectra therefore carry information about the local temperature, undistorted by the local flow velocity at the time of emission~\cite{vanHees:2007th,Geurts:2022xmk,Du:2025dot,Churchill:2023zkk,Churchill:2023vpt,Santini:2011zw,Endres:2016tkg,Linnyk:2011hz,Nayak:2009xrg,Goes-Hirayama:2025lhs,Jorge:2025wwp,Wu:2025iix}.
Given that hadronic observables  are more strongly affected by the late-stage evolution and freeze-out conditions and that real photons are also influenced by both the local temperature and medium flow \cite{vanHees:2011vb,Shen:2013vja,Du:2025dot}, dileptons complete the palette of multi-messenger observables which, together, can provide quantitative information on the space-time dynamics of heavy-ion collisions. 

The same penetrating nature also makes the interpretation of dilepton measurements challenging~\cite{NA60:2006ymb,NA60:2007lzy,NA60:2008ctj,NA60:2008dcb,PHENIX:2015vek,STAR:2013pwb,STAR:2015tnn,STAR:2015zal,STAR:2023wta,STAR:2024bpc,ALICE:2018ael,ALICE:2023jef} as 
lepton pairs are produced throughout the collision, from the earliest stages, the QGP, and the hadronic phase. The measured spectrum contains contributions from both thermal and nonthermal sources. 
The latter is dominated by a ``hadronic cocktail'' which includes long-lived meson decays, as well as correlated heavy-flavor decays and Drell--Yan production. The relative importance of these contributions depends on the collision energy and dilepton invariant mass. We shall consider other non-thermal dilepton sources later in this paper. 

After subtracting the hadronic cocktail, the remaining signal (labelled ``excess'' in the rest of this paper)  can be studied as a function of invariant mass $M$, with
$M^2=\left(p_{\ell{^+}}+p_{{\ell^-}}\right)^2$,
to distinguish different production mechanisms.
In the low-mass region (LMR) ($M \lesssim m_\phi$), the electromagnetic spectral function is dominated by light vector mesons, particularly the $\rho$ meson. 
Due to its short lifetime, the $\rho$ is continuously regenerated and decays within the hadronic medium, where interactions with surrounding mesons and baryons strongly modify its spectral function~\cite{Rapp:1997fs,Rapp:1999us,vanHees:2007th}. 
In the intermediate-mass region (IMR) ($m_\phi \lesssim M \lesssim m_{J/\psi}$), the hadronic spectral strength receives contributions from a continuum of multipion states~\cite{vanHees:2007th,Ruppert:2007cr,Santini:2011zw,Endres:2014zua} and thermal radiation from the QGP becomes increasingly important there.

On the theoretical side, considerable progress has  been made in calculating dilepton emission from both hadronic and partonic matter.
Hadronic many-body calculations~\cite{Rapp:1997fs,Rapp:1999us,vanHees:2007th} include the medium modifications of the vector-meson spectral functions and provide a successful description of the broadened $\rho$ contribution observed at SPS and RHIC energies~\cite{NA60:2006ymb,NA60:2007lzy,NA60:2008ctj,NA60:2008dcb,PHENIX:2015vek,STAR:2013pwb,STAR:2015tnn,STAR:2015zal,STAR:2023wta,STAR:2024bpc}, and also have been used to interpret data at lower energy obtained by the HADES Collaboration \cite{Seck:2017zjr}. 
For the partonic phase, thermal dilepton emission rates have been derived at next-to-leading order (NLO) in the strong coupling, with Landau-Pomeranchuk-Migdal (LPM) resummation and finite baryon chemical potential consistently taken into account~\cite{Churchill:2023vpt}.

A quantitative calculation of dilepton production also requires a realistic description of the space-time evolution of the medium. Previous studies have employed microscopic transport models~\cite{Endres:2016tkg,Endres:2015egk,Endres:2015fna,Endres:2014zua,Linnyk:2011hz,Goes-Hirayama:2024aqz}, such as UrQMD, as well as macroscopic approaches based on expanding fireball models~\cite{Rapp:1999us,vanHees:2007th,Rapp:2014hha,Rapp:2013nxa,vanHees:2006ng,Rapp:1999zw} and relativistic hydrodynamics~\cite{Santini:2011zw,Banerjee:2026hlm,Wu:2025iix,Wu:2024pba,Churchill:2023zkk,Churchill:2023vpt,Nayak:2009xrg,Goes-Hirayama:2025lhs, Vujanovic:2019yih,Vujanovic:2017psb,Vujanovic:2016anq,Vujanovic:2013jpa}.
These approaches describe the collision geometry and collective expansion differently.
Fireball models generally rely on parametrized space-time evolutions and flow profiles. Coarse-grained transport calculations retain the three-dimensional evolution of the microscopic dynamics, but event averaging may, to some extent, limit the description of collective flow evolution~\cite{Endres:2015egk}.
In contrast, hydrodynamic simulations describe the expansion and development of collective flow from the initial collision geometry. The resulting flow evolution can be tested against the measured anisotropic flow of final-state charged hadrons~\cite{Gale:2013da, Shen:2020gef, Heinz:2024jwu,Heinz:2013th,Jeon:2015dfa,Du:2024wjm, Song:2010mg,Shen:2015msa, Ryu:2015vwa, Schenke:2020mbo,Wu:2021fjf}. An accurate description of this evolution is also important for that of dilepton anisotropic flow.

Experimentally, CERES, a collaboration dedicated to low-mass dielectron measurements at the CERN SPS, observed a dielectron spectrum larger than that attributable to hadron decays~\cite{CERES:1995vll}. Subsequently, the NA60 measurement of dimuons in In+In
collisions at $\sqrt{s_{\rm NN}}=17.3$~GeV marked an important
step in the study of thermal
dileptons~\cite{NA60:2006ymb,NA60:2007lzy,NA60:2008ctj,NA60:2008dcb}.
Its high statistics and mass resolution made it possible
to isolate the excess spectrum with unprecedented precision.
Dilepton measurements have since been extended to RHIC and
LHC energies by the PHENIX, STAR, and ALICE
Collaborations~\cite{PHENIX:2015vek,STAR:2013pwb,STAR:2015tnn,STAR:2015zal,STAR:2023wta,STAR:2024bpc,ALICE:2018ael,ALICE:2023jef}.
More recently, STAR reported the first measurement of
thermal-dielectron elliptic flow in isobar collisions at
$\sqrt{s_{\rm NN}}=200$~GeV~\cite{Bao:2026HP}.
These measurements, together with developments in dilepton
emission rates and multistage hydrodynamic models, motivate
us to revisit dilepton production and study its anisotropic
flow.

In this study, we aim to provide a comprehensive theoretical benchmark for dilepton observables at SPS and RHIC energies. We 
investigate dilepton spectra and elliptic flow in the LMR and IMR using the multistage iEBE-MUSIC framework. The event-by-event medium evolution is simulated with parameters obtained in a recent Bayesian analysis of
charged-hadron data at RHIC-BES energies~\cite{Jahan:2024wpj}. We combine NLO QGP emission rates~\cite{Churchill:2023vpt} with in-medium $\rho$ radiation from hadronic many-body theory and four-pion radiation, and include freeze-out and surviving primordial $\rho$ decays~\cite{Rapp:1997fs,Rapp:1999us,vanHees:2007th}. 
We first compare the invariant-mass and transverse-mass spectra and the extracted effective temperatures in In+In collisions at $\sqrt{s_{\rm NN}}=17.3$~GeV with the NA60 measurements. We then extend the calculation to isobar collisions at $\sqrt{s_{\rm NN}}=200$~GeV and compare the spectra and elliptic flow with the preliminary STAR data. We examine how nonthermal $\rho$ decays affect $T_{\rm eff}(M)$ and how four-pion radiation contributes to the IMR elliptic flow. Finally, we compare the spectra and elliptic flow obtained with LO and NLO QGP rates in isobar collisions, focusing on their differences at low mass and high $p_T$.

This paper is organized as follows: The thermal and nonthermal dilepton emission sources are presented in Sec.~\ref{sec:EM_sources}. In Sec.~\ref{sec:framework}, we briefly introduce the multistage model.
In Secs.~\ref{sec:SPS_results} and~\ref{sec:RHIC_results}, we present our results for dilepton invariant-mass and transverse-mass spectra, effective inverse slopes, and elliptic flow at SPS and RHIC energies, respectively. The sensitivity of dilepton spectra and elliptic flow to the QGP emission rate is examined in Sec.~\ref{sec:qgp_rate_sensitivity}.
A summary is provided in Sec.~\ref{sec:conclusion}.
Comparisons with charged-hadron observables are given in
Appendix~\ref{sec:calibration}.

\section{Dilepton emission}
\label{sec:EM_sources}
This section presents the thermal and nonthermal dilepton sources included in our calculation. We first describe thermal radiation from the QGP and hadronic medium, including the in-medium $\rho$ and four-pion contributions, and then discuss freeze-out and surviving primordial $\rho$ decays.

\subsection{Thermal dilepton emission}
\label{sec:thermal_dilepton}

The thermal dilepton emission rate from locally equilibrated
strongly interacting matter can be expressed in terms of the
retarded electromagnetic current-current correlation function.
For a thermal medium characterized by temperature $T$, baryon
chemical potential $\mu_{\rm B}$, and four-velocity $u^\mu$, the
differential emission rate per unit four-volume and
four-momentum is given by~\cite{McLerran:1984ay,Gale:1990pn,Weldon:1990iw,Kapusta:2023eix}

\begin{equation}
    \frac{dN_{\ell^+\ell^-}}{d^4x\,d^4p}
    =
    -\frac{\alpha_{\rm em}^2}{\pi^3 M^2}
    L\left(\frac{m_\ell}{M}\right)
    f_{\rm B}\left(\frac{q^0}{T}\right)
    \im \Pi_{\rm em}^{{\rm R}}
    (q^0,|\mathbf q|;T,\mu_{\rm B}),
    \label{eq:thermal_dilepton_rate}
\end{equation}
where $\alpha_{\rm em}$ is the electromagnetic fine-structure
constant.
We denote the dilepton four-momentum in the laboratory frame by $p^\mu$ and its components in the local rest frame (LRF) of a fluid cell by
$q^\mu=(q^0,\mathbf q)$. The invariant mass satisfies $M^2=p^\mu p_\mu=q^\mu q_\mu$. The LRF energy and
three-momentum magnitude are obtained as
\begin{equation}
    q^0=p\cdot u,
    \qquad
    |\mathbf q|=\sqrt{(p\cdot u)^2-M^2}.
    \label{eq:lrf_kinematics}
\end{equation}
Note that the invariant mass $M$ is
independent of the reference frame. The lepton phase-space
factor is given by
\begin{equation}
    L(x)
    =
    \left(1+2x^2\right)
    \sqrt{1-4x^2},
    \qquad
    x=\frac{m_\ell}{M},
    \label{eq:lepton_phase_space}
\end{equation}
where $m_\ell$ denotes the lepton mass, with
$m_e=0.511~{\rm MeV}$ and $m_\mu=105.6~{\rm MeV}$. The
Bose-Einstein distribution function is
\begin{equation}
    f_{\rm B}\left(\frac{q^0}{T}\right)
    =
    \frac{1}{\exp(q^0/T)-1}.
\end{equation}
In the LRF, the retarded electromagnetic current-current correlator is defined as
\begin{equation}
    \Pi_{\rm em}^{{\rm R},\mu\nu}(q)
    =
    i\int d^4x\,
    e^{iq\cdot x}
    \Theta(x^0)
    \left\langle
        \left[
            J_{\rm em}^{\mu}(x),
            J_{\rm em}^{\nu}(0)
        \right]
    \right\rangle ,
    \label{eq:em_correlator}
\end{equation}
where $J_{\rm em}^{\mu}$ denotes the electromagnetic current.
The electromagnetic spectral function entering
Eq.~(\ref{eq:thermal_dilepton_rate}) is defined in terms of
the trace of the retarded correlator as
\begin{equation}
    \im \Pi_{\rm em}^{{\rm R}}(q)
    \equiv
    \frac{1}{3}
    g_{\mu\nu}
    \im \Pi_{\rm em}^{{\rm R},\mu\nu}(q).
    \label{eq:em_scalar_correlator}
\end{equation}
The microscopic properties of the emitting medium are encoded
in the electromagnetic spectral function. Consequently,
Eq.~(\ref{eq:thermal_dilepton_rate}) provides a common
framework for describing thermal dilepton emission from both
hadronic and partonic matter. In the following, we specify
the electromagnetic spectral functions employed for the
partonic phase, the in-medium $\rho$ contribution, and the
multipion contribution.

\subsubsection{Partonic emission}
\label{sec:partonic_emission}

In the QGP phase, the electromagnetic current is
carried by quarks and antiquarks. For three light quark
flavors, the QGP contribution to the electromagnetic spectral
function can be written as
\begin{align}
    \left.
    \im \Pi_{\rm em}^{{\rm R}}
    \right|_{\rm QGP}
    =
    -\frac{C_{\rm em}}{3}\,
    \rho_V ,
    \label{eq:Pi_QGP}
\end{align}
where
$C_{\rm em}=\sum_{f=u,d,s}Q_f^2=2/3$
is the sum of the squared electric charges of the light quarks.

Earlier phenomenological studies employed hard-thermal-loop (HTL) improved QGP dilepton rates at vanishing quark chemical potential~\cite{Braaten:1990wp,Rapp:2002mm,Rapp:2009yu,Rapp:2013nxa}.
The HTL treatment includes thermal quark and gluon masses, $m_{q,g}^{\rm th}\sim gT$, with $g$ the QCD coupling, and Bremsstrahlung contributions.
In the present work, we employ the complete NLO vector spectral function $\rho_V$ at finite baryon chemical potential~\cite{Churchill:2023zkk,Churchill:2023vpt}. The calculation includes the strict one- and two-loop perturbative contributions, consistently matched to the LPM-resummed result to account for multiple soft scatterings near the light cone. The NLO corrections become particularly important toward low invariant masses, where they enhance the thermal dilepton emission rate relative to the leading-order result~\cite{Churchill:2023zkk,Churchill:2023vpt}.

\subsubsection{In-medium $\rho$ meson emission}
\label{sec:inmedium_rho}

In the hadronic phase, the low-mass electromagnetic spectral
function is governed by the light vector mesons $\rho$,
$\omega$, and $\phi$ within the vector-dominance model
(VDM). Among these states, the isovector $\rho$ channel
provides the dominant contribution in the mass region of
interest. In the present calculation, we therefore retain only
the $\rho$-meson contribution, while the contributions from
the $\omega$ and $\phi$ mesons are neglected. Within VDM,
the $\rho$ contribution to the electromagnetic spectral
function is related to the in-medium $\rho$ propagator
according to
\begin{equation}
    \im \Pi_{\rm em}^{{\rm R}}
    \big|_{\rho}
    =
    \frac{\left(m_\rho^{(0)}\right)^4}{g_\rho^2}
    \im D_\rho^{{\rm R}},
    \label{eq:Pi_rho}
\end{equation}
where $m_\rho^{(0)}=0.853$~GeV is the bare mass of the $\rho$-meson and $g_\rho=5.9$ is the $\rho\pi\pi$ coupling constant.
The bare mass differs from the physical $\rho$ mass because of the vacuum pion-cloud self-energy~\cite{vanHees:2007th}. The in-medium $\rho$ propagator is given by
\begin{equation}
    D_\rho^{{\rm R}}
    =
    \frac{1}{
        M^2-\left(m_\rho^{(0)}\right)^2
        -\Sigma_\rho
    }.
    \label{eq:rho_propagator}
\end{equation}

In vacuum, the electromagnetic spectral function is well
constrained by measurements of $e^+e^-$ annihilation into
hadrons. In a thermal medium, however, the spectral function
acquires a nontrivial dependence on temperature and baryon
density and therefore requires the description of
the relevant medium effects. In the present work, we employ
the in-medium $\rho$ spectral function obtained from hadronic
many-body theory~\cite{Rapp:1997fs,Rapp:1999us,vanHees:2007th}. In this framework, the
in-medium $\rho$ self-energy is decomposed as
\begin{equation}
    \Sigma_\rho
    =
    \Sigma_{\rho\pi\pi}
    +
    \Sigma_{\rho M}
    +
    \Sigma_{\rho B},
    \label{eq:rho_selfenergy}
\end{equation}
where $\Sigma_{\rho\pi\pi}$ accounts for medium modifications
of the pion cloud, while $\Sigma_{\rho M}$ and
$\Sigma_{\rho B}$ describe direct interactions of the
$\rho$ meson with mesons and baryons in the surrounding
medium, respectively.

\subsubsection{Multipion annihilation}
\label{sec:multipion}

At invariant masses above the $\phi$-meson region
($M>1$~GeV), the hadronic electromagnetic spectral function
is no longer dominated by individual vector-meson resonances.
Instead, overlapping hadronic states give rise to a broad
multipion continuum~\cite{vanHees:2007th,Ruppert:2007cr,Li:1998ma,Santini:2011zw,Endres:2014zua}. In the mass range considered here, the
four-pion channel provides the leading contribution to this
continuum. We therefore include the four-pion contribution
using a parametrization of the corresponding vacuum
vector-isovector spectral function constrained by hadronic
$\tau$-decay data from the ALEPH Collaboration~\cite{ALEPH:1998rgl}.

\subsection{Nonthermal dilepton sources}
In addition to thermal radiation from the evolving medium,
the measured dilepton spectrum contains contributions from
nonthermal production mechanisms and hadronic decays.
Dileptons from long-lived hadrons constitute the conventional
hadronic decay cocktail. These contributions, including those
from the $\pi^0$, $\eta$, $\omega$, and $\phi$ mesons, can be
constrained and subtracted experimentally using their vacuum
decay properties.

The $\rho$ meson requires special treatment. 
As the hadronic medium becomes increasingly dilute at late
times, the continuous thermal description eventually
terminates at kinetic freeze-out. A finite population of
$\rho$ mesons is then emitted from the kinetic freeze-out
hypersurface and can subsequently decay into dileptons~\cite{vanHees:2007th}.
Another distinct nonthermal contribution arises from primordial
$\rho$ mesons produced in the initial hard scatterings~\cite{vanHees:2007th}, which
may escape the medium without reaching thermal equilibrium
and subsequently contribute through their dilepton decays.
We therefore include both the freeze-out and primordial
$\rho$ contributions introduced in
Ref.~\cite{vanHees:2007th}, in addition to the in-medium thermal
radiation discussed above. Their implementations are
described in the following subsections.

\subsubsection{Freeze-out $\rho$ mesons}
\label{sec:freezeout_rho}

To evaluate the freeze-out contribution, we determine the
$\rho$-meson distribution on the kinetic freeze-out
hypersurface using the Cooper--Frye prescription with the finite width of the $\rho$ meson. Then the
resulting $\rho$ yield is converted into the
dilepton yield through the corresponding electromagnetic
branching ratio.

Specifically, we first write down the generalized Cooper--Frye formula on the kinetic
freeze-out hypersurface $\Sigma_{\rm fo}$ for particles in species $s$ with a finite decay width.
\begin{equation}
    \frac{dN_s^{\rm fo}}{d^4p}
    =
    \int_{\Sigma_{\rm fo}}
    \frac{d_s}{(2\pi)^3}\,
    p^\mu d\sigma_\mu\,
    \frac{A_s(p)}{\pi}\,
    f\!\left(
        \frac{p\cdot u(x)}
             {T_{\rm fo}(x)}
    \right).
    \label{eq:s_fo_CF}
\end{equation}
Here, $d_s$ denotes the degeneracy factor for species $s$,
and $d\sigma_\mu$ is the hypersurface element. 
$u^\mu(x)$ is the local fluid four-velocity. The spectral function $A_s(p)$ is related to the retarded propagator $D_s^{\rm R}(p)$ as
\begin{equation}
    A_s(p) = -2\,{\rm Im}\,D_s^{\rm R}(p).
\end{equation}
Note that for on-shell particles with $p^0>0$, 
\begin{equation}
    D_s^{\rm R}(p) = \frac{1}{p^2-m^2+ip^0\epsilon}
    \quad\Rightarrow\quad
    A_s(p) = 2\pi\delta(p^2-m^2).
\end{equation}
The standard Cooper--Frye formula for on-shell particles can therefore be consistently recovered by integrating $p^0$ out on both sides of Eq.~\eqref{eq:s_fo_CF}.
\par
Multiplying Eq.~\eqref{eq:s_fo_CF} by the branching ratio of $\rho \to \ell^+ \ell^-$ mode gives the dilepton yields from freeze-out $\rho$ meson decays
\begin{align}
    \frac{dN_{\ell^+\ell^-}^{\rm fo}}{d^4p}
    &=
    \int_{\Sigma_{\rm fo}}
    \frac{d_\rho}{(2\pi)^3}\,
    p^\mu d\sigma_\mu\,
    \frac{A_\rho(p)}{\pi}\,
    f_{\rm B}\!\left(
        \frac{p\cdot u(x)}
             {T_{\rm fo}(x)}
    \right) \notag \\
    &\quad\times
    {\rm BR}_{\rho\to\ell^+\ell^-}(M),
    \label{eq:rho_fo_dilepton}
\end{align}
where $d_\rho=3$ denotes the spin degeneracy and 
\begin{equation}
    {\rm BR}_{\rho\to\ell^+\ell^-}(M)
    =
    \frac{\Gamma_{\rho\to\ell^+\ell^-}(M)}
         {\Gamma_\rho^{\rm fo}}.
    \label{eq:rho_fo_BR}
\end{equation}

Within VDM~\cite{Rapp:1999ej}, the partial decay width is given by
\begin{equation}
    \Gamma_{\rho\to\ell^+\ell^-}(M)
    =
    \frac{4\pi\alpha_{\rm em}^2}{3}
    \frac{m_\rho^4}{g_\rho^2 M^3}
    L\left(\frac{m_\ell}{M}\right),
    \label{eq:rho_ll_width}
\end{equation}
where $g_\rho=5.9$ is the $\rho\pi\pi$ coupling constant, fitted to the vacuum $\rho\to\pi\pi$ decay width. It enters the VDM $\rho$--photon coupling factor as $m_\rho^4/g_\rho^2$. 
The $L(m_\ell/M)$ is the lepton phase-space factor defined in
Eq.~(\ref{eq:lepton_phase_space}).

For the spectral function and the total width in
Eq.~(\ref{eq:rho_fo_dilepton}), we follow the setup of Ref.~\cite{vanHees:2007th}. The vacuum form of the
$\rho$ self-energy is employed with an effective width that
accounts for the residual broadening at decoupling. We use
$\Gamma_\rho^{\rm fo}\simeq260~{\rm MeV}$~\cite{vanHees:2007th},
which corresponds to the full width at half maximum of the
in-medium $\rho$ spectral function under freeze-out
conditions.

Here, if we consider the isochronous freeze-out condition, $d\sigma^\mu=(d^3x,\mathbf{0})$, such that $p^\mu d\sigma_\mu=p^0d^3x$,
Eq.~(\ref{eq:rho_fo_dilepton}) reduces to the freeze-out
$\rho$ expression of Ref.~\cite{vanHees:2007th}. Relative to the
continuous in-medium thermal emission rate, the resulting
freeze-out contribution contains an additional Lorentz factor $\gamma=p^0/M$.
This factor reflects the time dilation of the lifetime
of a freeze-out $\rho$ meson.

\subsubsection{Primordial $\rho$ mesons}
\label{sec:primordial_rho}

Another nonthermal source considered in this work is the
primordial $\rho$ component produced in the initial hard nucleon--nucleon collisions. In contrast to the schematic
jet-quenching treatment of Ref.~\cite{vanHees:2007th}, we estimate this
contribution using a PYTHIA+SMASH framework.

The initial primordial $\rho$ distribution in $pp$ collisions
is constructed from the vacuum $\rho$ spectral function and
the transverse-momentum spectrum obtained from PYTHIA~\cite{Sjostrand:2014zea},
\begin{equation}
    \frac{dN_\rho^{pp}}
         {M dM\,p_T dp_T\,d\phi\,dy}
    =
    \frac{1}{2\pi}
    \frac{A_\rho^{\rm vac}(M)}{\pi}
    \frac{dN_\rho^{pp}}
         {p_T dp_T\,dy},
    \label{eq:rho_prim_pp}
\end{equation}
where the factor $1/(2\pi)$ assumes an isotropic azimuthal
distribution. In the PYTHIA calculation, both soft and hard
QCD processes are included, with hadronization enabled and
hadronic decays switched off.

For nucleus--nucleus collisions, the primordial $\rho$ yield is assumed to scale with the number of binary collisions, $N_{\rm coll}$, and the production positions are sampled from the corresponding binary-collision points. 

Motivated by the small effective cross section
used for the primordial $\rho$ mesons in the QGP stage in
Ref.~\cite{vanHees:2007th}, we neglect both their interactions with and
energy loss in the QGP medium during the hydrodynamic
evolution. The primordial $\rho$ mesons therefore free-stream
and are transferred to SMASH~\cite{SMASH:2016zqf} at their last outward
crossing of the switching hypersurface.

To isolate the surviving primordial component, we retain only
the initially produced $\rho$ mesons that undergo no hadronic
collisions before their decay. Denoting their distribution
after the SMASH evolution by
$dN_{\rho,\rm surv}^{\rm prim}/d^4p$, the corresponding
dilepton spectrum is
\begin{equation}
    \frac{dN_{\ell^+\ell^-}^{\rm prim}}{d^4p}
    =
    \frac{dN_{\rho,\rm surv}^{\rm prim}}{d^4p}
    {\rm BR}_{\rho\to\ell^+\ell^-}(M),
    \label{eq:rho_prim_dilepton}
\end{equation}
where the mass-dependent branching ratio is the same as that
introduced for the freeze-out $\rho$ contribution.

\section{Multistage model}
\label{sec:framework}

To provide a realistic space-time evolution of relativistic heavy-ion collisions at SPS and RHIC-BES energies, we employ the multistage iEBE-MUSIC framework. The initial conditions for the energy density, flow velocity, and net-baryon density are generated using a 3D Monte Carlo Glauber model~\cite{Shen:2017bsr, Shen:2017fnn, Shen:2018pty, Shen:2022oyg}, which incorporates a classical string deceleration mechanism to describe the dynamical deposition of energy and baryon number into the medium. The subsequent evolution of the QGP medium is simulated using second-order viscous hydrodynamics model~\cite{Schenke:2010nt, Schenke:2010rr, Paquet:2015lta, Denicol:2018wdp} based on the Denicol-Niemi-Molnar-Rischke (DNMR) formalism~\cite{Denicol:2012cn}. The specific shear and bulk viscosity depend on temperature  and on baryon chemical potential~\cite{Jahan:2024wpj, Shen:2023awv}. The hydrodynamic evolution employs the NEOS-BQS equation of state~\cite{Monnai:2019hkn}, which describes a crossover transition between the QGP and hadronic phases.

Throughout this work, we employ the maximum-a-posteriori (MAP) parameter set obtained from a recent Bayesian analysis which provides a good description of charged-hadron observables over a broad range of RHIC-BES energies in Au+Au collisions. The detailed parameter values are given in Ref.~\cite{Jahan:2024wpj}, and no additional modification is done to accommodate the electromagnetic observable.  In that procedure, once the local energy density decreases to the switching value $\epsilon_{\rm sw}=0.3775~\rm GeV/fm^3$ \cite{Jahan:2025cbp}, hadrons are sampled from the switching hypersurface according to the Cooper-Frye prescription~\cite{Cooper:1974mv}. Their subsequent hadronic rescatterings, resonance formation, and decays are simulated with the SMASH transport model~\cite{SMASH:2016zqf}. 
The original calibration employed UrQMD for the hadronic stage~\cite{Jahan:2024wpj}. Replacing UrQMD with SMASH maintains the overall quality of the description of hadronic observables. The corresponding comparisons with data are presented in Appendix~\ref{sec:calibration}.

For the thermal dilepton calculation, we continue the hydrodynamic evolution below the switching energy density $\epsilon_{\rm sw}=0.3775~{\rm GeV/fm^3}$ down to
$\epsilon_{\rm fo}=0.05~{\rm GeV/fm^3}$ to simulate  emission from the late-stage hadronic ensemble \cite{Gotz:2021dco}. We use the QGP emission rate for $\epsilon>\epsilon_{\rm sw}$ and the in-medium $\rho$ and four-pion rates for $\epsilon_{\rm fo}<\epsilon<\epsilon_{\rm sw}$.
The switching energy density is also taken as the chemical freeze-out condition for the hadronic emission. Thermal emission ends at $\epsilon_{\rm fo}$, where the freeze-out $\rho$ contribution is evaluated on the corresponding hypersurface.
At $n_{\rm B}=0$, this switching energy density $\epsilon_{\rm sw}$ corresponds to roughly $T\simeq157$~MeV, slightly below the transition temperature
of $170$~MeV used in Ref.~\cite{Rapp:2013nxa}.
The choice of switching energy density introduces
a model uncertainty in the transition region, which could be quantified in future
Bayesian analyses including dilepton observables.

Below chemical freeze-out, the approximate conservation of effective hadron abundances leads to finite chemical potentials~\cite{Rapp:2002fc}. We account for this effect in the hadronic emission through the pion chemical potential $\mu_\pi$. As an approximation, we retain the SMASH-compatible hadronic equation of state for the hydrodynamic evolution and use the resulting local
temperature to determine $\mu_\pi(T)$ along the chemically frozen hadronic trajectory of Ref.~\cite{Rapp:2002fc}. In the Boltzmann approximation, the hadronic emission
rates include the pion fugacity factor
\begin{equation}
    z_\pi^n=\exp\left(\frac{n\mu_\pi}{T}\right),
    \label{eq:pion_fugacity}
\end{equation}
with $n=2$ for the in-medium $\rho$ contribution and
$n=4$ for four-pion radiation~\cite{vanHees:2007th}.

\section{Dilepton observables in In+In collisions at SPS}
\label{sec:SPS_results}
\subsection{Spectra and effective temperatures}

\begin{figure}[htb]
	\centering
	\includegraphics[width=1\linewidth]{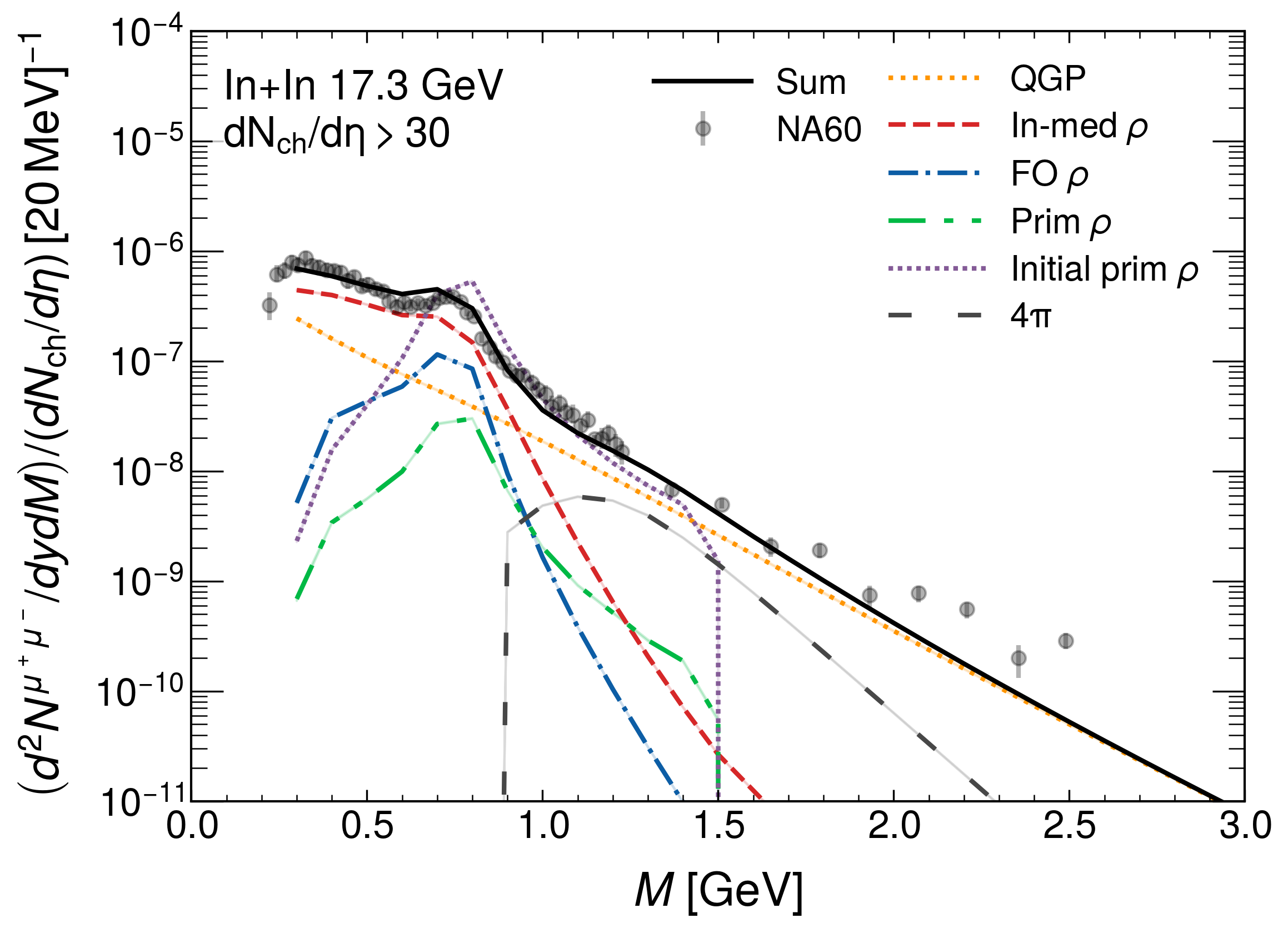}
 \caption{The $p_T$-integrated excess dimuon invariant-mass spectrum in In+In collisions at $\sqrt{s_{\rm NN}}=17.3$~GeV is shown, for $dN_{\rm ch}/d\eta>30$.  The QGP, in-medium $\rho$, four-pion, freeze-out $\rho$, and surviving primordial $\rho$ contributions are shown, together with their sum (black solid line) and data from the NA60 Collaboration~\cite{NA60:2008dcb}. The QGP contribution is evaluated at NLO in the strong coupling. The initial primordial $\rho$ spectrum (purple dotted line) is shown for comparison and is excluded from the total yield. }
	\label{fig:InIn17p3_dilepton_dNdM_normalized}
\end{figure}

Figure~\ref{fig:InIn17p3_dilepton_dNdM_normalized} shows the
$p_T$-integrated excess dimuon invariant-mass spectrum in
In+In collisions at $\sqrt{s_{\rm NN}}=17.3$~GeV for events with
$dN_{\rm ch}/d\eta>30$. The spectrum is normalized by
$dN_{\rm ch}/d\eta$. The calculation includes QGP radiation,
in-medium $\rho$ and four-pion radiation, and freeze-out and
surviving primordial $\rho$ contributions. The individual
sources and their sum are compared with  NA60 data.
The QGP contribution is evaluated with the NLO rate. In the
LMR, the hadronic decay cocktail has been subtracted from the
experimental spectrum, while the $\rho$ contribution is
retained. In the IMR, the Drell--Yan and correlated open-charm
contributions are also subtracted~\cite{NA60:2008dcb}. The initial
primordial $\rho$ spectrum is shown separately here and in subsequent figures only to illustrate
its suppression during the subsequent SMASH evolution. 

The calculation provides a good overall description of the
measured spectrum across the LMR and part of the
IMR. In the LMR, $0.2<M<1.0$~GeV, the in-medium $\rho$
represents the dominant contribution, consistent with previous
studies~\cite{Endres:2016tkg,Endres:2015egk,Endres:2015fna,Endres:2014zua,Linnyk:2011hz,Rapp:1999us,vanHees:2007th,Rapp:2014hha,Rapp:2013nxa,vanHees:2006ng,Rapp:1999zw,Santini:2011zw}. Interactions of the $\rho$ with surrounding
mesons and baryons produce a broadened spectral function that
is essential for reproducing the signal observed by NA60.
At $M\lesssim0.5$~GeV, QGP radiation remains subleading
relative to the in-medium $\rho$, but is not negligible. The effects of the LO and NLO QGP emission rates are discussed further in Sec.~\ref{sec:qgp_rate_sensitivity}. 

\begin{figure*}[htb]
	\centering
    \includegraphics[width=0.45\linewidth]{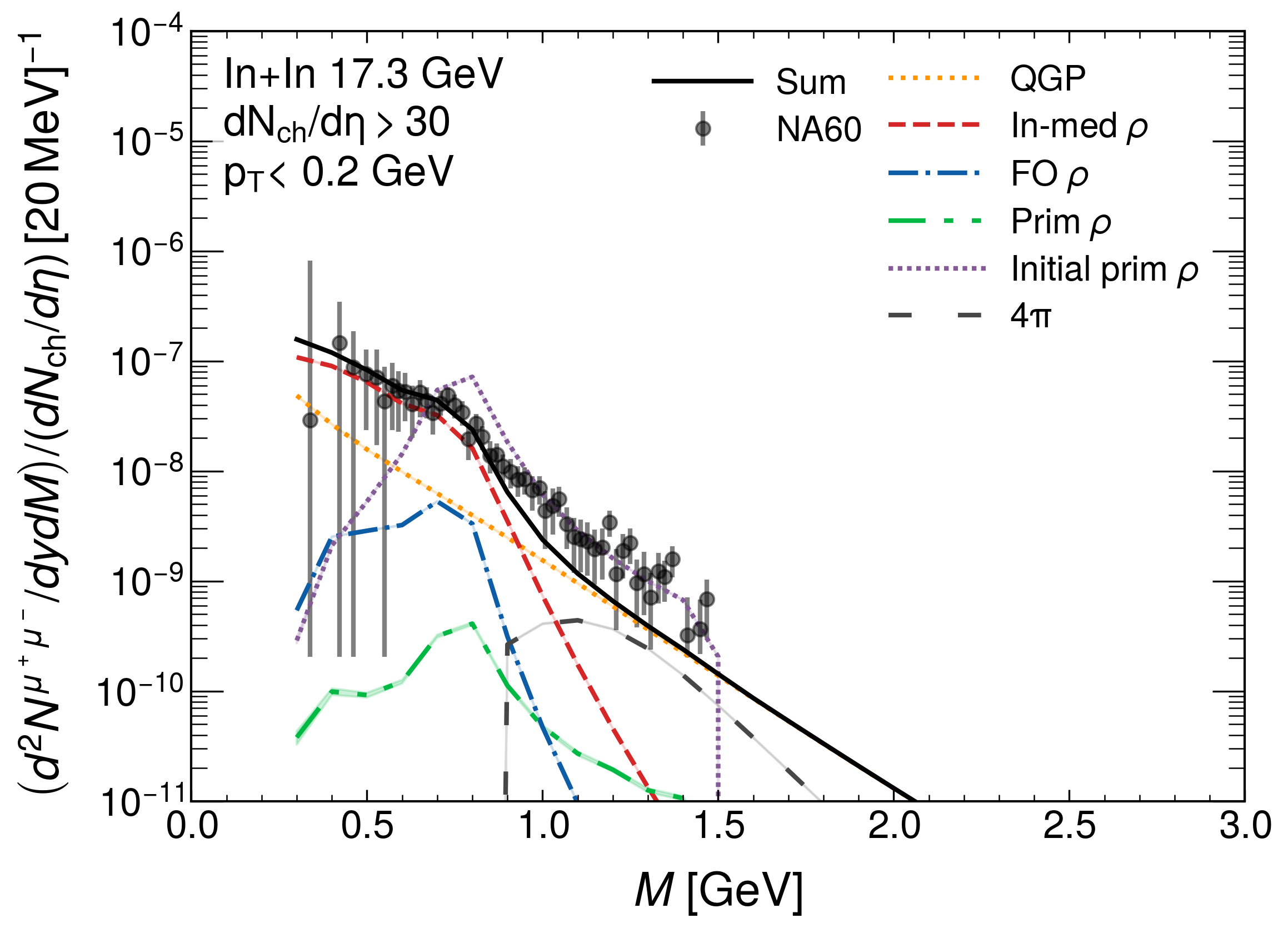}
	\includegraphics[width=0.45\linewidth]{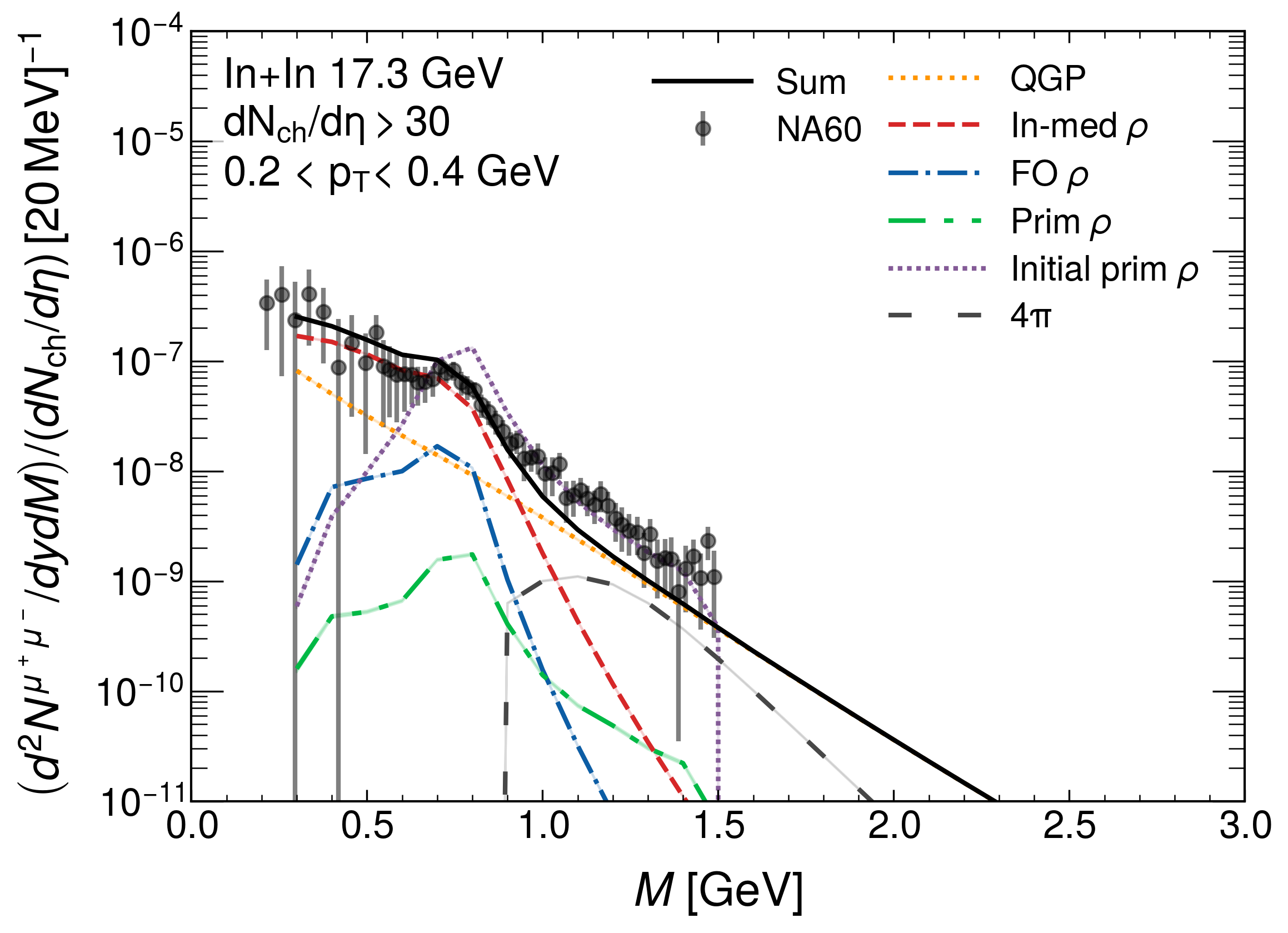}
    \includegraphics[width=0.45\linewidth]{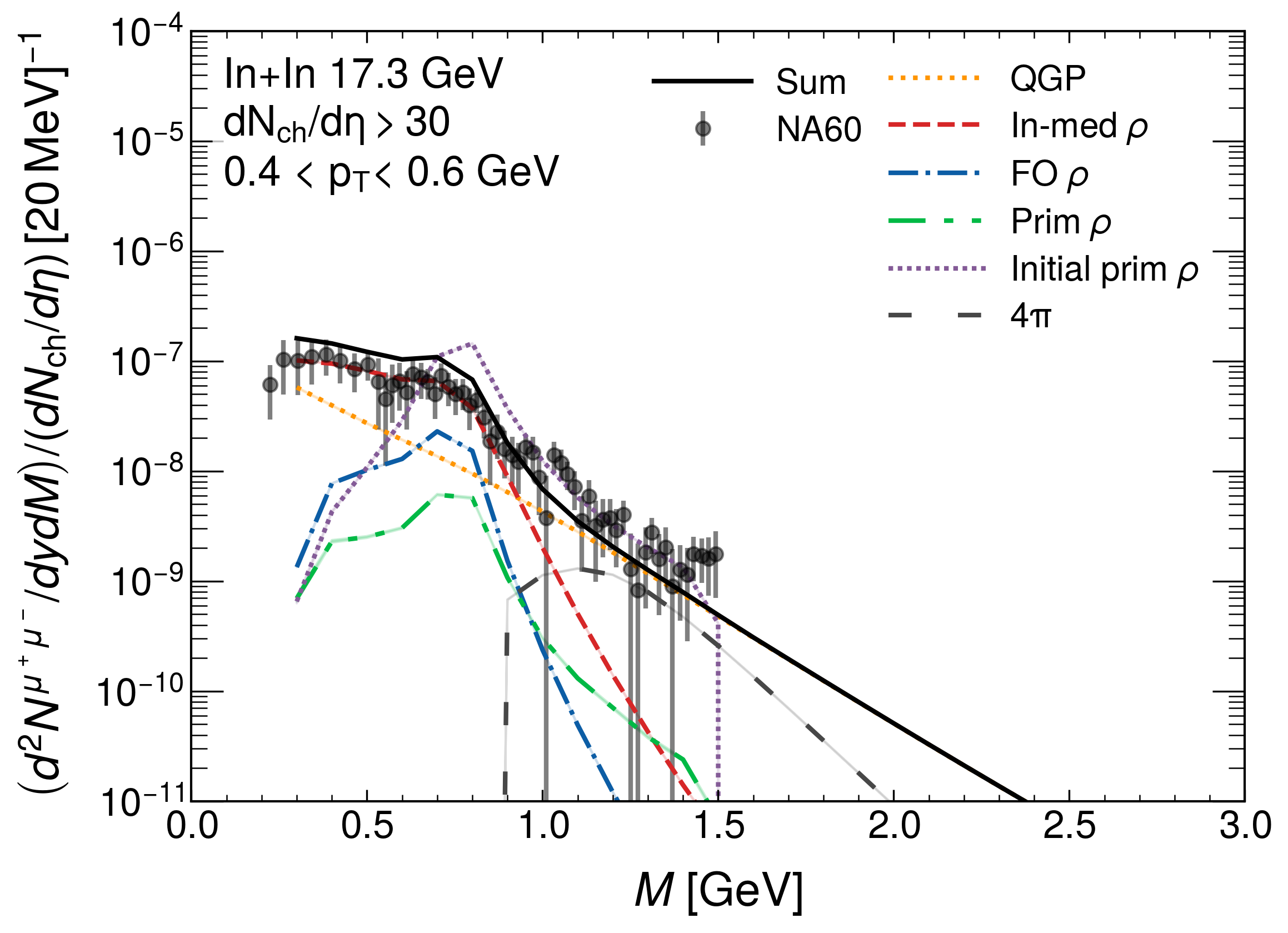}
    \includegraphics[width=0.45\linewidth]{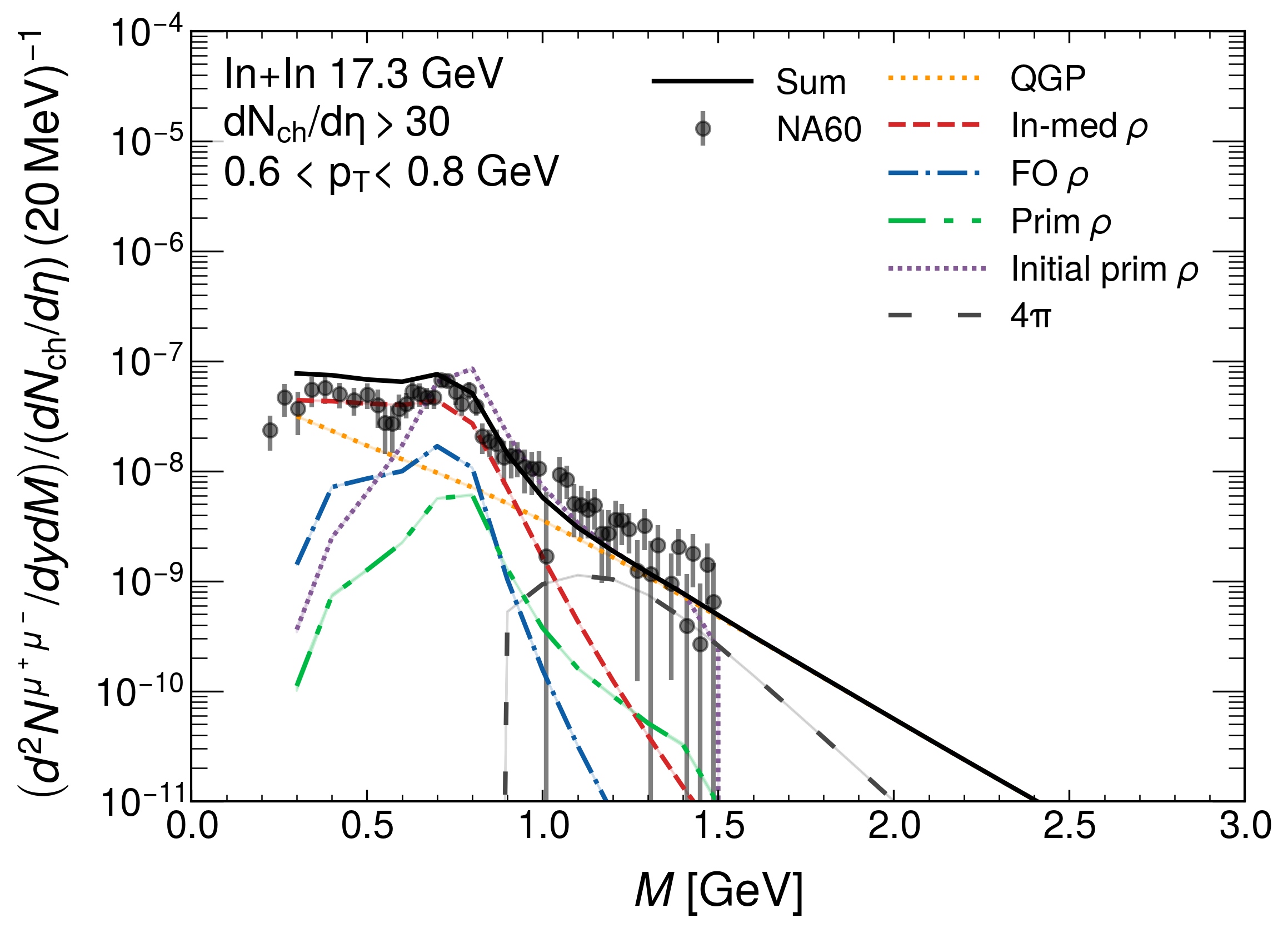}
    \includegraphics[width=0.45\linewidth]{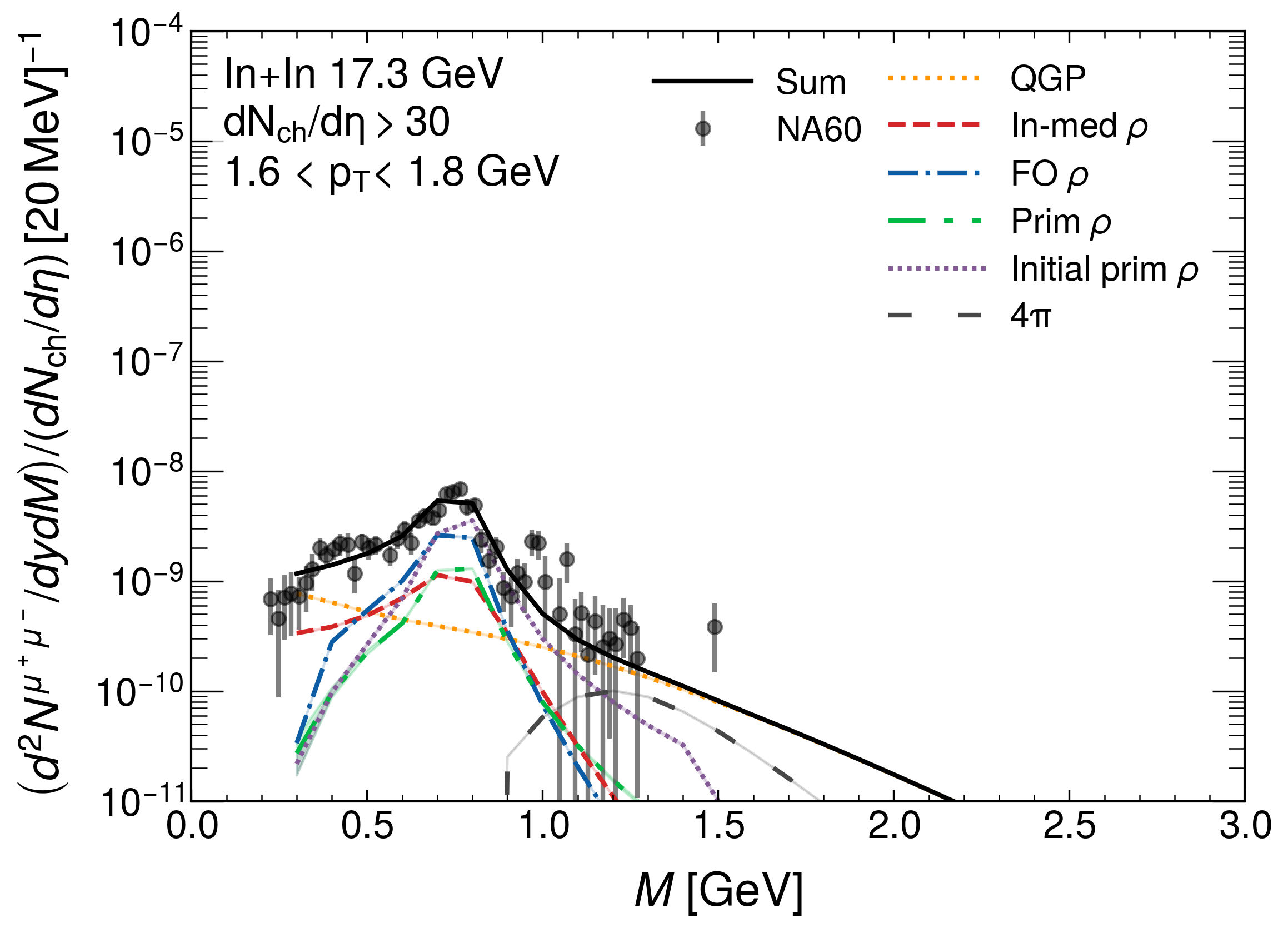}
    \includegraphics[width=0.45\linewidth]{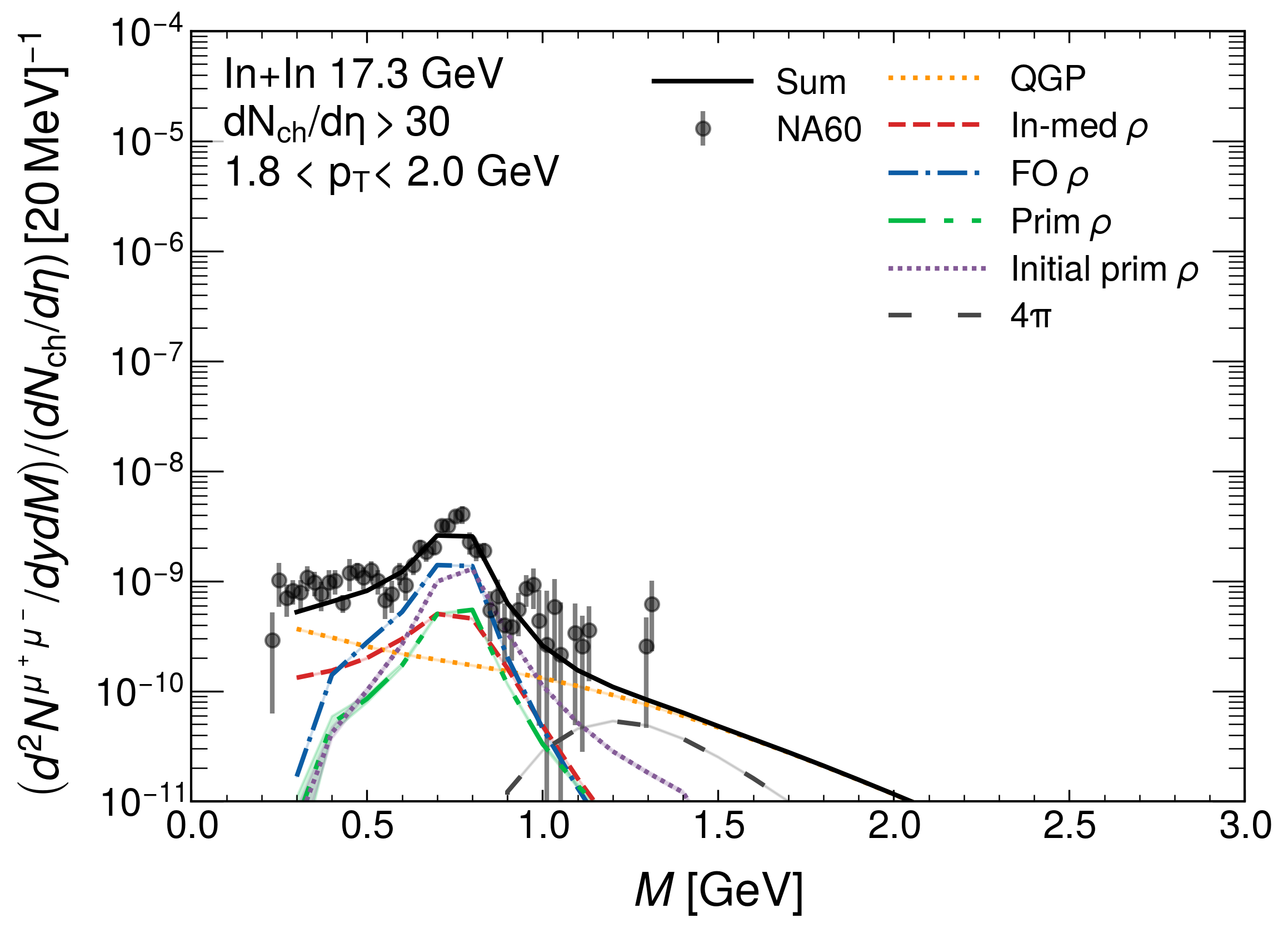}
 \caption{Same caption as that for  Fig.~\ref{fig:InIn17p3_dilepton_dNdM_normalized},
but for six selected $p_T$ intervals. The NA60 data
are taken from Ref.~\cite{NA60:2008ctj}. }
	\label{fig:dNdM_normalized_pT}
\end{figure*}

Near the $\rho$-meson mass, the in-medium $\rho$ provides the largest contribution, followed by the freeze-out $\rho$. The QGP radiation and the surviving primordial $\rho$ are smaller but comparable in magnitude. The nonthermal $\rho$ sources therefore make a non-negligible contribution to the yield in this region. The initial primordial $\rho$ spectrum alone would overestimate the measured yield near the $\rho$ mass. After the hadronic evolution, only a small fraction of the initially produced primordial $\rho$ mesons escape the hadronic medium without undergoing any collisions, leading to a strong suppression of this component. The freeze-out and surviving primordial $\rho$ contributions have narrower mass distributions than the in-medium component because their spectra are described using vacuum-like $\rho$ spectral functions.

As the invariant mass increases, the $\rho$-meson contributions decrease rapidly and QGP radiation becomes dominant in the IMR. This region is particularly sensitive to the early, high-temperature stage of the evolution~\cite{Wu:2024pba,Coquet:2021lca}, although four-pion radiation remains appreciable. For $1.5<M<2.5$~GeV, the calculation tends to underestimate the NA60 data. At SPS energies, the finite nuclear overlap time can lengthen the pre-equilibrium stage and increase its contribution to the dilepton yield. Radiation from this stage is absent from the present calculation and may partly account for this discrepancy.

These results indicate that a quantitative and unified description of the dilepton invariant-mass spectrum from the LMR to the IMR requires both an accurate description of the space-time evolution and a comprehensive treatment of the relevant thermal and nonthermal sources.

Figure~\ref{fig:dNdM_normalized_pT} presents the excess dimuon invariant-mass spectra in selected $p_T$ intervals, providing a more differential view of the source composition than the $p_T$-integrated spectrum shown in Fig.~\ref{fig:InIn17p3_dilepton_dNdM_normalized}. Overall, the calculation provides a good description of the NA60 data from the lowest interval, $p_T<0.2$~GeV, to the highest, $1.8<p_T<2.0$~GeV.

At low $p_T$, the LMR is dominated by the in-medium $\rho$ contribution, whose mass distribution is strongly broadened and has substantial strength below the $\rho$ pole. As $p_T$ increases, the strength below the pole becomes less prominent relative to the peak, and the total spectrum develops a more pronounced $\rho$-like shape. In our source decomposition, this change reflects a decreasing relative contribution from the in-medium $\rho$ and an increasing importance of the nonthermal $\rho$ components. In particular, for $p_T>1.6$~GeV, the freeze-out $\rho$ contribution exceeds the in-medium component around the $\rho$-meson mass and becomes the largest contribution in the pole region.

\begin{figure*}[htb]
	\centering
    \includegraphics[width=0.45\linewidth]{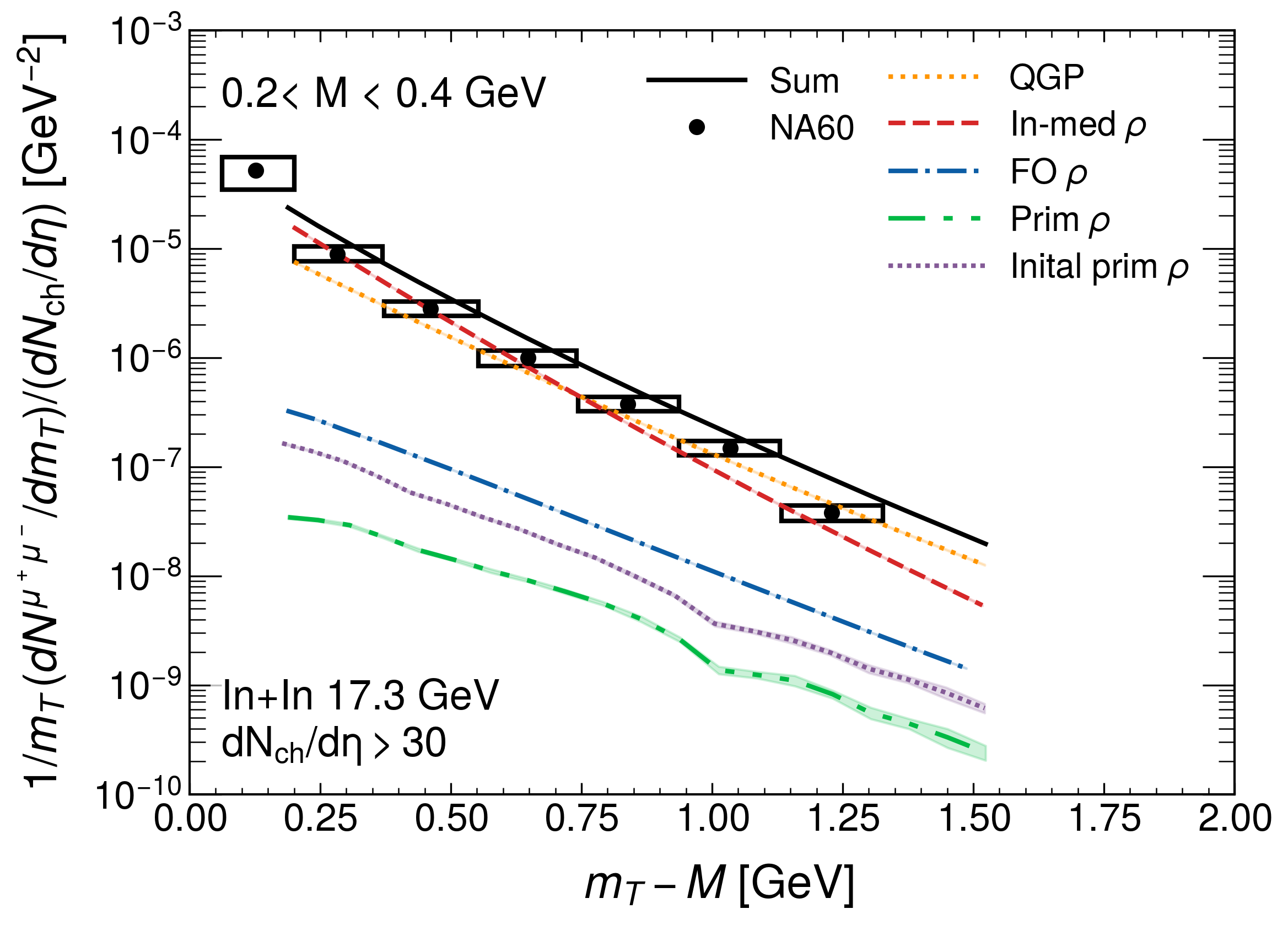}
	\includegraphics[width=0.45\linewidth]{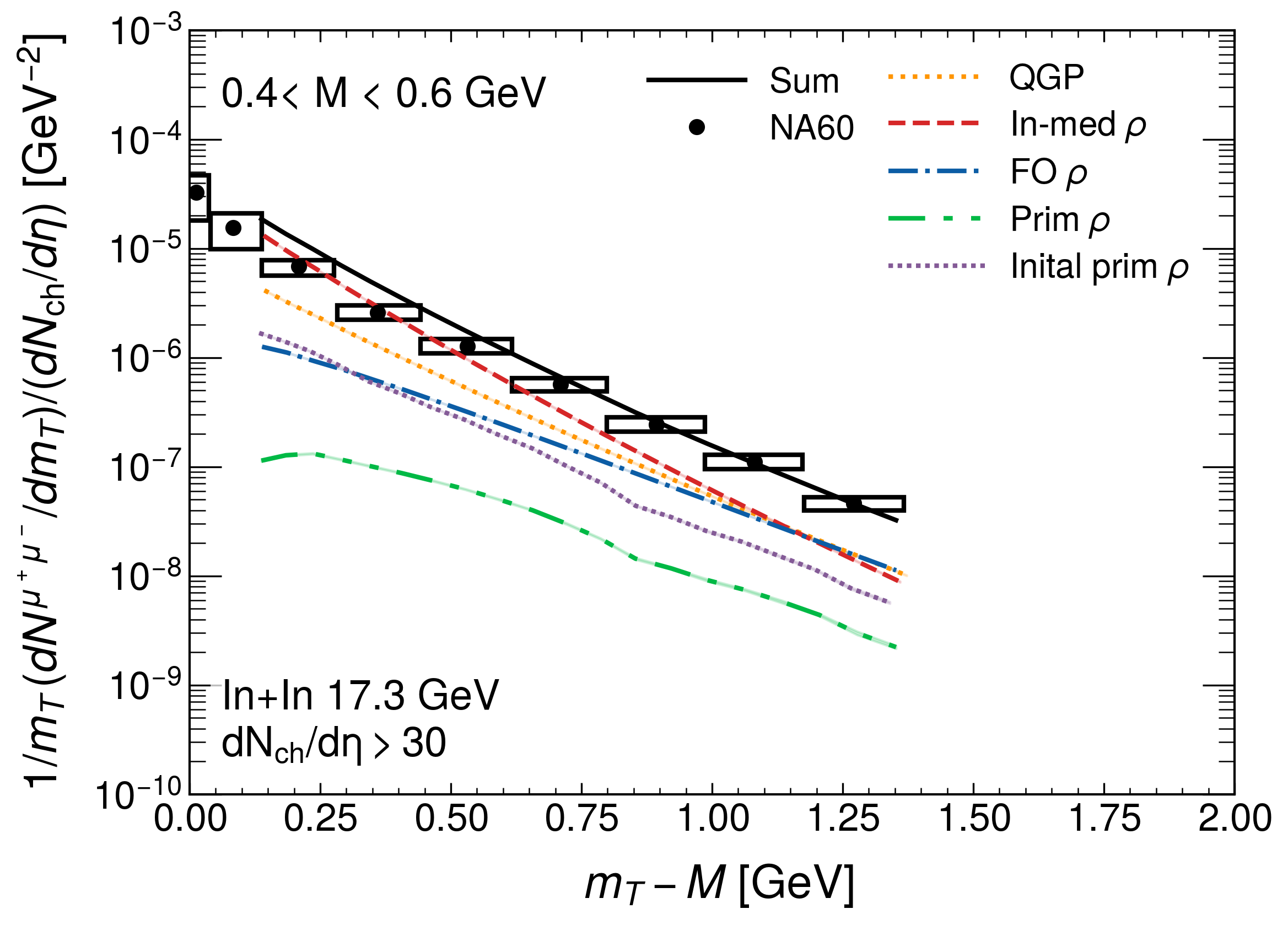}
    \includegraphics[width=0.45\linewidth]{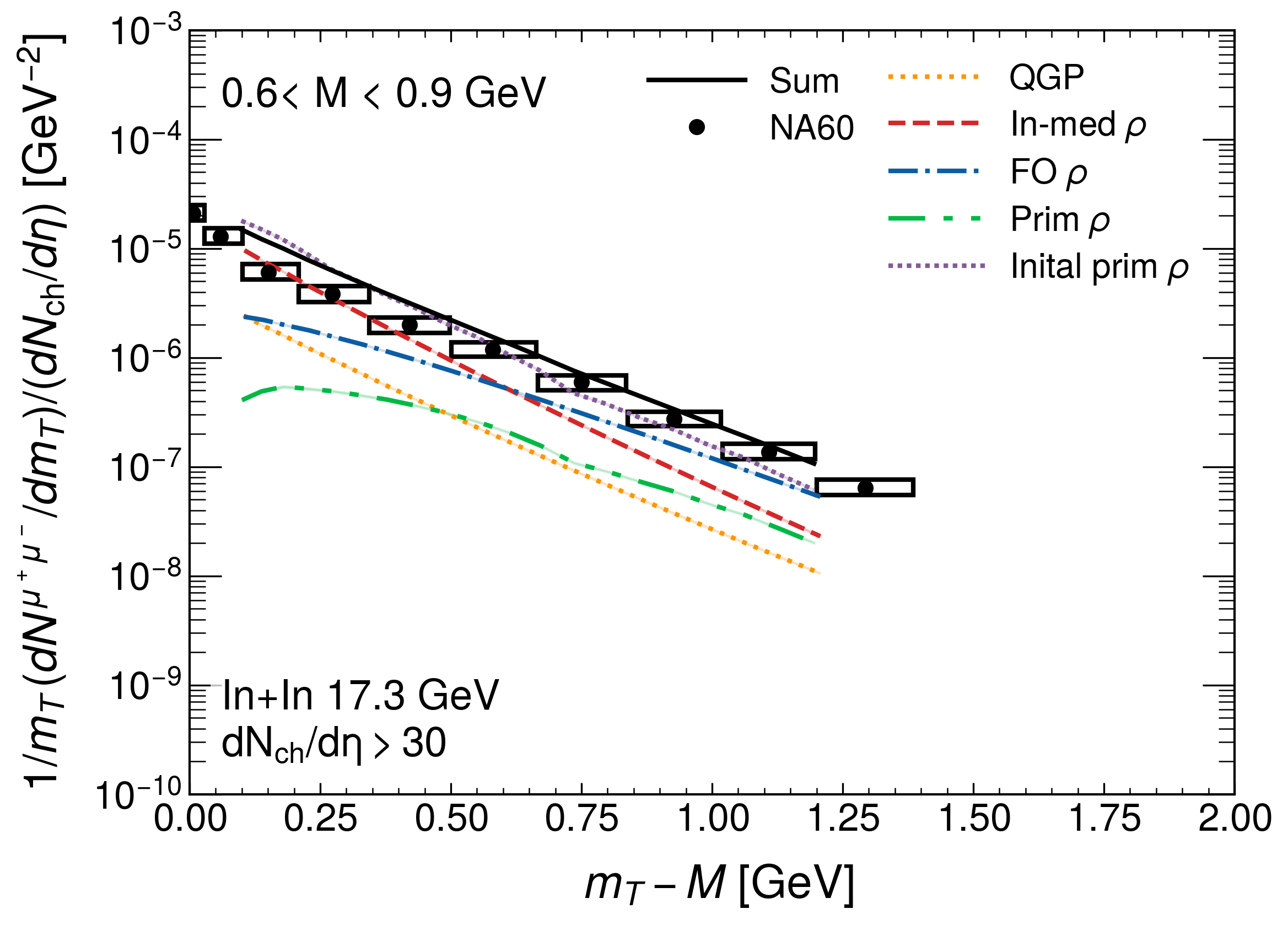}
    \includegraphics[width=0.45\linewidth]{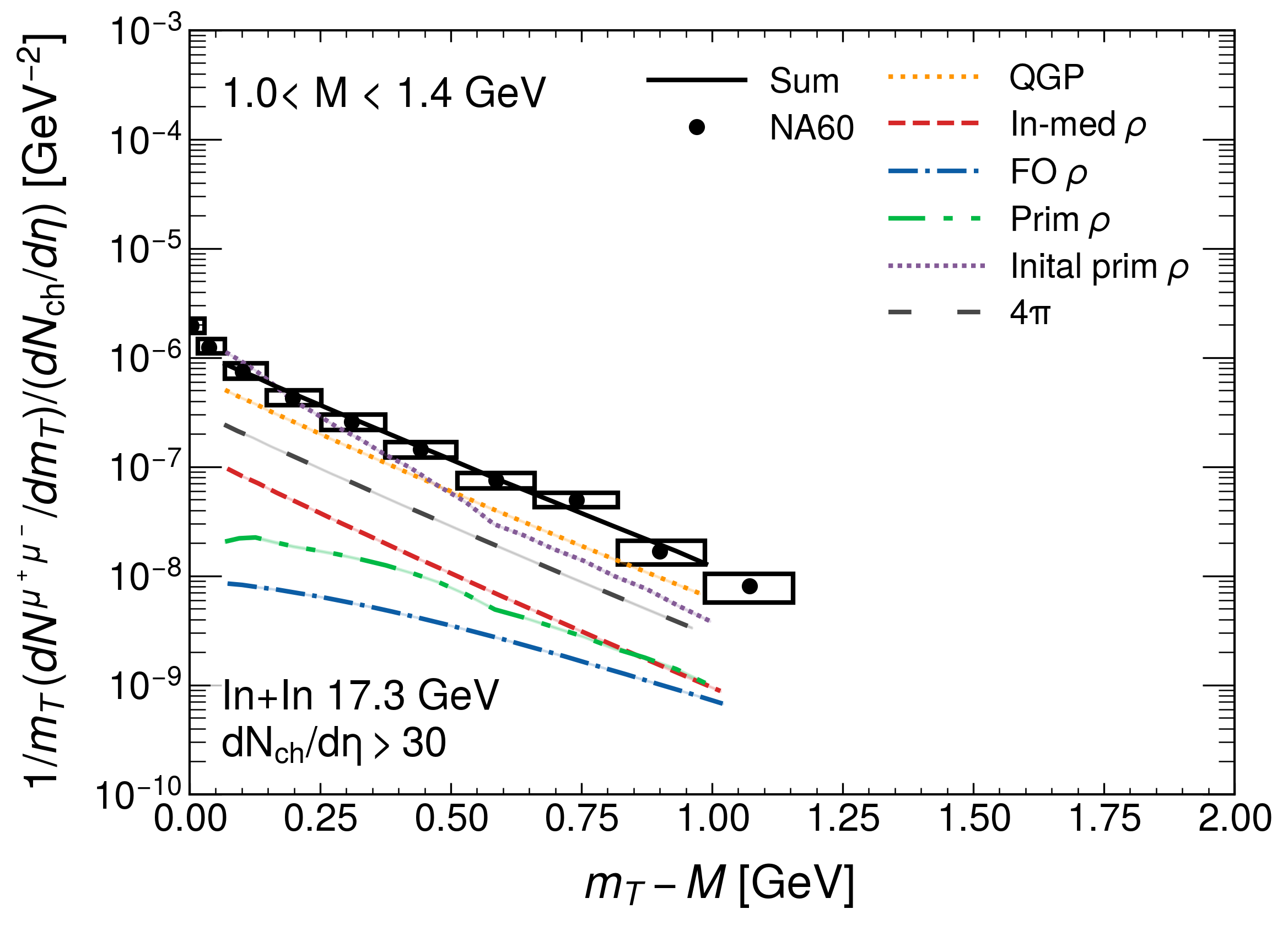}
 \caption{Excess dimuon transverse-mass spectra in four invariant-mass intervals in In+In collisions at $\sqrt{s_{\rm NN}}=17.3$~GeV with $dN_{\rm ch}/d\eta>30$. The NA60 data are taken from Ref.~\cite{NA60:2008ctj}.}
	\label{fig:dilepton_mT_spectrum}
\end{figure*}

The surviving primordial $\rho$ contribution also gains relative importance as $p_T$ increases. At low $p_T$, this component is strongly suppressed by hadronic interactions. Within our transport treatment, low-$p_T$ primordial $\rho$ mesons are less likely to escape the hadronic medium without a collision, whereas their survival probability increases toward higher $p_T$. Consequently, the surviving primordial contribution becomes comparable in magnitude to the in-medium $\rho$ contribution in the highest-$p_T$ intervals. Therefore, nonthermal $\rho$ decays make an important contribution, particularly around the $\rho$-meson mass.

In the IMR, QGP radiation remains the dominant source over the full $p_T$ range considered here, while four-pion radiation provides the leading hadronic contribution. The invariant-mass spectrum in this region becomes progressively flatter as $p_T$ increases. Selecting higher-$p_T$ pairs favors harder dilepton radiation and can increase the relative weight of emission from hotter stages. However, the slope of a $p_T$-selected mass spectrum cannot be interpreted directly as a single medium temperature, because the selection also introduces kinematic and collective-flow effects.

Another feature appears at low invariant mass and high $p_T$. 
For $1.6<p_T<1.8$~GeV and $1.8<p_T<2.0$~GeV,
QGP radiation becomes the largest contribution at sufficiently low $M$, providing a region with enhanced sensitivity to the partonic emission rate. Since NLO corrections substantially enhance the partonic dilepton rate at low invariant mass relative to the leading-order result, the choice of QGP rate becomes particularly relevant in this region.

Figure~\ref{fig:dilepton_mT_spectrum} shows the excess dimuon transverse-mass spectra, with $m_T = \sqrt{M^2+p_T^2}$, in four invariant-mass intervals for In+In collisions with $dN_{\rm ch}/d\eta>30$. The individual dilepton sources and their sum are compared with the NA60 data. Overall, the calculated spectra describe the data within the experimental uncertainties in all four mass intervals.

In the two lowest mass intervals, the in-medium $\rho$ provides the largest contribution at small $m_T-M$. 
Its spectrum falls more steeply than the QGP spectrum, so QGP radiation exceeds the in-medium $\rho$ at larger $m_T-M$.
For $0.6<M<0.9$~GeV, the freeze-out $\rho$ becomes the largest contribution at high $m_T-M$.
In the highest mass interval, $1.0<M<1.4$~GeV, QGP radiation dominates over most of the $m_T-M$ range, while four-pion radiation provides the largest hadronic contribution. 
The $\rho$-meson sources contribute much less in this mass interval. Comparison of the initial and surviving primordial $\rho$ spectra shows that hadronic scattering strongly suppresses the surviving
yield, particularly at small $m_T-M$.

For thermal radiation, the different slopes of the $m_T$ spectra provide additional information on the emission temperature and collective radial flow~\cite{vanHees:2007th}. We extract the effective temperature $T_{\rm eff}$ by fitting the spectra with
\begin{equation}
    \frac{1}{m_T}\frac{dN}{dm_T}
    =
    C\exp\left(-\frac{m_T}{T_{\rm eff}}\right),
    \label{eq:Teff_fit}
\end{equation}
using the same fit range as the NA60 analysis~\cite{NA60:2008ctj},
$0.4<p_T<1.8$~GeV.

\begin{figure}[htb]
	\centering
	\includegraphics[width=1\linewidth]{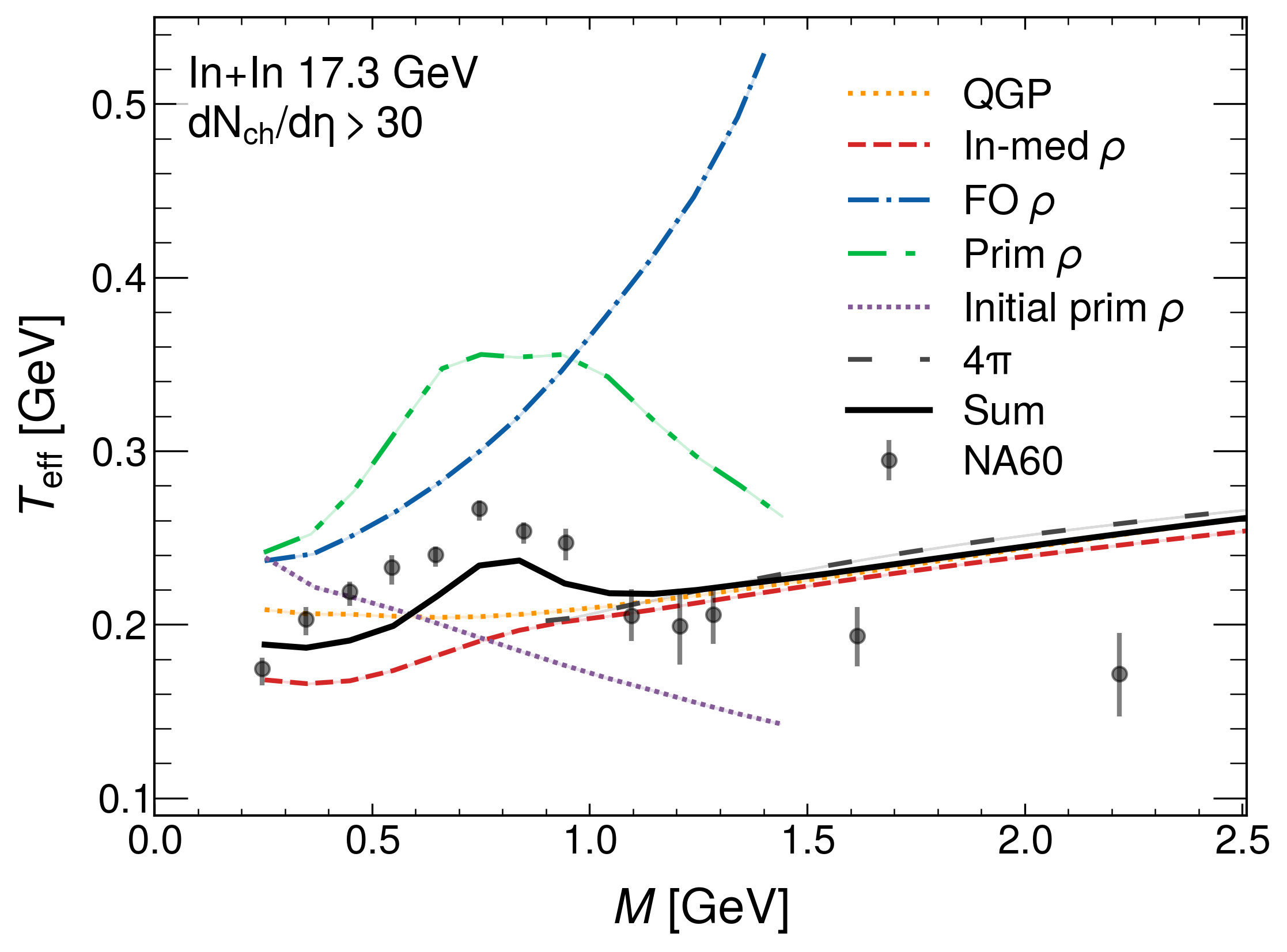}
 \caption{Effective temperature $T_{\rm eff}$ as a function of invariant mass in In+In collisions at $\sqrt{s_{\rm NN}}=17.3$~GeV with $dN_{\rm ch}/d\eta>30$. The $m_T$ spectra of the individual sources and their sum are fitted using Eq.~(\ref{eq:Teff_fit}) over $0.4<p_T<1.8$~GeV. The NA60 data are taken from Ref.~\cite{NA60:2008ctj}. }
	\label{fig:Teff_vs_M}
\end{figure}

Figure~\ref{fig:Teff_vs_M} shows $T_{\rm eff}$ as a function of $M$ for the individual dilepton sources and their sum, compared with the NA60 data. The total $T_{\rm eff}$ is extracted by fitting the sum of the source $m_T$ spectra with Eq.~(\ref{eq:Teff_fit}).  

The mass dependence of $T_{\rm eff}$ for thermal radiation can be understood using the small-$p_T$ approximation. For $M \gg p_T \gg T$  and weak transverse flow, the effective temperature approximately follows
\begin{equation}
    T_{\rm eff}
    \simeq
    T+\frac{M}{2}\left\langle v_\perp^2\right\rangle ,
    \label{eq:Teff_lowpt}
\end{equation}
where $T$ is the emission temperature and $\langle v_\perp^2\rangle$ is the mean squared transverse
flow velocity~\cite{vanHees:2007th}. 
Since our fit range, $0.4<p_T<1.8$~GeV, extends beyond this limit, we only use Eq.~(\ref{eq:Teff_lowpt}) as a qualitative guide. The QGP radiation originates mainly from the earlier, hotter stage of the collision, when less radial flow has developed. Its higher emission temperature gives a sizable $T_{\rm eff}$, while the weaker flow leads to a mild increase with $M$. 
The in-medium $\rho$, in contrast, is emitted mainly during the hadronic evolution, where the temperature is lower but radial flow is stronger. Its $T_{\rm eff}$ therefore rises more rapidly with $M$. Four-pion radiation shows a similar trend because it also originates from the hadronic
stage.

The freeze-out $\rho$ exhibits the strongest increase of $T_{\rm eff}$ with $M$. By kinetic freeze-out, transverse flow has reached its largest value, producing a strong mass-dependent blue shift. Relative to continuous in-medium emission, the freeze-out dilepton spectrum also contains
the additional Lorentz factor $p^0/M$ discussed in Sec.~\ref{sec:freezeout_rho}, which further hardens the spectrum~\cite{Ruppert:2007cr,vanHees:2007th}.

The primordial $\rho$ contribution is nonthermal and is not described by Eq.~(\ref{eq:Teff_lowpt}). In the initial primordial sample, the same PYTHIA $p_T$ distribution is used for all sampled masses. At fixed $M$, the spectra satisfy $(1/p_T)dN/dp_T=(1/m_T)dN/dm_T$.
If $(1/p_T)dN/dp_T$ is locally approximated by $C\exp(-p_T/T_p)$, the local effective temperature of the $m_T$ spectrum satisfies
\begin{equation}
    T_{\rm eff,loc}^{\rm init}
    \simeq
    T_p\frac{p_T}{m_T}.
    \label{eq:Teff_prim_init}
\end{equation}
As $M$ increases, the common $p_T$ fit range corresponds to a narrower $m_T$ interval. The same relative decrease of the spectrum then occurs over a smaller $m_T$ interval, giving a steeper $m_T$ spectrum. This helps explain the decrease of the fitted $T_{\rm eff}^{\rm init}(M)$ shown in Fig.~\ref{fig:Teff_vs_M}.

The subsequent hadronic evolution changes this mass dependence. Low-$p_T$ primordial $\rho$ mesons have a smaller probability of surviving without a hadronic collision and are preferentially removed from the surviving sample, as reflected in Fig.~\ref{fig:dilepton_mT_spectrum}. The resulting $p_T$ spectrum is harder than the initial one, raising  $T_{\rm eff}$. The broad maximum of the surviving primordial contribution reflects the competition between this momentum-dependent suppression and the decreasing $T_{\rm eff}$ of the initial primordial spectrum.

None of the thermal sources alone produces the clear maximum around the $\rho$ mass observed by NA60. The freeze-out and surviving primordial $\rho$ contributions have relatively large effective temperatures in this region and raise the total effective temperature around the $\rho$ pole. 
The total calculation develops a similar peak structure, although its magnitude remains below the experimental data. Since the effective temperature in this mass region is strongly affected by the late-stage transverse expansion, larger radial flow would increase the calculated $T_{\rm eff}$.
However, this interpretation is not unique. Previous SPS studies have shown that the equation of state and hadrochemical evolution can also modify the effective temperature~\cite{vanHees:2007th}, while the treatment of the freeze-out stage provides an additional uncertainty.
Understanding the remaining difference with the NA60 data in the LMR may require a more detailed study of the late hadronic evolution.

\begin{figure*}[htb]
	\centering
    \includegraphics[width=0.45\linewidth]{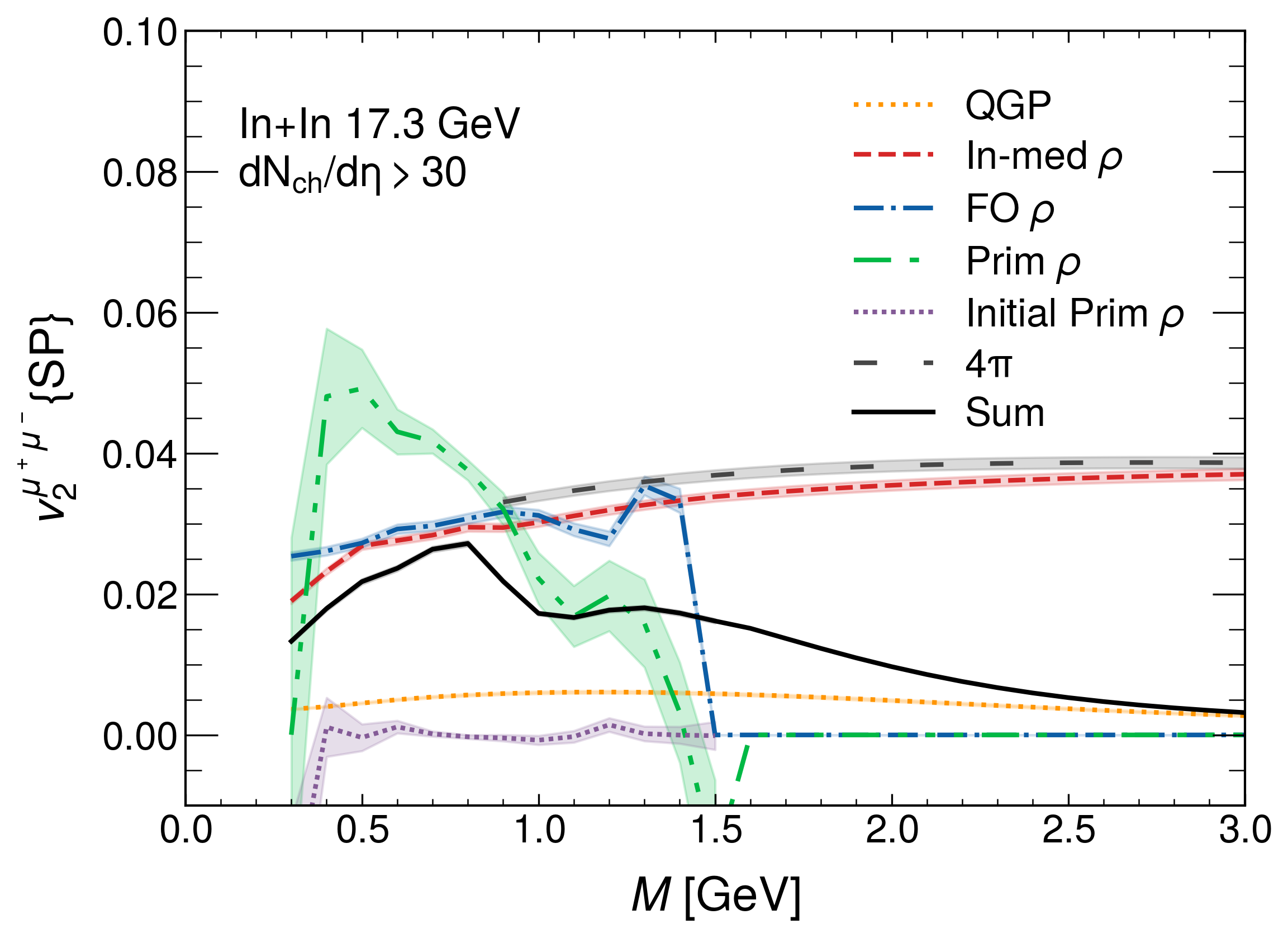}
	\includegraphics[width=0.45\linewidth]{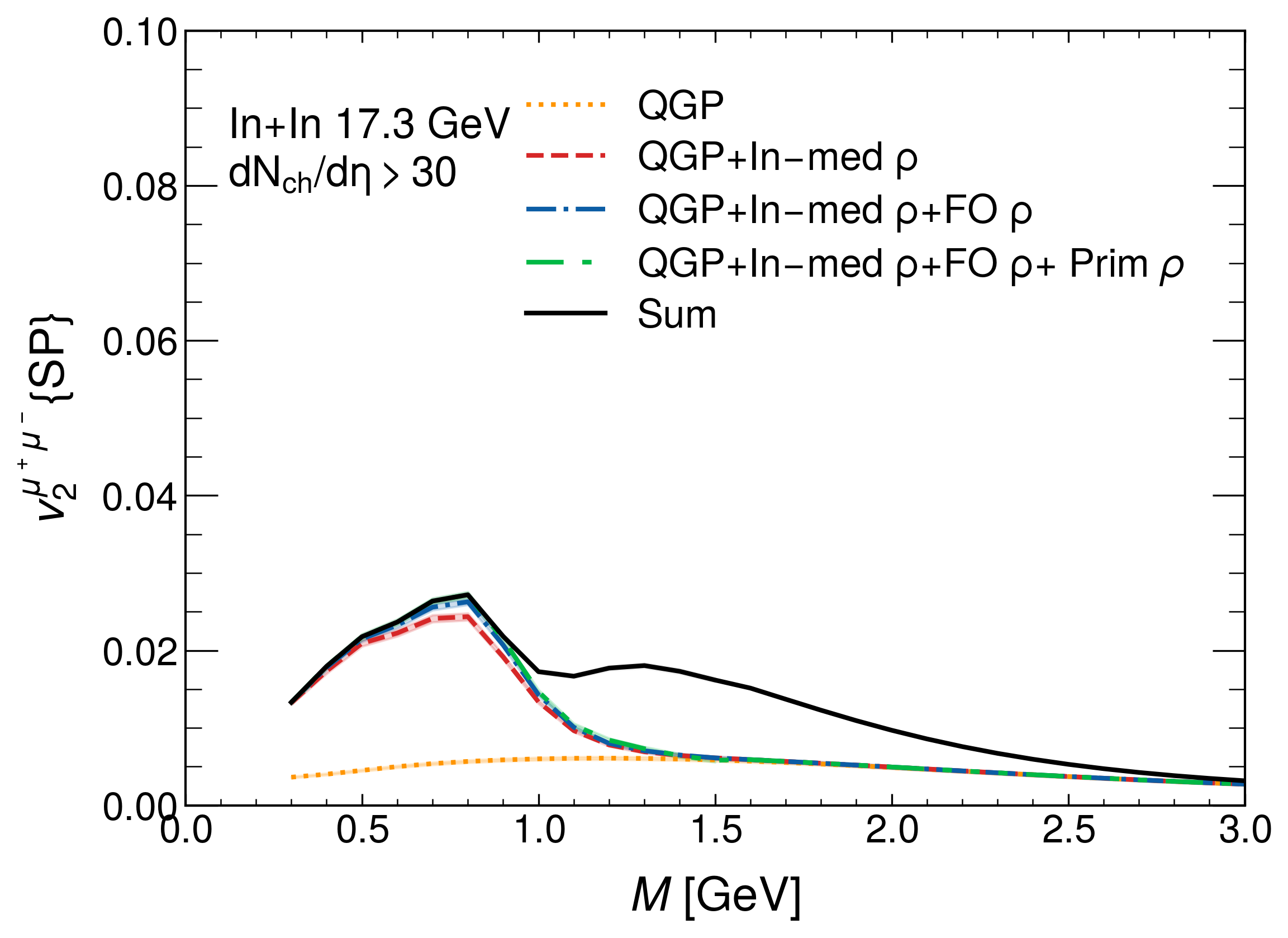}
 \caption{
Dilepton elliptic flow $v_2\{\mathrm{SP}\}$ as a function of invariant mass for $p_T>0.2$~GeV in In+In collisions at $\sqrt{s_{\rm NN}}=17.3$~GeV with $dN_{\rm ch}/d\eta>30$. 
The left panel shows the individual sources and their total. The right panel shows the combined $v_2$ as the hadronic and nonthermal sources are successively added to QGP radiation. 
In the right panel, the difference between the green dash-dotted curve and the sum (black solid curve) is due to
four-pion radiation.}
	\label{fig:InIn17p3_dilepton_v2_SP}
\end{figure*}

In the IMR, $1.5<M<2.5$~GeV, the dilepton yield in our calculation is dominated by QGP radiation, and the total $T_{\rm eff}$ closely follows the QGP contribution. The calculated effective temperature increases over this mass range, whereas the NA60 data show a much weaker dependence on $M$. 
With increasing invariant mass, emission from cooler stages is more strongly suppressed, and the QGP
yield receives a larger contribution from earlier, hotter times. Less transverse flow has developed at these early times, reducing the radial-flow blue shift. 
In our calculation, the higher emission temperature has the larger effect, producing the increase in $T_{\rm eff}$ with $M$.

Pre-equilibrium dilepton radiation, which is not included in the present calculation, may further modify both the IMR yield and the extracted $T_{\rm eff}$.
Its effect on $T_{\rm eff}$ cannot be inferred from the small early-time radial flow alone, since the anisotropic quark momentum distribution and the evolving quark abundance also affect the dilepton spectrum~\cite{Wu:2024pba,Coquet:2021lca}.
A calculation including pre-equilibrium radiation is needed to assess its effect on the comparison with the NA60 data.

The weaker mass dependence of the IMR $T_{\rm eff}$ measured by NA60 may reflect uncertainties in the early evolution of baryon-rich matter. 
The equation of state at finite baryon chemical potential and bulk viscosity both influence the temperature history and the development of transverse flow~\cite{Monnai:2019hkn,Vujanovic:2019yih,Shen:2023awv}.
Comparing calculations with varied equations of state and transport coefficients to dilepton and hadronic observables could help clarify the origin of the remaining difference in $T_{\rm eff}$.

\subsection{Dilepton elliptic flow}

An important advantage of the event-by-event multistage hybrid framework is that fluctuations of the initial collision geometry are propagated through the hydrodynamic evolution and converted into final-state momentum anisotropies. 
This allows the dilepton elliptic flow to be evaluated with the scalar-product method using charged hadrons as the reference: 
\begin{equation}
    v_2^{\ell^{+}\ell^{-}}\{{\rm SP}\}(M)
    =
    \frac{
    \left\langle
    v_2^{\ell^+\ell^-}(M)v_2^h
    \cos\left[
    2\left(\Psi_2^{\ell^+\ell^-}(M)-\Psi_2^h\right)
    \right]
    \right\rangle
    }{
    \sqrt{\left\langle (v_2^h)^2\right\rangle}
    },
    \label{eq:dilepton_v2_SP}
\end{equation}
where the charged-hadron reference flow is evaluated with $3<\eta<4$ and $0.2<p_T<3.0$~GeV, while dileptons with $p_T>0.2$~GeV are included. 
Figure~\ref{fig:InIn17p3_dilepton_v2_SP} shows the dilepton  $v_2$ in In+In collisions at $\sqrt{s_{\rm NN}}=17.3$~GeV with $dN_{\rm ch}/d\eta>30$. The left panel shows the elliptic flow of the individual sources,  whereas the right panel shows how their combined $v_2$ changes as the contributions are successively added.

The QGP contribution has the smallest elliptic flow over most of the mass range. This is expected because QGP radiation originates from the early stage of the collision, before a sizable momentum anisotropy has developed. The in-medium $\rho$ and four-pion contributions have larger $v_2$, reflecting the stronger collective flow during the hadronic evolution. The freeze-out $\rho$ has a sizable $v_2$ comparable to that of the in-medium $\rho$ in the LMR.

The primordial $\rho$ shows a different behavior. Before the hadronic evolution, its elliptic flow is nearly zero, as expected for production from the initial binary collisions with an azimuthally isotropic momentum distribution. After the SMASH evolution, the surviving primordial component exhibits a sizable positive $v_2$. In our calculation, the surviving primordial $\rho$ mesons undergo no hadronic collisions, so this anisotropy does not arise from momentum transfer from the medium. 
Instead, it is generated by the azimuthal dependence of the survival probability: primordial $\rho$ mesons propagating along different directions traverse different amounts of hadronic matter and have different probabilities of escaping without a collision. This mechanism is analogous to the path-length-dependent suppression that generates a positive high-$p_T$ anisotropy for hard probes.

The cumulative results in the right panel show how the individual sources affect the total dilepton $v_2$. 
In the LMR, adding the in-medium $\rho$ produces the largest change, consistent with its dominant contribution to the dilepton yield. 
The freeze-out and surviving primordial $\rho$ components also have sizable $v_2$ around the
$\rho$ mass, but their smaller yields limit their effects on the total flow.

A qualitatively different situation occurs in the IMR. Although QGP radiation dominates the dilepton yield in this mass region, its elliptic flow is small. The four-pion contribution has a smaller yield but a substantially larger $v_2$, and consequently produces a sizable increase in the
combined elliptic flow. This increase is visible in the separation between the green curve and the black solid curve for $1<M<2.5$~GeV in the right panel of Fig.~\ref{fig:InIn17p3_dilepton_v2_SP}.
The four-pion channel forms an important part of the hadronic electromagnetic continuum in the IMR. The in-medium behavior of this continuum has been studied in connection with vector--axial-vector mixing and chiral symmetry restoration~\cite{vanHees:2007th}. In the present calculation, the sizable four-pion $v_2$ reflects the collective evolution of the hadronic phase.  A quantitative
extraction of the QGP anisotropy requires control of the four-pion contribution, even when its yield is subleading.

\section{Dilepton observables in isobar collisions at RHIC}
\label{sec:RHIC_results}
In the previous section, we studied dilepton observables in In+In collisions at SPS energies and obtained a good  description of the NA60 data.
We now turn to RHIC energies and consider the isobar systems Ru+Ru and Zr+Zr at $\sqrt{s_{\rm NN}}=200$~GeV. 
Since the calculated dilepton observables in the two isobar systems are nearly identical,
we present only the Ru+Ru results below. The In and Ru nuclei have comparable sizes.

\begin{figure}[htb]
	\centering
    \includegraphics[width=1.0\linewidth]{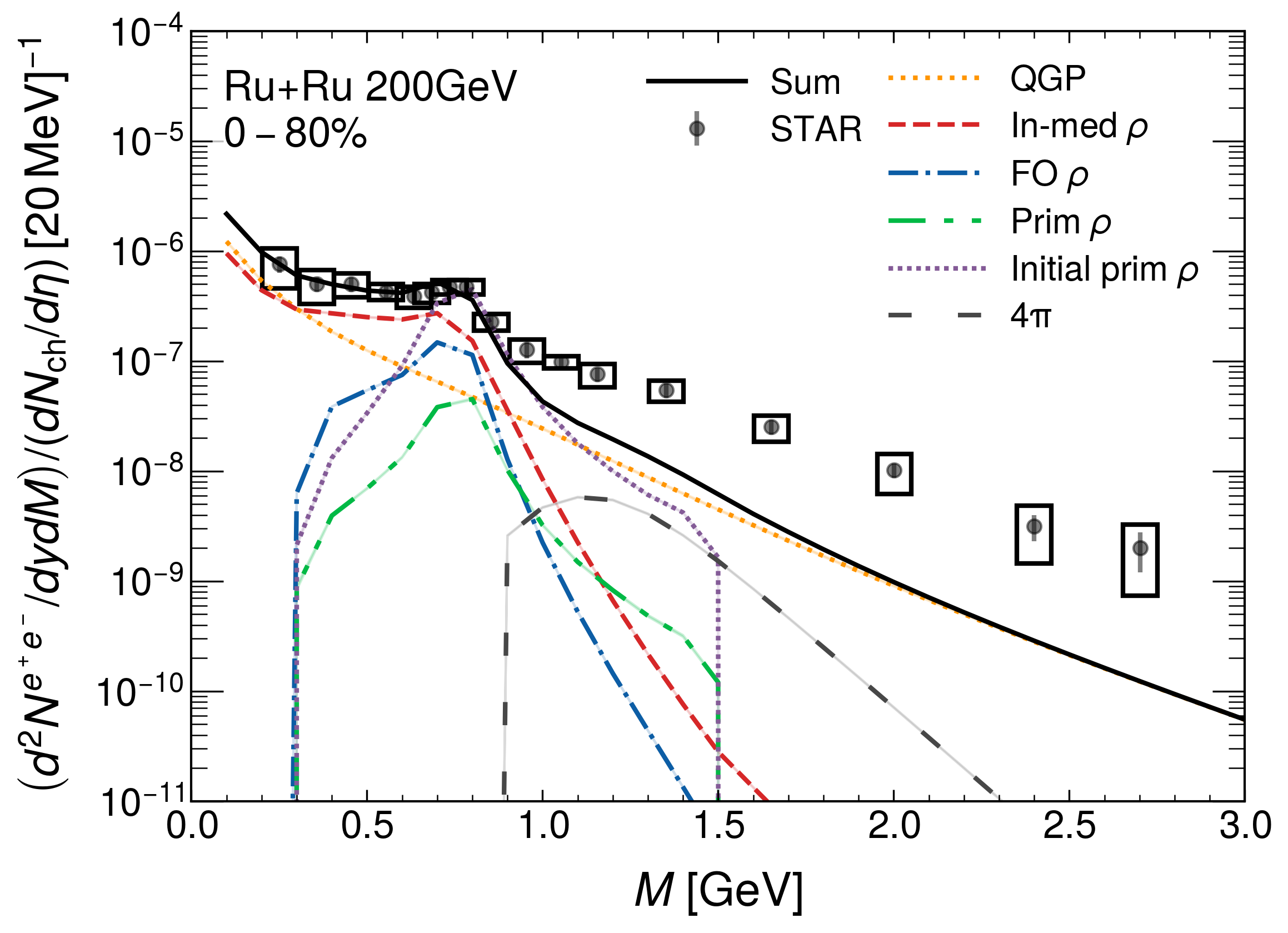}
\caption{
The $p_T$-integrated excess dielectron invariant-mass spectrum is shown, for Ru+Ru collisions at $\sqrt{s_{\rm NN}}=200$~GeV in the 0--80\% centrality window. The individual contributions and their sum are compared with the preliminary STAR isobar data~\cite{Luo:QM2025}.
}
	\label{fig:RuRu200_dielectron_dNdM_normalized}
\end{figure}

\begin{figure}[htb]
	\centering
    \includegraphics[width=1.0\linewidth]{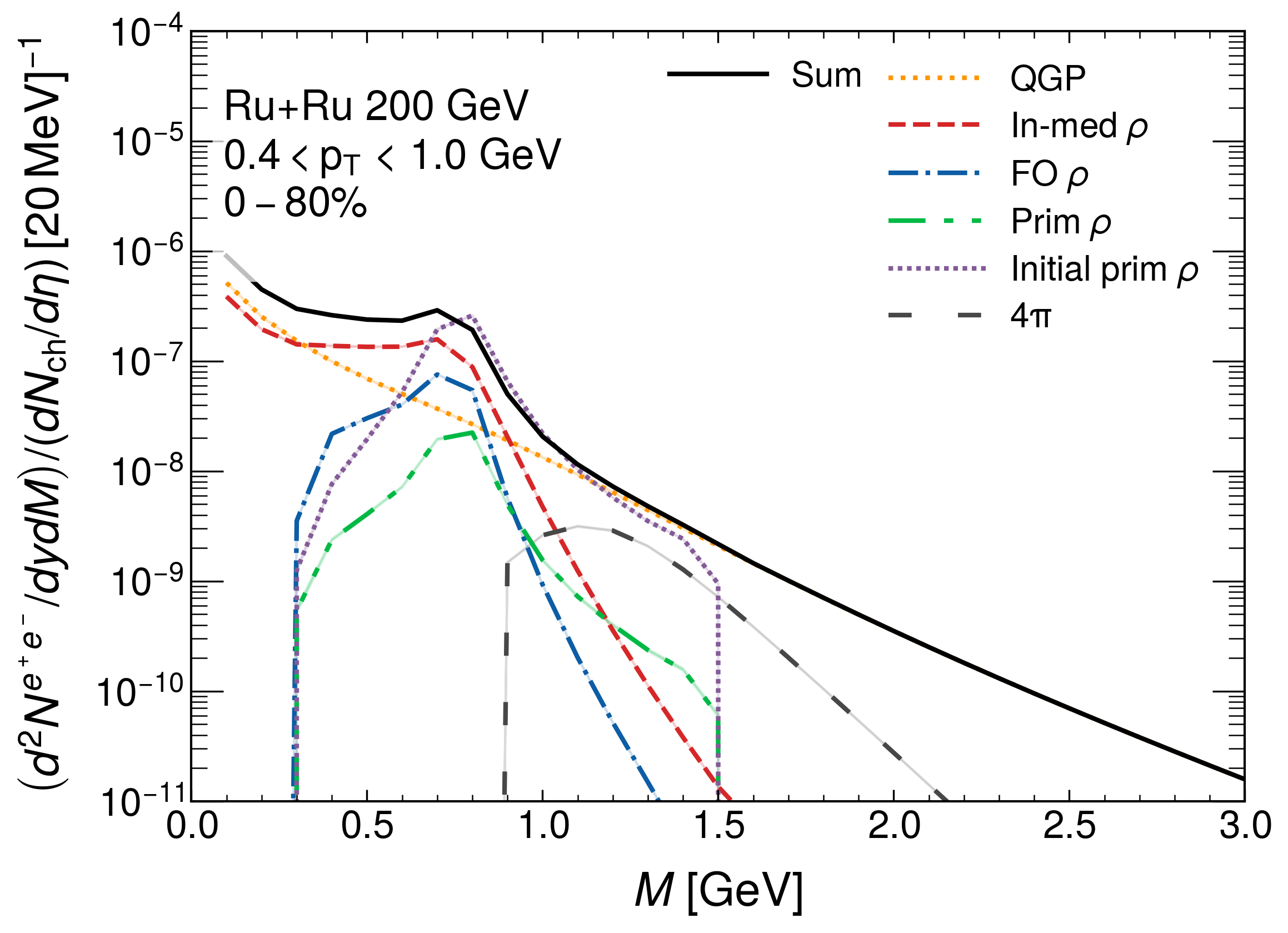}
    \includegraphics[width=1.0\linewidth]{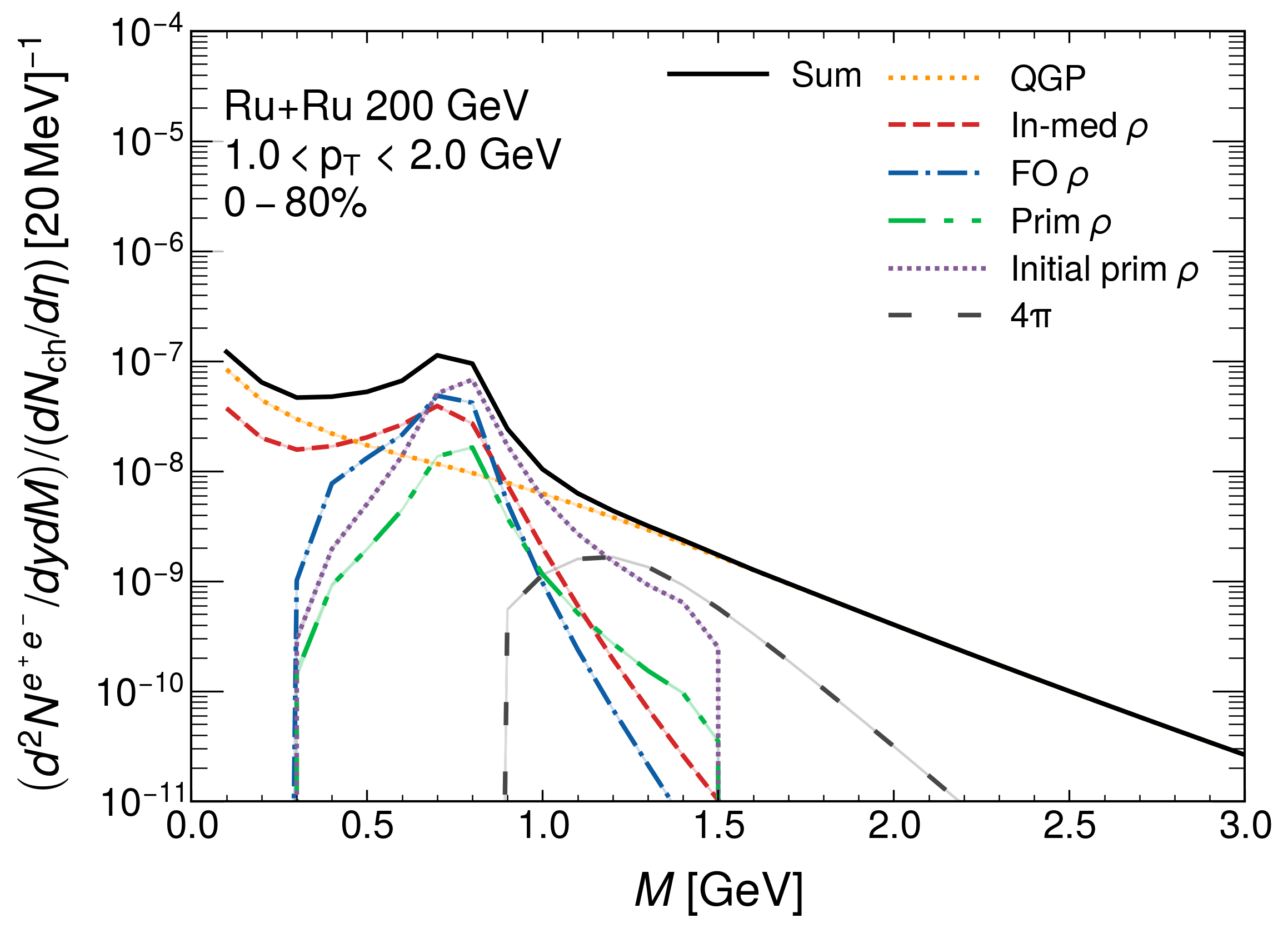}
    \includegraphics[width=1.0\linewidth]{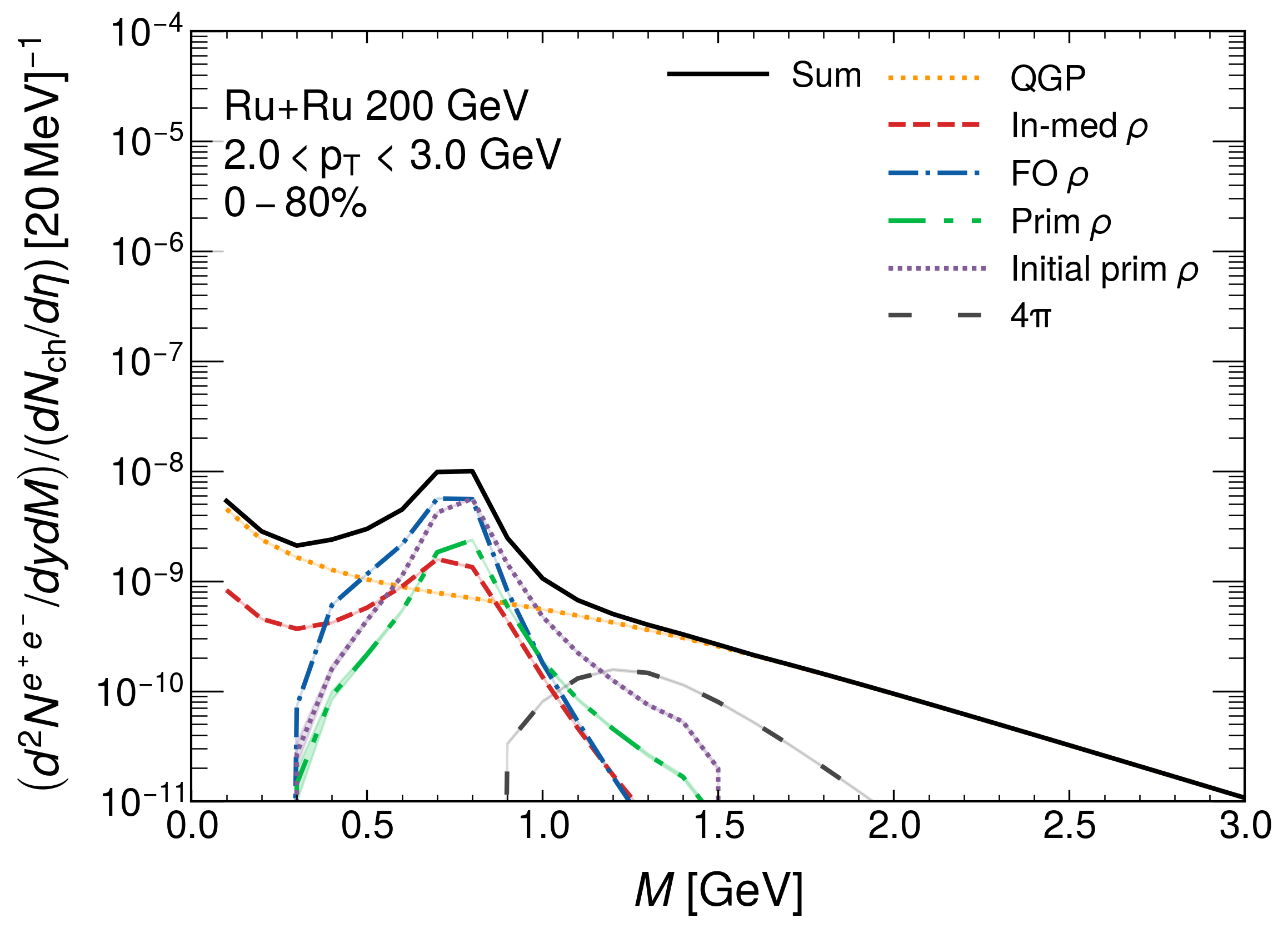}
\caption{
The excess dielectron invariant-mass spectra in 0--80\% Ru+Ru
collisions at $\sqrt{s_{\rm NN}}=200$~GeV for
$0.4<p_T<1.0$~GeV (top panel), $1.0<p_T<2.0$~GeV (middle panel),
and $2.0<p_T<3.0$~GeV (bottom panel).
}
	\label{fig:RuRu200_dilepton_dNdM_normalized_pT}
\end{figure}

Figure~\ref{fig:RuRu200_dielectron_dNdM_normalized} shows the $p_T$-integrated excess dielectron invariant-mass spectrum in 0--80\% Ru+Ru collisions at $\sqrt{s_{\rm NN}}=200$~GeV. 
The individual contributions and their sum are compared with the preliminary STAR isobar data \cite{Luo:QM2025}. As in
Fig.~\ref{fig:InIn17p3_dilepton_dNdM_normalized}, the spectrum is normalized by $dN_{\rm ch}/d\eta$, putting the In+In and Ru+Ru spectra on the same footing.

\begin{figure}[htb]
	\centering
    \includegraphics[width=1.0\linewidth]{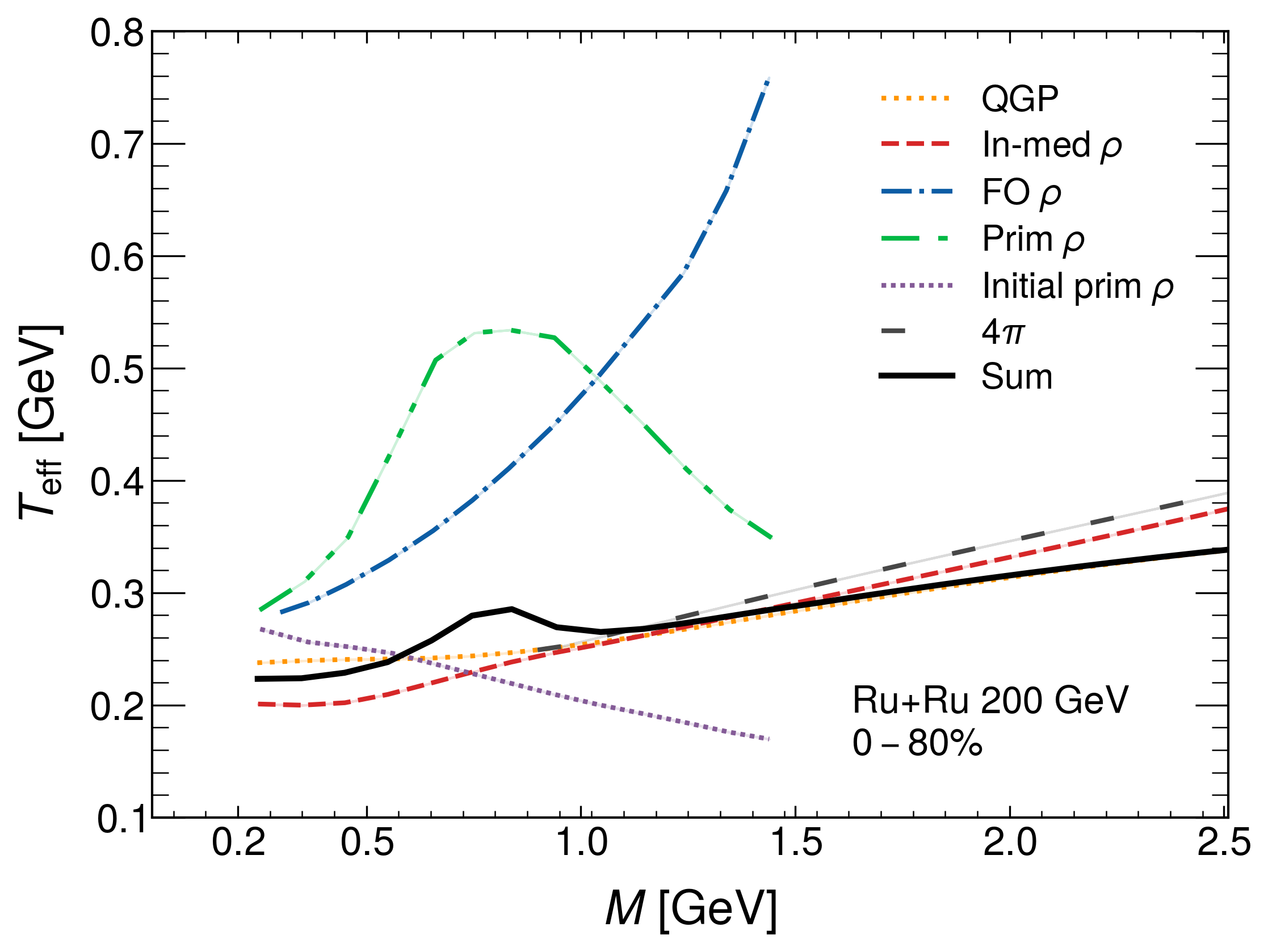}
\caption{
Effective temperature $T_{\rm eff}$ as a function of invariant mass in 0--80\% Ru+Ru collisions at $\sqrt{s_{\rm NN}}=200$~GeV. The $m_T$ spectra of the individual sources and their sum are fitted using Eq.~(\ref{eq:Teff_fit}) over $0.4<p_T<1.8$~GeV.
}
	\label{fig:RuRu200_dilepton_Teff_vs_M}
\end{figure}
\begin{figure}[htb]
	\centering
    \includegraphics[width=0.95\linewidth]{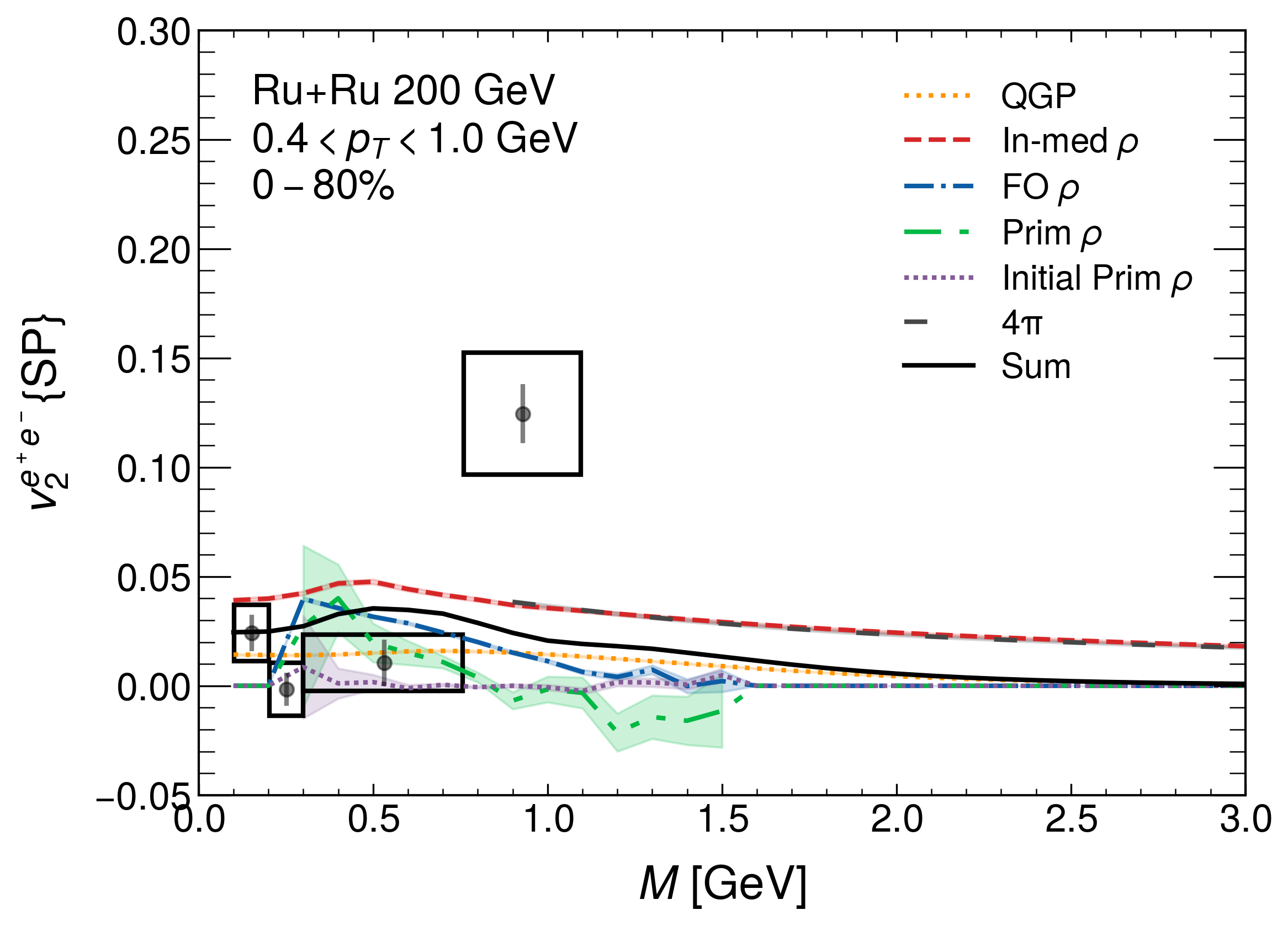}
    \includegraphics[width=0.95\linewidth]{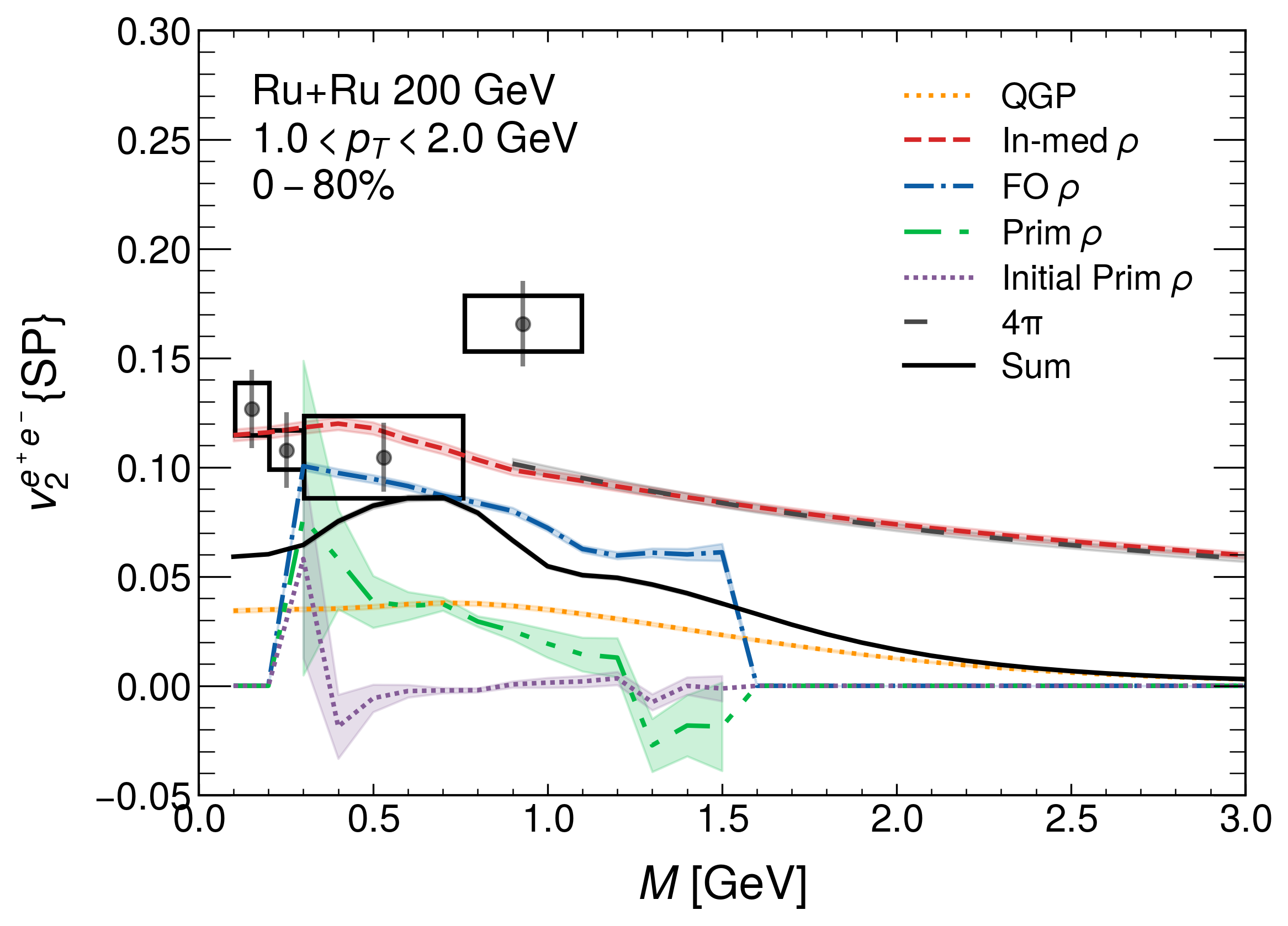}
    \hspace*{0.02\linewidth}
    \includegraphics[width=0.92\linewidth]{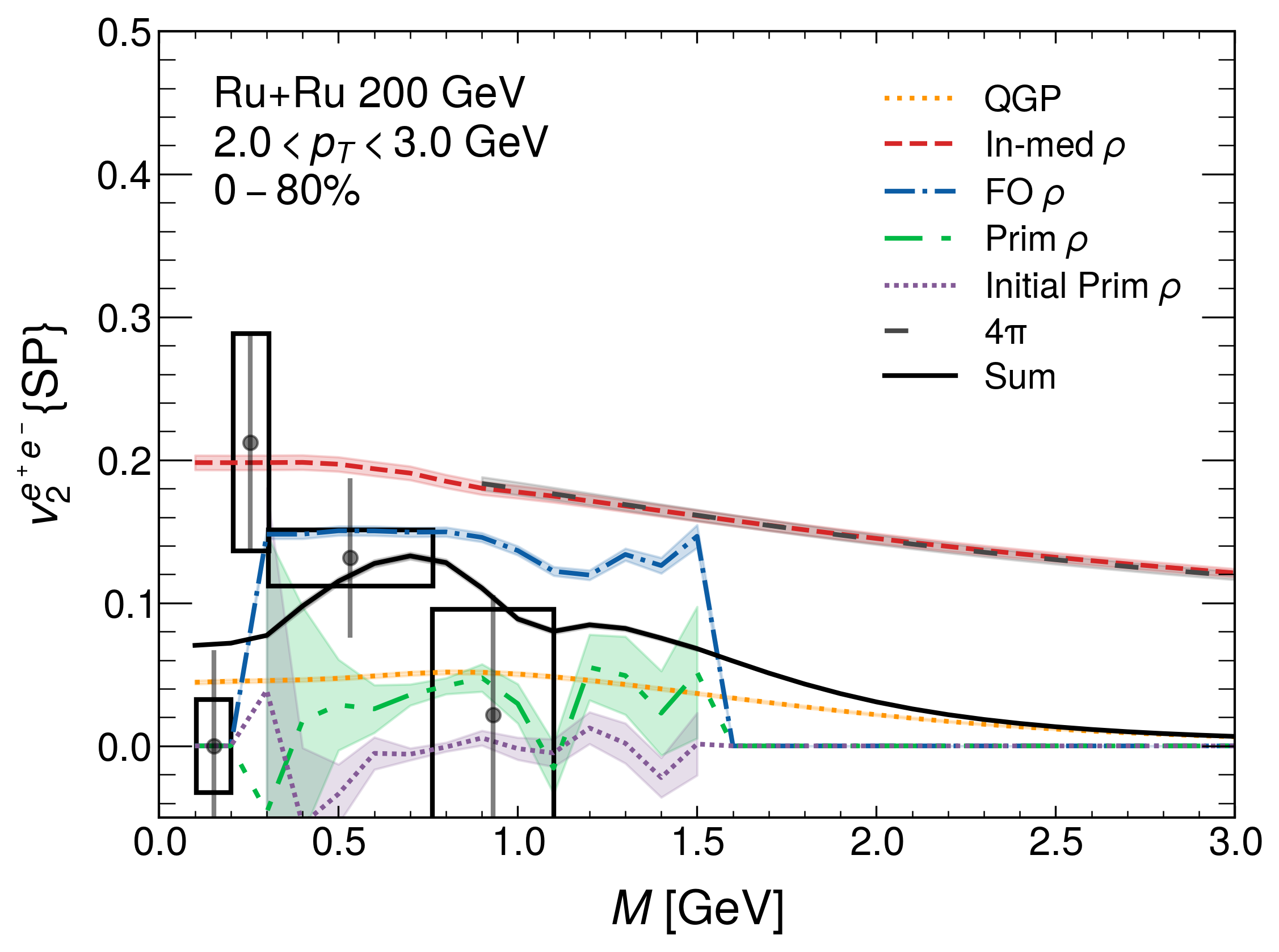}
 \caption{ Dielectron elliptic flow $v_2\{\mathrm{SP}\}$ as a function of invariant mass in 0--80\% Ru+Ru collisions at $\sqrt{s_{\rm NN}}=200$~GeV for $0.4<p_T<1.0$~GeV (top), $1.0<p_T<2.0$~GeV (middle), and $2.0<p_T<3.0$~GeV (bottom). The individual sources and the total are compared with the preliminary STAR isobar data~\cite{Bao:2026HP}.
}
	\label{fig:RuRu200_dielectron_v2_SP_vs_M_varypT}
\end{figure}
The overall source hierarchy remains qualitatively similar
to that in In+In collisions, while the relative importance
of QGP radiation increases at RHIC energies. The higher
collision energy produces a hotter and longer-lived QGP
phase, leading to a larger partonic
contribution. In particular, below the $\rho$-meson mass,
the QGP contribution becomes comparable to the in-medium
$\rho$ contribution. The change is also clear in the IMR, where QGP radiation
dominates more strongly over the four-pion contribution
than at SPS energies. 
The $p_T$-differential spectra in Fig.~\ref{fig:RuRu200_dilepton_dNdM_normalized_pT}
further show that QGP radiation contributes a larger fraction of the low-mass yield in the higher-$p_T$ intervals.

The current model calculation describes the preliminary STAR excess in the LMR, including the $\rho$-meson region, within the experimental uncertainties. However, at higher invariant mass it falls below the data, and the difference persists throughout the IMR. 
As discussed for In+In collisions, pre-equilibrium radiation and viscous corrections to the thermal emission rates may also affect the IMR yield here. This elucidation will wait for future investigations.

Figure~\ref{fig:RuRu200_dilepton_Teff_vs_M} shows the effective temperature extracted from the dilepton $m_T$ spectra as a function of invariant mass in 0--80\% Ru+Ru collisions at $\sqrt{s_{\rm NN}}=200$~GeV, including the individual dilepton sources and their sum. Compared with the
In+In results in Fig.~\ref{fig:Teff_vs_M}, the qualitative ordering of the different contributions is similar. We note that the centrality selections are different, with
the In+In calculation corresponding to the more central events, so we compare the two systems
qualitatively.

Despite this centrality difference, the effective temperatures tend to be larger at RHIC energies. For the QGP contribution, this mainly reflects the higher temperatures reached in Ru+Ru collisions, while the larger $T_{\rm eff}$ of the hadronic and freeze-out contributions is also influenced by the stronger transverse expansion.
This effect is particularly evident for the freeze-out $\rho$. 
The primordial $\rho$ follows the same survival mechanism discussed for In+In collisions, while a larger separation between the initial and surviving components is observed in Ru+Ru collisions. 

The total $T_{\rm eff}$ again develops a  maximum around the $\rho$-meson mass and reaches a larger value than in In+In collisions. At larger invariant mass, the total result approaches the QGP contribution, whose effective
temperature continues to increase with $M$. 
STAR has already extracted average temperatures of the medium from thermal dielectron invariant-mass spectra in isobar collisions~\cite{Luo:QM2025}. Extracting and measuring $T_{\rm eff}$ from the corresponding $m_T$ spectra would further provide complementary information on the transverse expansion of the medium.

Turning to dilepton elliptic flow, Fig.~\ref{fig:RuRu200_dielectron_v2_SP_vs_M_varypT} compares the results of our calculations with the preliminary STAR data in three transverse-momentum intervals, $0.4<p_T<1.0$~GeV, $1.0<p_T<2.0$~GeV, and $2.0<p_T<3.0$~GeV. 
For $0.4<p_T<1.0$~GeV, the calculation is consistent with
the data at low invariant mass within the experimental
uncertainties, but falls below the data around the
$\rho$-meson region. For $1.0<p_T<2.0$~GeV, the calculated
$v_2$ is generally smaller than the measurement. The STAR
data remain approximately constant at low mass and then
increase toward the $\rho$ region, whereas the calculated
$v_2$ reaches a maximum around the $\rho$ mass and
subsequently decreases. For $2.0<p_T<3.0$~GeV, the
calculation is consistent with the data within the currently
large experimental uncertainties. In general, the elliptic
flow of the thermal sources increases with pair transverse
momentum.

The mass dependence of the calculated $v_2$ can be understood from the source composition in the corresponding $p_T$ intervals shown in
Fig.~\ref{fig:RuRu200_dilepton_dNdM_normalized_pT}. For
$0.4<p_T<1.0$~GeV, the in-medium $\rho$ dominates the yield
around the $\rho$ mass and has the largest $v_2$ among the
main sources. The QGP and freeze-out $\rho$ contributions
have smaller $v_2$, bringing the total below the in-medium
$\rho$ result. Around $M\simeq0.2$~GeV, the QGP and in-medium
$\rho$ yields become comparable, and the total $v_2$ lies
between their individual values.

The relative contributions of the different sources change
with increasing $p_T$. Around the $\rho$-meson mass, the
nonthermal $\rho$ components become increasingly important,
particularly the freeze-out $\rho$. As a consequence, the
total elliptic flow shifts toward the freeze-out $\rho$
result in this mass region.

In the LMR  and starting from $p_T>1$~GeV, QGP
radiation becomes the leading contribution to the dilepton
yield. 
Similar behavior was found in earlier calculations
using HTL-improved QGP rates~\cite{Rapp:2002mm,Rapp:2009yu,Rapp:2013nxa}.
Since its elliptic flow is smaller than that of the
hadronic sources, the increasing QGP fraction lowers the
total $v_2$. 

As in In+In collisions, the initial primordial $\rho$ has a
$v_2$ consistent with zero in all three $p_T$ intervals,
while the surviving component develops a finite anisotropy.
The small surviving yield limits its effect on the total
$v_2$.

In the IMR, QGP radiation dominates the yield in all three
$p_T$ intervals. As in In+In collisions, the four-pion
contribution has a smaller yield but a larger $v_2$, raising
the total above the QGP result. Its impact is smaller than
at SPS energies because the relative QGP contribution is
larger at RHIC. 

\section{Sensitivity to the QGP dilepton emission rate}
\label{sec:qgp_rate_sensitivity}

The source analysis above shows that QGP radiation contributes sizably to the low-mass dilepton yield at high $p_T$ in both collision systems. 
The QGP radiation also accounts for a larger fraction of the yield in Ru+Ru collisions at $\sqrt{s_{\rm NN}}=200$~GeV than in In+In collisions at $\sqrt{s_{\rm NN}}=17.3$~GeV. 
Therefore, we focus on Ru+Ru to examine how the QGP emission rate affects the invariant-mass
spectrum and elliptic flow. 
The LO calculation uses the one-loop Born contribution, while the NLO rate includes the two-loop corrections with LPM~\cite{Churchill:2023vpt}.
The medium evolution and all hadronic and nonthermal contributions are held fixed, so differences between the total results arise from the choice of QGP rate.

\begin{figure}[htb]
	\centering
    \includegraphics[width=1.0\linewidth]{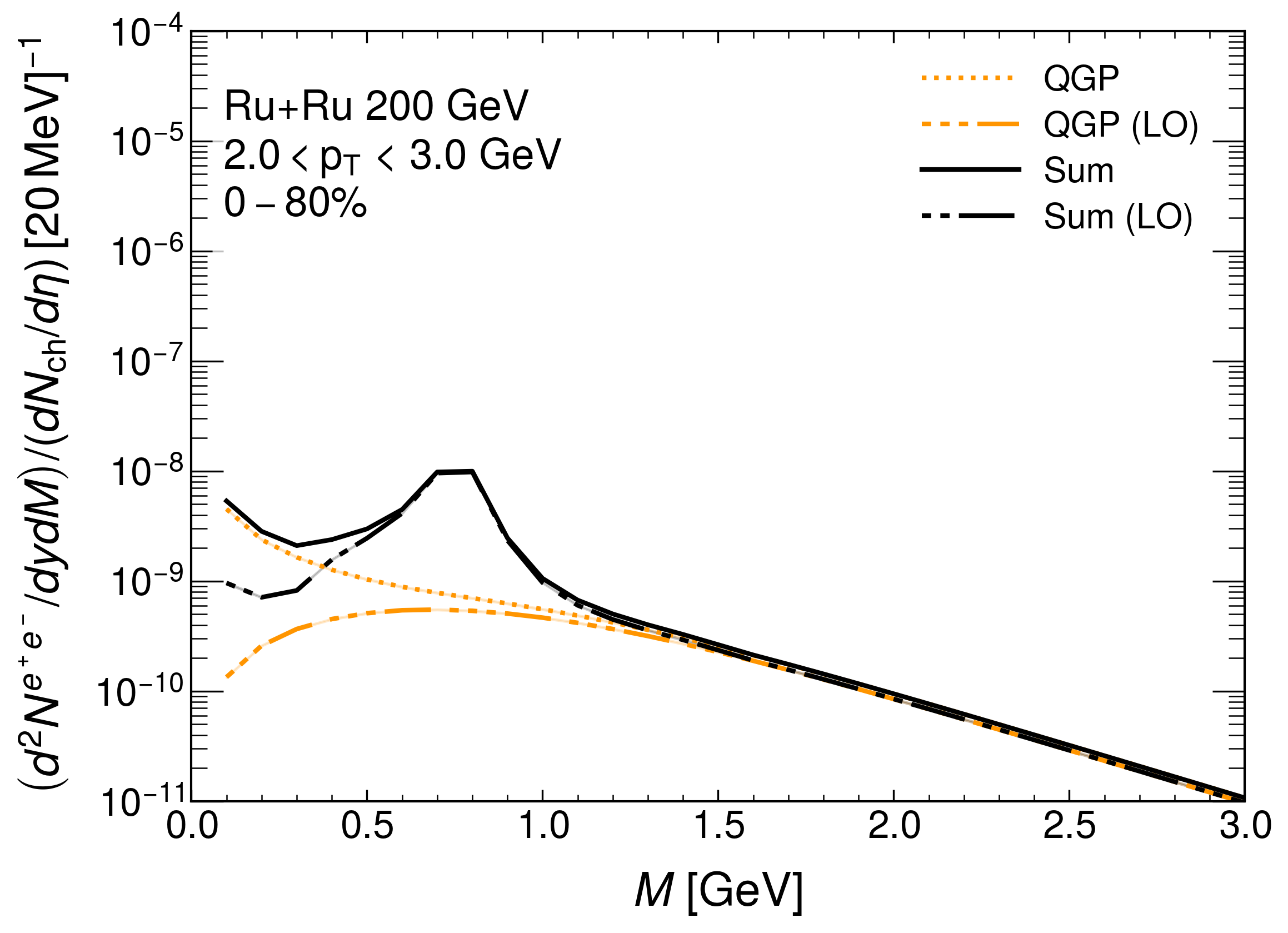}
     \hspace*{0.035\linewidth}
    \includegraphics[width=0.96\linewidth]{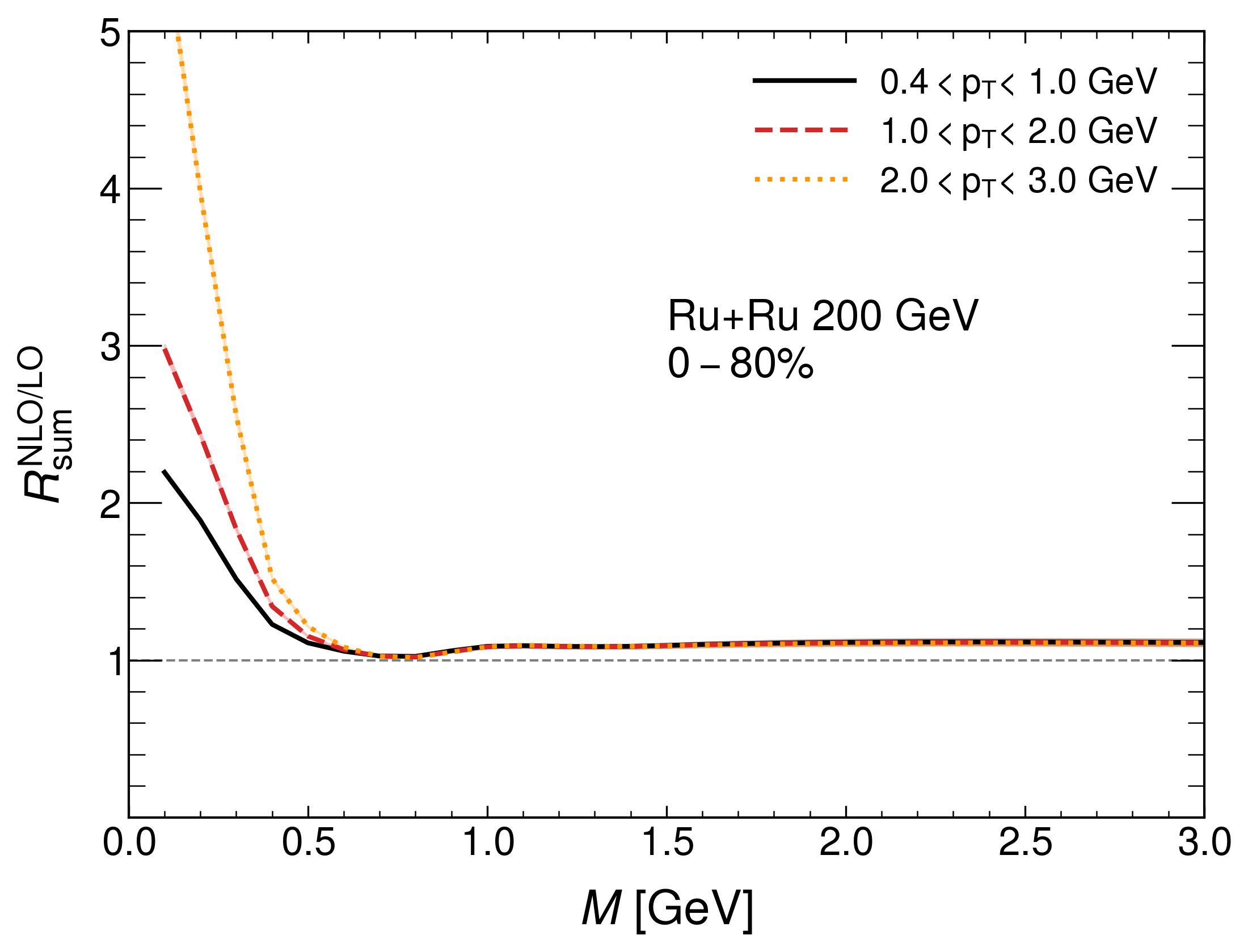}
\caption{
Excess dielectron invariant-mass spectra obtained with LO (labelled as such in the legend) and NLO (unlabelled) QGP rates in 0--80\% Ru+Ru collisions at $\sqrt{s_{\rm NN}}=200$~GeV. The upper panel shows the QGP
and total spectra for $2.0<p_T<3.0$~GeV. The lower panel shows the NLO-to-LO ratio of the total yields, $R_{\rm sum}^{\rm NLO/LO}$, in three selected $p_T$
intervals. The hadronic and nonthermal contributions are identical in the two calculations.
}
	\label{fig:total_yield_NLO_over_LO_pT}
\end{figure}

Figure~\ref{fig:total_yield_NLO_over_LO_pT} compares the
QGP and total invariant-mass spectra obtained with the LO
and NLO QGP rates for $2.0<p_T<3.0$~GeV in the upper panel.
The lower panel shows the ratio of the NLO to LO total
yields in three $p_T$ intervals. At low $M$,
the NLO rate enhances QGP radiation and consequently
increases the total yield. A similar low-mass enhancement
was found in the $p_T$-integrated calculation of
Ref.~\cite{Churchill:2023vpt}. Here, the differential
ratios show that the relative increase in the total yield
grows with $p_T$ and is largest in the highest interval.
Around the $\rho$ mass, the ratio approaches unity.

\begin{figure}[htb]
	\centering
    \includegraphics[width=1.0\linewidth]{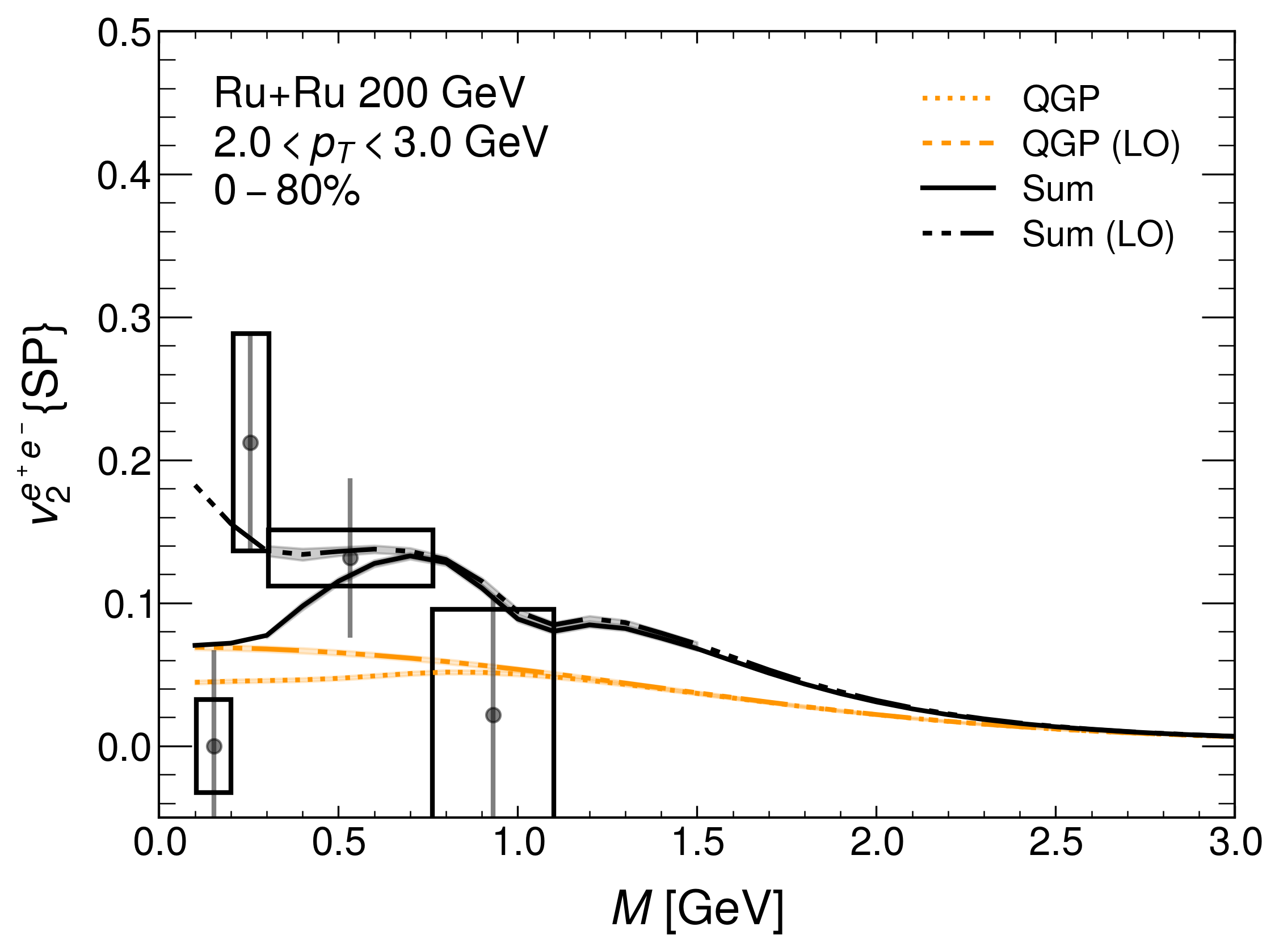}
    \includegraphics[width=1.0\linewidth]{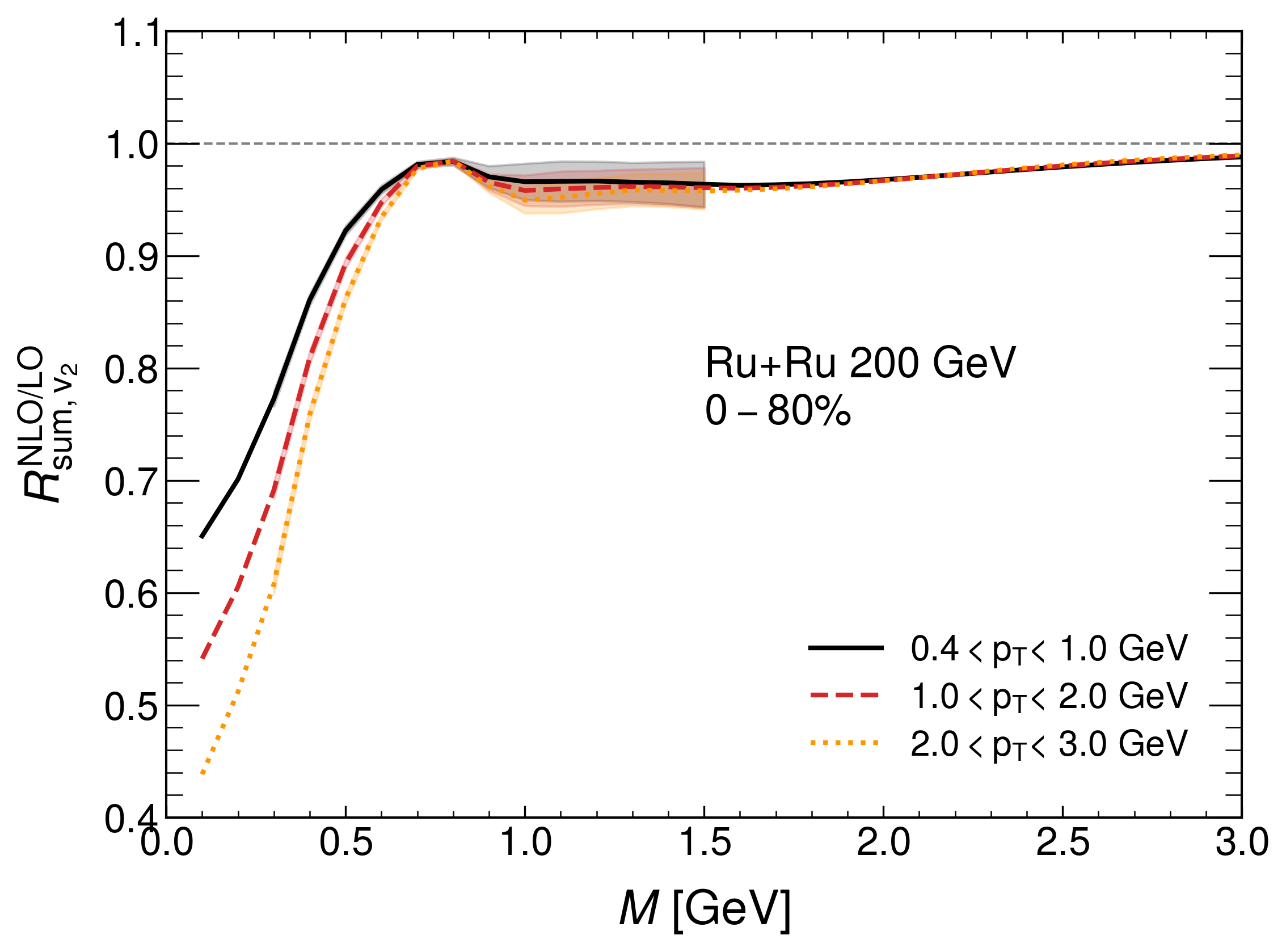}
\caption{
Dielectron elliptic flow $v_2\{\mathrm{SP}\}$ obtained
with LO and NLO QGP rates in 0--80\% Ru+Ru collisions
at $\sqrt{s_{\rm NN}}=200$~GeV. The upper panel shows the
QGP and total results for $2.0<p_T<3.0$~GeV, together
with the preliminary STAR isobar data~\cite{Bao:2026HP}.
The lower panel shows the NLO-to-LO ratio of the total
$v_2$, $R_{{\rm sum},v_2}^{\rm NLO/LO}$, in three
selected $p_T$ intervals.
}
	\label{fig:v2_NLO_over_LO_pT}
\end{figure}
\begin{figure}[htb]
	\centering
    \includegraphics[width=0.9\linewidth]{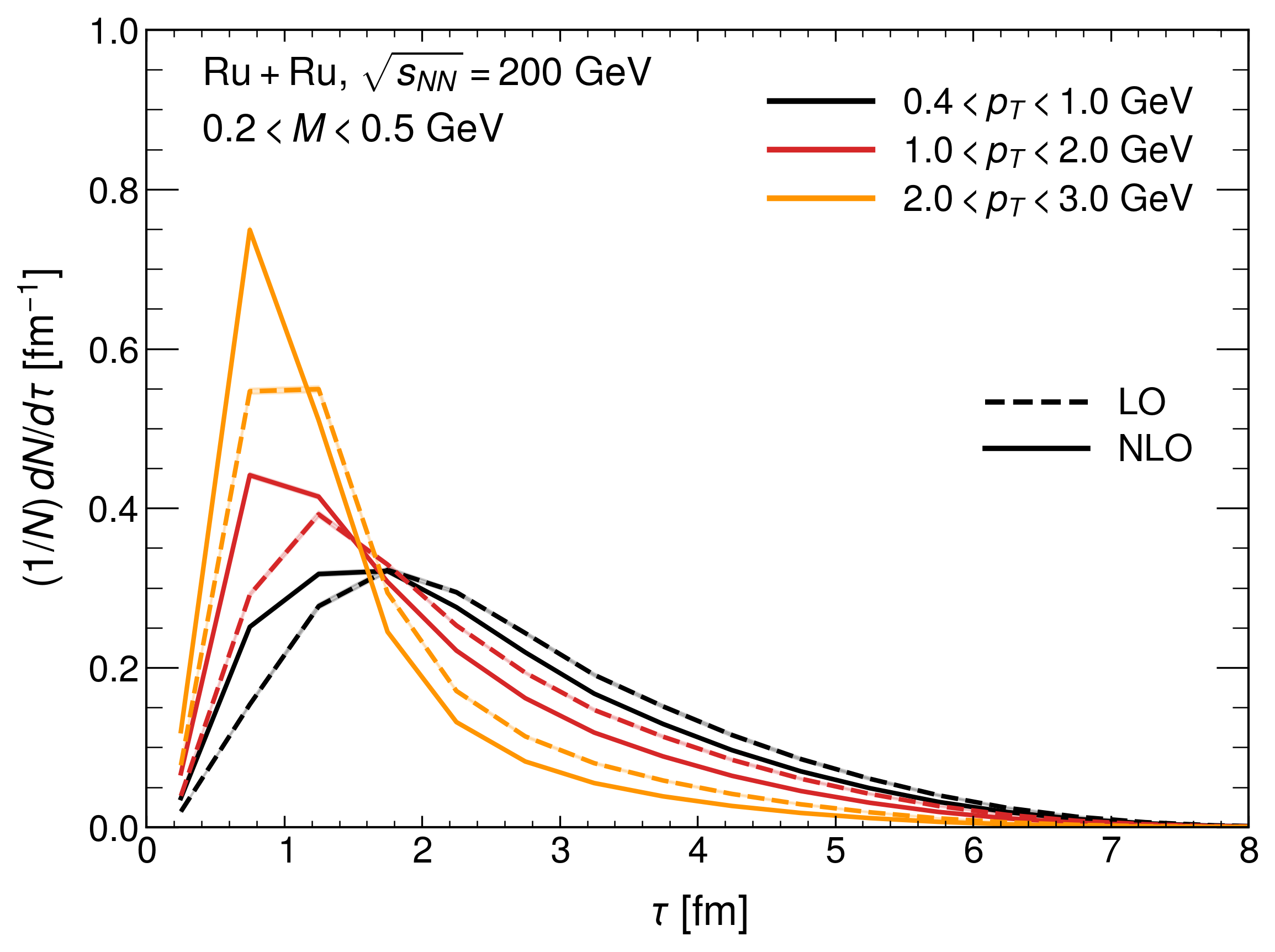}
\caption{
Normalized emission-time distributions $(1/N)dN/d\tau$
of QGP dielectrons with $0.2<M<0.5$~GeV in
0--80\% Ru+Ru collisions at $\sqrt{s_{\rm NN}}=200$~GeV
for three $p_T$ intervals. Solid and dashed curves show the NLO and LO results, respectively. 
}
	\label{fig:RuRu200_tau_distribution_LO_NLO}
\end{figure}

The effect of the rate on the elliptic flow is shown in Fig.~\ref{fig:v2_NLO_over_LO_pT}. At low $M$, the NLO
calculation gives a smaller QGP $v_2$ than the LO
calculation. The total $v_2$ is also reduced, with the
largest difference among the selected intervals occurring
at $2.0<p_T<3.0$~GeV. Around the $\rho$ mass and in the
IMR, the two complete results become much closer. Thus the
same low-mass region in which the NLO rate most strongly
increases the yield also shows the greatest sensitivity in
the elliptic flow.

The lower QGP $v_2$ with the NLO rate can be understood
from the time distribution of QGP emission. At fixed small
$M$, the hotter QGP at earlier times corresponds to a
smaller $M/T$. For low-mass pairs at high $p_T$, the virtual
photon lies near the light cone, where the Born
$q\bar q\to\gamma^*$ contribution becomes less important
relative to QCD corrections, including $qg\to q\gamma^*$
scattering and collinear emission treated with LPM
resummation~\cite{Ghiglieri:2014kma,Jackson:2019mop,Churchill:2023vpt}.

To illustrate this point, Fig.~\ref{fig:RuRu200_tau_distribution_LO_NLO} shows the normalized emission-time distributions $(1/N)dN/d\tau$ of QGP dielectrons with $0.2<M<0.5$~GeV in the same
three $p_T$ intervals. Compared with LO, the NLO rate gives greater relative weight to emission at early times. This difference is particularly clear for $2.0<p_T<3.0$~GeV. Since anisotropic flow is less developed at early times, the increased early-time contribution lowers the QGP $v_2$. Finally, the NLO rate at low $M$ increases the QGP fraction of the yield, further reducing the total $v_2$.
These results suggest for the first time to our knowledge that low-$M$ dileptons at high $p_T$ may provide insight into NLO corrections to the QGP emission rate.

\section{Conclusion}
\label{sec:conclusion}

In this work, we have studied dilepton spectra and elliptic flows in
In+In collisions at $\sqrt{s_{\rm NN}}=17.3$~GeV and isobar collisions at
$\sqrt{s_{\rm NN}}=200$~GeV using the $(3+1)$D iEBE-MUSIC framework.
The event-by-event medium evolution was obtained from a Bayesian analysis
of charged-hadron data at RHIC-BES energies. We included NLO QGP
emission at finite baryon chemical potential, in-medium $\rho$ and
four-pion radiation, and freeze-out and primordial $\rho$ decays.

We obtain a good description of the NA60 invariant-mass and transverse-mass spectra. The broadened in-medium $\rho$ dominates the LMR at low $p_T$, while nonthermal $\rho$ decays become increasingly important near the $\rho$ mass at higher $p_T$.
QGP radiation dominates the IMR yield. In Ru+Ru collisions, the low-mass spectrum, including the $\rho$ region, is consistent with the preliminary STAR data within the experimental uncertainties. The IMR yield remains below the measured excess in both systems, with a modest difference for $1.5<M<2.5$~GeV in In+In.

We also extracted the effective temperature $T_{\rm eff}$ from the transverse-mass spectra of individual sources. For thermal radiation, $T_{\rm eff}$ reflects both the emission temperature and radial flow. The freeze-out $\rho$ has a relatively large $T_{\rm eff}$ near the $\rho$ mass. For primordial $\rho$, hadronic interactions will remove low-$p_T$ mesons from the surviving sample, leaving a harder spectrum and increasing its effective temperature. The freeze-out and surviving primordial $\rho$ contributions raise the effective temperature of the summed spectrum near $m_\rho$ and help produce
a maximum $T_{\rm eff}$ in both systems. However, for In+In collisions analyzed within our multistage model, the maximum remains below the NA60 data, while $T_{\rm eff}$ increases more rapidly with $M$ in the IMR than observed. This fact will fuel further developments. 

In the IMR, four-pion radiation has a smaller yield but a larger $v_2$ than QGP radiation and raises the total elliptic flow, particularly in In+In collisions.
This effect is less significant in Ru+Ru because of the larger QGP fraction of the yield. The surviving primordial $\rho$ develops a positive $v_2$, but its small yield limits its effect on the total flow. For $0.4<p_T<1.0$~GeV, our model prediction of low-mass dielectron $v_2$ agrees with the preliminary STAR data at low $M$ within experimental uncertainties, although being smaller near $m_\rho$. For $1.0<p_T<2.0$~GeV, $v_2$ is generally below the data, while for $2.0<p_T<3.0$~GeV it agrees with the data within the large experimental uncertainties.

We further compared dilepton spectra and elliptic flows in Ru+Ru collisions using LO and NLO QGP emission rates, with the medium evolution and all other sources held fixed. Using the NLO production rate increases the total low-mass yield and reduces the total $v_2$. The largest change is observed at $2.0<p_T<3.0$~GeV. At fixed low $M$, the NLO rate gives a greater relative weight to the hotter early QGP, where $M/T$ is smaller and anisotropic flow is less developed, which lowers the QGP dilepton $v_2$.
Since NLO channels increase the QGP fraction, the  total flow is further reduced. These results suggest that low-mass dilepton spectra and elliptic flow at high $p_T$
may provide insights into NLO corrections to the QGP emission rate.

Finally, this study combines up-to-date partonic and hadronic dilepton rates with a state-of-the-art simulation of the medium evolution, providing a comprehensive theoretical benchmark for dilepton observables at SPS and RHIC energies. The remaining differences with the measured spectra and elliptic flows suggest that a description constrained by hadronic observables alone may not be sufficient to simultaneously reproduce the dilepton data. Including electromagnetic probes in the Bayesian quantification, which currently mainly focuses on hadronic observables, can help constrain the transport coefficients with a better precision.

\begin{acknowledgments}
We are grateful to I. Tserruya for discussions (CG) and for valuable comments on a previous draft of this paper (CG and XYW). This work is supported in part by the Natural Sciences and Engineering Research Council of Canada  (NSERC) [SAPIN-2026-00047 (CG) and SAPIN-2024-00026 (SJ)], and in part by US National Science Foundation (NSF) under grant numbers OAC-2004571 and PHY-2514775 (RR). Numerical simulations presented in this work were performed on the Rorqual supercomputer system, managed by Calcul Qu\'ebec and by the Digital Research Alliance of Canada. 
\end{acknowledgments}

\appendix

\section{Model calibration}
\label{sec:calibration}
\begin{figure}[htb]
	\centering
    \includegraphics[width=0.9\linewidth]{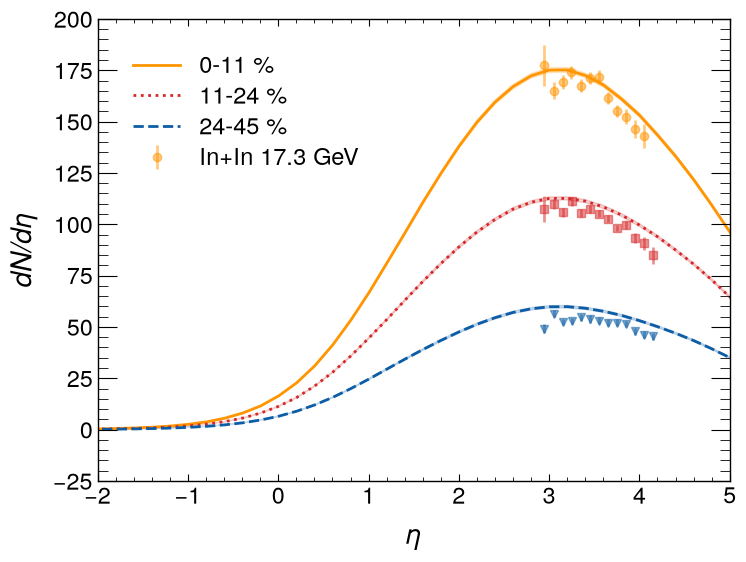}
 \caption{Charged-hadron pseudorapidity distributions $dN_{\rm ch}/d\eta$
in the laboratory frame for In+In collisions at
$\sqrt{s_{\rm NN}}=17.3$~GeV in the 0--11\%, 11--24\%, and
24--45\% centrality intervals. The curves show the model
calculations, and the symbols represent the experimental
data~\cite{NA60:2005qgv}. }
	\label{fig:InIn17p3_charged_hadron_dNdeta}
\end{figure}

\begin{figure}[thb]
	\centering
    \includegraphics[width=0.85\linewidth]{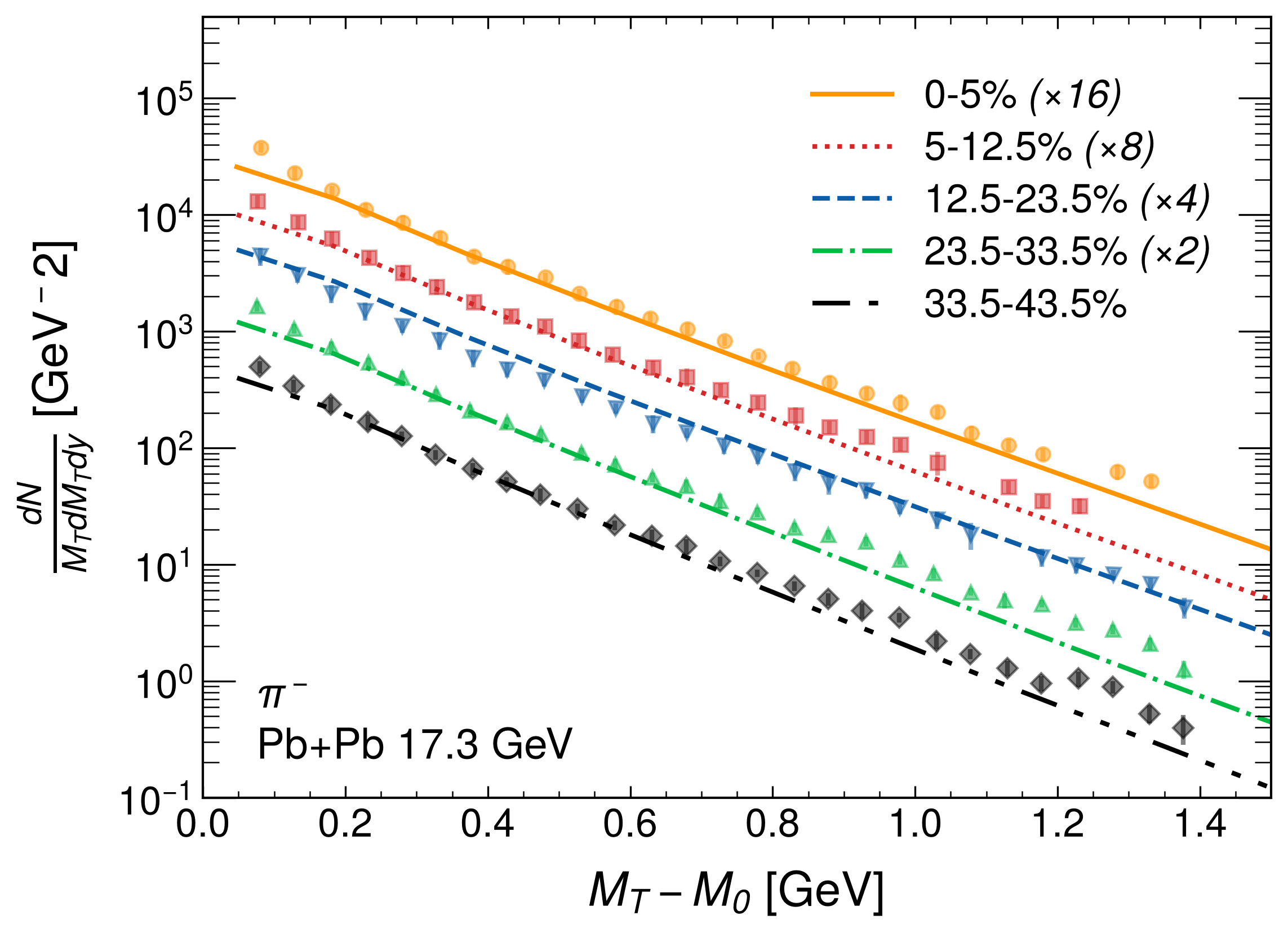}
    \includegraphics[width=0.85\linewidth]{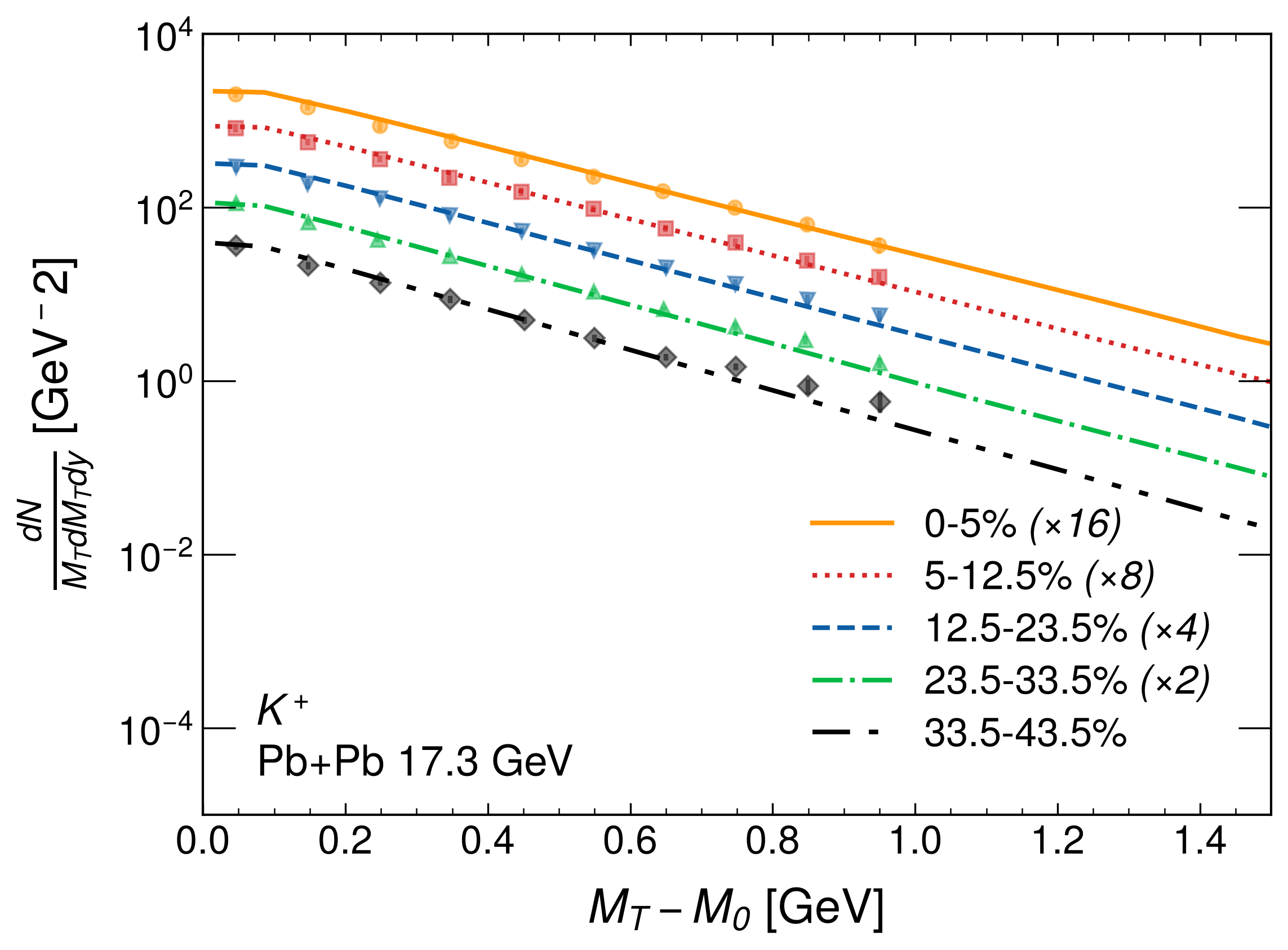}
    
\caption{
Transverse-mass spectra of $\pi^-$ (top) and  $K^+$
(bottom) at midrapidity in Pb+Pb collisions at
$\sqrt{s_{\rm NN}}=17.3$~GeV for five centrality intervals.
The curves show the model calculations, and the symbols
represent the NA49 data~\cite{NA49:2012rsi}.
The spectra are scaled by the factors shown in the legends.
}
	\label{fig:PbPb17p3_Kplus_mT_spectrum}
\end{figure}

\begin{figure}[htb]
	\centering
    \includegraphics[width=0.85\linewidth]{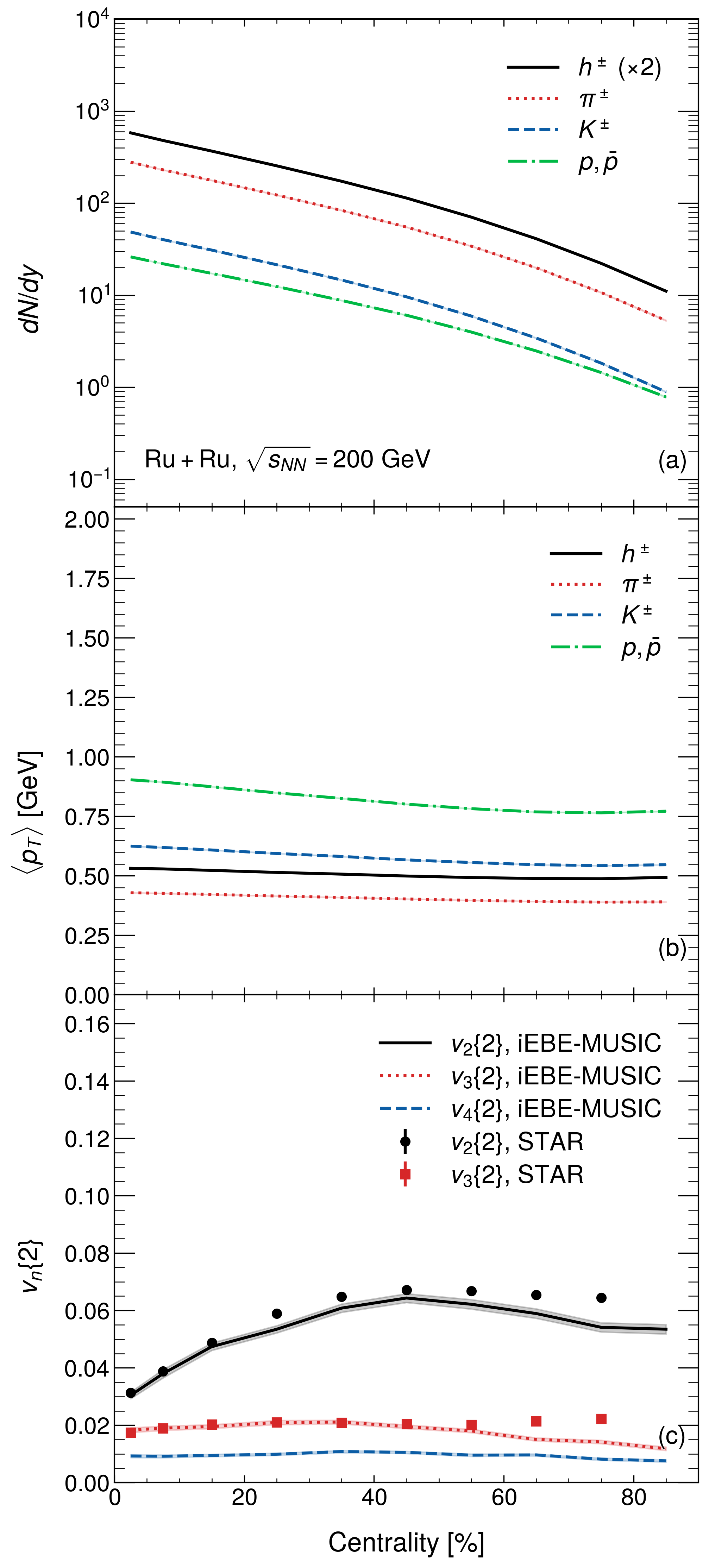}
\caption{
(a) Particle yields $dN/dy$, (b) average transverse momenta
$\langle p_T\rangle$, and (c) momentum anisotropy coefficients
$v_n\{2\}$ for $n=2,3,4$ as functions of collision centrality
in Ru+Ru collisions at $\sqrt{s_{\rm NN}}=200$~GeV.
Panels (a) and (b) show identified particles and charged
hadrons, while panel (c) refers only to charged hadrons.
The data in panel (c) are from the STAR Collaboration~\cite{STAR:2021mii}.
}
	\label{fig:RuRu200_bulk_observables}
\end{figure}
In this appendix, we compare hadronic observables with available measurements at SPS and RHIC energies using the MAP parameter set from the Bayesian analysis of charged-hadron observables in Au+Au collisions at RHIC-BES energies~\cite{Jahan:2024wpj}, as described in Sec.~\ref{sec:framework}.
Figure~\ref{fig:InIn17p3_charged_hadron_dNdeta} compares the charged-hadron pseudorapidity distributions in the lab frame for In+In collisions at $\sqrt{s_{\rm NN}}=17.3$~GeV with data from the NA60 Collaboration~\cite{NA60:2005qgv}.
Figure~\ref{fig:PbPb17p3_Kplus_mT_spectrum} shows the $\pi^-$ and $K^+$   transverse-mass spectra in Pb+Pb collisions at the same energy, providing an additional comparison with the NA49 data at SPS energies~\cite{NA49:2012rsi}.
Figure~\ref{fig:RuRu200_bulk_observables} presents particle yields, mean transverse momenta, and charged-hadron flow coefficients in Ru+Ru collisions at $\sqrt{s_{\rm NN}}=200$~GeV. The $v_2\{2\}$ and $v_3\{2\}$ results are compared with the STAR
data~\cite{STAR:2021mii}. Overall, these comparisons demonstrate the predictive power of the model for hadronic observables in collision systems that were not included in the Bayesian calibration. The consistent description of these observables sets our dilepton studies on a firm foundation.

\bibliography{inspire, non-inspire}

@article{STAR:2015tnn,
    author = "Adamczyk, L. and others",
    collaboration = "STAR",
    title = "{Measurements of Dielectron Production in Au$+$Au Collisions at $\sqrt{s_{\rm NN}}$ = 200 GeV from the STAR Experiment}",
    eprint = "1504.01317",
    archivePrefix = "arXiv",
    primaryClass = "hep-ex",
    doi = "10.1103/PhysRevC.92.024912",
    journal = "Phys. Rev. C",
    volume = "92",
    number = "2",
    pages = "024912",
    year = "2015"
}

@article{STAR:2013pwb,
    author = "Adamczyk, L. and others",
    collaboration = "STAR",
    title = "{Dielectron Mass Spectra from Au+Au Collisions at $\sqrt{s_{\rm NN}}$ = 200 GeV}",
    eprint = "1312.7397",
    archivePrefix = "arXiv",
    primaryClass = "hep-ex",
    doi = "10.1103/PhysRevLett.113.022301",
    journal = "Phys. Rev. Lett.",
    volume = "113",
    number = "2",
    pages = "022301",
    year = "2014",
    note = "[Addendum: Phys.Rev.Lett. 113, 049903 (2014)]"
}

@article{PHENIX:2015vek,
    author = "Adare, A. and others",
    collaboration = "PHENIX",
    title = "{Dielectron production in Au$+$Au collisions at $\sqrt{s_{NN}}$=200 GeV}",
    eprint = "1509.04667",
    archivePrefix = "arXiv",
    primaryClass = "nucl-ex",
    doi = "10.1103/PhysRevC.93.014904",
    journal = "Phys. Rev. C",
    volume = "93",
    number = "1",
    pages = "014904",
    year = "2016"
}

@article{NA60:2007lzy,
    author = "Arnaldi, R. and others",
    collaboration = "NA60",
    title = "{Evidence for radial flow of thermal dileptons in high-energy nuclear collisions}",
    eprint = "0711.1816",
    archivePrefix = "arXiv",
    primaryClass = "nucl-ex",
    doi = "10.1103/PhysRevLett.100.022302",
    journal = "Phys. Rev. Lett.",
    volume = "100",
    pages = "022302",
    year = "2008"
}

@article{NA60:2008dcb,
    author = "Arnaldi, R and others",
    collaboration = "NA60",
    title = "{Evidence for the production of thermal-like muon pairs with masses above 1-GeV/c**2 in 158-A-GeV Indium-Indium Collisions}",
    eprint = "0810.3204",
    archivePrefix = "arXiv",
    primaryClass = "nucl-ex",
    doi = "10.1140/epjc/s10052-008-0857-2",
    journal = "Eur. Phys. J. C",
    volume = "59",
    pages = "607--623",
    year = "2009"
}

@article{CERES:1995vll,
    author = "Agakichiev, G. and others",
    collaboration = "CERES",
    title = "{Enhanced production of low mass electron pairs in 200-GeV/u S - Au collisions at the CERN SPS}",
    reportNumber = "CERN-PPE-95-026, CERN-PPE-95-26",
    doi = "10.1103/PhysRevLett.75.1272",
    journal = "Phys. Rev. Lett.",
    volume = "75",
    pages = "1272--1275",
    year = "1995"
}

@article{Harris:2023tti,
    author = {Harris, John W. and M{\"u}ller, Berndt},
    title = "{''QGP Signatures'' Revisited}",
    eprint = "2308.05743",
    archivePrefix = "arXiv",
    primaryClass = "hep-ph",
    doi = "10.1140/epjc/s10052-024-12533-y",
    journal = "Eur. Phys. J. C",
    volume = "84",
    number = "3",
    pages = "247",
    year = "2024"
}

@article{Bailhache:2025kwa,
    author = {Bailhache, R. and Appelsh{\"a}user, H.},
    title = "{Dileptons at Colliders as Probes of the Quark{\textendash}Gluon Plasma}",
    eprint = "2512.10597",
    archivePrefix = "arXiv",
    primaryClass = "nucl-ex",
    doi = "10.1146/annurev-nucl-121423-100858",
    journal = "Ann. Rev. Nucl. Part. Sci.",
    volume = "75",
    number = "1",
    pages = "463--486",
    year = "2025"
}

@article{Rapp:1999ej,
    author = "Rapp, R. and Wambach, J.",
    title = "{Chiral symmetry restoration and dileptons in relativistic heavy ion collisions}",
    eprint = "hep-ph/9909229",
    archivePrefix = "arXiv",
    reportNumber = "SUNY-NTG-99-29",
    doi = "10.1007/0-306-47101-9_1",
    journal = "Adv. Nucl. Phys.",
    volume = "25",
    pages = "1",
    year = "2000"
}

@article{Goes-Hirayama:2025lhs,
    author = "G{\'o}es-Hirayama, Renan and Paul{\'\i}nyov{\'a}, Zuzana and Egger, Joscha and Karpenko, Iurii and Elfner, Hannah",
    title = "{Imprints of dynamic fluidization on dilepton production}",
    eprint = "2510.02862",
    archivePrefix = "arXiv",
    primaryClass = "nucl-th",
    doi = "10.1051/epjconf/202636407005",
    journal = "EPJ Web Conf.",
    volume = "364",
    pages = "07005",
    year = "2026"
}

@article{Nayak:2009xrg,
    author = "Nayak, Jajati K. and Alam, Jan-e and Hirano, Tetsufumi and Sarkar, Sourav and Sinha, Bikash",
    title = "{Muon pairs from In+In collision at SPS energy}",
    eprint = "0902.0446",
    archivePrefix = "arXiv",
    primaryClass = "nucl-th",
    doi = "10.1103/PhysRevC.85.064906",
    journal = "Phys. Rev. C",
    volume = "85",
    pages = "064906",
    year = "2012"
}

@article{Linnyk:2011hz,
    author = "Linnyk, O. and Bratkovskaya, E. L. and Ozvenchuk, V. and Cassing, W. and Ko, C. M.",
    title = "{Dilepton production in nucleus-nucleus collisions at top SPS energy within the Parton-Hadron-String Dynamics (PHSD) transport approach}",
    eprint = "1107.3402",
    archivePrefix = "arXiv",
    primaryClass = "nucl-th",
    doi = "10.1103/PhysRevC.84.054917",
    journal = "Phys. Rev. C",
    volume = "84",
    pages = "054917",
    year = "2011"
}

@article{Endres:2016tkg,
    author = "Endres, Stephan and van Hees, Hendrik and Bleicher, Marcus",
    title = "{Energy, centrality and momentum dependence of dielectron production at collider energies in a coarse-grained transport approach}",
    eprint = "1604.06415",
    archivePrefix = "arXiv",
    primaryClass = "nucl-th",
    doi = "10.1103/PhysRevC.94.024912",
    journal = "Phys. Rev. C",
    volume = "94",
    number = "2",
    pages = "024912",
    year = "2016"
}

@article{Banerjee:2026hlm,
    author = "Banerjee, Soham and Singh, Mayank and Gale, Charles and Jeon, Sangyong and Pablos, Daniel and Paquet, Jean-Fran{\c{c}}ois",
    title = "{Electromagnetic probes of concurrent minijet-hydrodynamics evolution}",
    eprint = "2609.39939",
    archivePrefix = "arXiv",
    primaryClass = "hep-ph",
    month = "9",
    year = "2026"
}

@article{Endres:2015egk,
    author = "Endres, Stephan and van Hees, Hendrik and Bleicher, Marcus",
    title = "{Photon and dilepton production at the Facility for Antiproton and Ion Research and beam-energy scan at the Relativistic Heavy-Ion Collider using coarse-grained microscopic transport simulations}",
    eprint = "1512.06549",
    archivePrefix = "arXiv",
    primaryClass = "nucl-th",
    doi = "10.1103/PhysRevC.93.054901",
    journal = "Phys. Rev. C",
    volume = "93",
    number = "5",
    pages = "054901",
    year = "2016"
}

@article{Endres:2015fna,
    author = "Endres, Stephan and van Hees, Hendrik and Weil, Janus and Bleicher, Marcus",
    title = "{Dilepton production and reaction dynamics in heavy-ion collisions at SIS energies from coarse-grained transport simulations}",
    eprint = "1505.06131",
    archivePrefix = "arXiv",
    primaryClass = "nucl-th",
    doi = "10.1103/PhysRevC.92.014911",
    journal = "Phys. Rev. C",
    volume = "92",
    number = "1",
    pages = "014911",
    year = "2015"
}

@article{Endres:2014zua,
    author = "Endres, Stephan and van Hees, Hendrik and Weil, Janus and Bleicher, Marcus",
    title = "{Coarse-graining approach for dilepton production at energies available at the CERN Super Proton Synchrotron}",
    eprint = "1412.1965",
    archivePrefix = "arXiv",
    primaryClass = "nucl-th",
    doi = "10.1103/PhysRevC.91.054911",
    journal = "Phys. Rev. C",
    volume = "91",
    number = "5",
    pages = "054911",
    year = "2015"
}

@article{Santini:2011zw,
    author = "Santini, E. and Steinheimer, J. and Bleicher, M. and Schramm, S.",
    title = "{Dimuon radiation at the CERN SPS within a (3+1)d hydrodynamic+cascade model}",
    eprint = "1102.4574",
    archivePrefix = "arXiv",
    primaryClass = "nucl-th",
    doi = "10.1103/PhysRevC.84.014901",
    journal = "Phys. Rev. C",
    volume = "84",
    pages = "014901",
    year = "2011"
}

@article{Du:2025dot,
    author = "Du, Lipei and Heinz, Ulrich",
    title = "{Electromagnetic Tomography of Radial Flow in the Quark-Gluon Plasma}",
    eprint = "2505.20752",
    archivePrefix = "arXiv",
    primaryClass = "hep-ph",
    doi = "10.1103/ptyq-cs9m",
    journal = "Phys. Rev. Lett.",
    volume = "136",
    number = "10",
    pages = "102301",
    year = "2026"
}

@article{NA49:2012rsi,
    author = "Anticic, T. and others",
    collaboration = "NA49",
    title = "{System-size and centrality dependence of charged kaon and pion production in nucleus-nucleus collisions at 40A GeV and158A GeV beam energy}",
    eprint = "1207.0348",
    archivePrefix = "arXiv",
    primaryClass = "nucl-ex",
    doi = "10.1103/PhysRevC.86.054903",
    journal = "Phys. Rev. C",
    volume = "86",
    pages = "054903",
    year = "2012"
}

@inproceedings{Gale:2025ome,
    author = "Gale, Charles",
    title = "{Electromagnetic Radiation from High-Energy Nuclear Collisions}",
    eprint = "2502.13938",
    archivePrefix = "arXiv",
    primaryClass = "nucl-th",
    month = "2",
    year = "2025"
}

@inproceedings{Heinz:2024jwu,
    author = {Heinz, Ulrich and Schenke, Bj{\"o}rn},
    title = "{Hydrodynamic Description of the Quark-Gluon Plasma}",
    eprint = "2412.19393",
    archivePrefix = "arXiv",
    primaryClass = "nucl-th",
    month = "12",
    year = "2024"
}

@article{NA60:2006ymb,
    author = "Arnaldi, R. and others",
    collaboration = "NA60",
    title = "{First measurement of the rho spectral function in high-energy nuclear collisions}",
    eprint = "nucl-ex/0605007",
    archivePrefix = "arXiv",
    doi = "10.1103/PhysRevLett.96.162302",
    journal = "Phys. Rev. Lett.",
    volume = "96",
    pages = "162302",
    year = "2006"
}

@article{Wu:2021fjf,
    author = "Wu, Xiang-Yu and Qin, Guang-You and Pang, Long-Gang and Wang, Xin-Nian",
    title = "{(3+1)-D viscous hydrodynamics at finite net baryon density: Identified particle spectra, anisotropic flows, and flow fluctuations across energies relevant to the beam-energy scan at RHIC}",
    eprint = "2107.04949",
    archivePrefix = "arXiv",
    primaryClass = "hep-ph",
    doi = "10.1103/PhysRevC.105.034909",
    journal = "Phys. Rev. C",
    volume = "105",
    number = "3",
    pages = "034909",
    year = "2022"
}

@article{Gale:2013da,
    author = "Gale, Charles and Jeon, Sangyong and Schenke, Bjoern",
    title = "{Hydrodynamic Modeling of Heavy-Ion Collisions}",
    eprint = "1301.5893",
    archivePrefix = "arXiv",
    primaryClass = "nucl-th",
    doi = "10.1142/S0217751X13400113",
    journal = "Int. J. Mod. Phys. A",
    volume = "28",
    pages = "1340011",
    year = "2013"
}

@article{Shen:2020gef,
    author = "Shen, Chun",
    editor = "Liu, Feng and Wang, Enke and Wang, Xin-Nian and Xu, Nu and Zhang, Ben-Wei",
    title = "{Studying QGP with flow: A theory overview}",
    eprint = "2001.11858",
    archivePrefix = "arXiv",
    primaryClass = "nucl-th",
    doi = "10.1016/j.nuclphysa.2020.121788",
    journal = "Nucl. Phys. A",
    volume = "1005",
    pages = "121788",
    year = "2021"
}

@article{Song:2010mg,
    author = "Song, Huichao and Bass, Steffen A. and Heinz, Ulrich and Hirano, Tetsufumi and Shen, Chun",
    title = "{200 A GeV Au+Au collisions serve a nearly perfect quark-gluon liquid}",
    eprint = "1011.2783",
    archivePrefix = "arXiv",
    primaryClass = "nucl-th",
    doi = "10.1103/PhysRevLett.106.192301",
    journal = "Phys. Rev. Lett.",
    volume = "106",
    pages = "192301",
    year = "2011",
    note = "[Erratum: Phys.Rev.Lett. 109, 139904 (2012)]"
}

@article{Shen:2015msa,
    author = "Shen, Chun and Heinz, Ulrich",
    title = "{The road to precision: Extraction of the specific shear viscosity of the quark-gluon plasma}",
    eprint = "1507.01558",
    archivePrefix = "arXiv",
    primaryClass = "nucl-th",
    doi = "10.1080/10619127.2015.1006502",
    journal = "Nucl. Phys. News",
    volume = "25",
    number = "2",
    pages = "6--11",
    year = "2015"
}

@article{Ryu:2015vwa,
    author = "Ryu, S. and Paquet, J. -F. and Shen, C. and Denicol, G. S. and Schenke, B. and Jeon, S. and Gale, C.",
    title = "{Importance of the Bulk Viscosity of QCD in Ultrarelativistic Heavy-Ion Collisions}",
    eprint = "1502.01675",
    archivePrefix = "arXiv",
    primaryClass = "nucl-th",
    doi = "10.1103/PhysRevLett.115.132301",
    journal = "Phys. Rev. Lett.",
    volume = "115",
    number = "13",
    pages = "132301",
    year = "2015"
}

@article{Schenke:2020mbo,
    author = "Schenke, Bjoern and Shen, Chun and Tribedy, Prithwish",
    title = "{Running the gamut of high energy nuclear collisions}",
    eprint = "2005.14682",
    archivePrefix = "arXiv",
    primaryClass = "nucl-th",
    doi = "10.1103/PhysRevC.102.044905",
    journal = "Phys. Rev. C",
    volume = "102",
    number = "4",
    pages = "044905",
    year = "2020"
}

@article{Ruppert:2007cr,
    author = "Ruppert, Jorg and Gale, Charles and Renk, Thorsten and Lichard, Peter and Kapusta, Joseph I.",
    title = "{Low mass dimuons produced in relativistic nuclear collisions}",
    eprint = "0706.1934",
    archivePrefix = "arXiv",
    primaryClass = "hep-ph",
    doi = "10.1103/PhysRevLett.100.162301",
    journal = "Phys. Rev. Lett.",
    volume = "100",
    pages = "162301",
    year = "2008"
}

@article{Paquet:2015lta,
    author = {Paquet, Jean-Fran{\c{c}}ois and Shen, Chun and Denicol, Gabriel S. and Luzum, Matthew and Schenke, Bj{\"o}rn and Jeon, Sangyong and Gale, Charles},
    title = "{Production of photons in relativistic heavy-ion collisions}",
    eprint = "1509.06738",
    archivePrefix = "arXiv",
    primaryClass = "hep-ph",
    doi = "10.1103/PhysRevC.93.044906",
    journal = "Phys. Rev. C",
    volume = "93",
    number = "4",
    pages = "044906",
    year = "2016"
}

@article{Bazavov:2009zn,
    author = "Bazavov, A. and others",
    title = "{Equation of state and QCD transition at finite temperature}",
    eprint = "0903.4379",
    archivePrefix = "arXiv",
    primaryClass = "hep-lat",
    reportNumber = "CU-TP-1189",
    doi = "10.1103/PhysRevD.80.014504",
    journal = "Phys. Rev. D",
    volume = "80",
    pages = "014504",
    year = "2009"
}

@article{Busza:2018rrf,
    author = "Busza, Wit and Rajagopal, Krishna and van der Schee, Wilke",
    title = "{Heavy Ion Collisions: The Big Picture, and the Big Questions}",
    eprint = "1802.04801",
    archivePrefix = "arXiv",
    primaryClass = "hep-ph",
    reportNumber = "MIT-CTP-4892, MIT-CTP/4892",
    doi = "10.1146/annurev-nucl-101917-020852",
    journal = "Ann. Rev. Nucl. Part. Sci.",
    volume = "68",
    pages = "339--376",
    year = "2018"
}

@article{vanHees:2011vb,
    author = "van Hees, Hendrik and Gale, Charles and Rapp, Ralf",
    title = "{Thermal Photons and Collective Flow at the Relativistic Heavy-Ion Collider}",
    eprint = "1108.2131",
    archivePrefix = "arXiv",
    primaryClass = "hep-ph",
    doi = "10.1103/PhysRevC.84.054906",
    journal = "Phys. Rev. C",
    volume = "84",
    pages = "054906",
    year = "2011"
}

@article{Gale:2003iz,
    author = "Gale, Charles and Haglin, Kevin L.",
    editor = "Hwa, Rudolph C. and Wang, Xin-Nian",
    title = "{Electromagnetic radiation from relativistic nuclear collisions}",
    eprint = "hep-ph/0306098",
    archivePrefix = "arXiv",
    pages = "364--429",
    month = "6",
    year = "2003"
}

@article{vanHees:2007th,
    author = "van Hees, Hendrik and Rapp, Ralf",
    title = "{Dilepton Radiation at the CERN Super Proton Synchrotron}",
    eprint = "0711.3444",
    archivePrefix = "arXiv",
    primaryClass = "hep-ph",
    doi = "10.1016/j.nuclphysa.2008.03.009",
    journal = "Nucl. Phys. A",
    volume = "806",
    pages = "339--387",
    year = "2008"
}

@article{Jorge:2025wwp,
    author = "Jorge, Adrian William Romero and Song, Taesoo and Zhou, Qi and Bratkovskaya, Elena",
    title = "{Electromagnetic emission from strongly interacting hadronic and partonic matter created in heavy-ion collisions}",
    eprint = "2503.05253",
    archivePrefix = "arXiv",
    primaryClass = "nucl-th",
    doi = "10.1103/PhysRevC.111.064904",
    journal = "Phys. Rev. C",
    volume = "111",
    number = "6",
    pages = "064904",
    year = "2025"
}

@article{Wu:2024pba,
    author = "Wu, Xiang-Yu and Du, Lipei and Gale, Charles and Jeon, Sangyong",
    title = "{Probing the equilibration of the QCD matter created in heavy-ion collisions with dileptons}",
    eprint = "2407.04156",
    archivePrefix = "arXiv",
    primaryClass = "nucl-th",
    doi = "10.1103/PhysRevC.110.054904",
    journal = "Phys. Rev. C",
    volume = "110",
    number = "5",
    pages = "054904",
    year = "2024"
}

@article{Coquet:2021lca,
    author = "Coquet, Maurice and Du, Xiaojian and Ollitrault, Jean-Yves and Schlichting, Soeren and Winn, Michael",
    title = "{Intermediate mass dileptons as pre-equilibrium probes in heavy ion collisions}",
    eprint = "2104.07622",
    archivePrefix = "arXiv",
    primaryClass = "nucl-th",
    doi = "10.1016/j.physletb.2021.136626",
    journal = "Phys. Lett. B",
    volume = "821",
    pages = "136626",
    year = "2021"
}

@article{Shen:2013vja,
    author = "Shen, Chun and Heinz, Ulrich W and Paquet, Jean-Francois and Gale, Charles",
    title = "{Thermal photons as a quark-gluon plasma thermometer reexamined}",
    eprint = "1308.2440",
    archivePrefix = "arXiv",
    primaryClass = "nucl-th",
    doi = "10.1103/PhysRevC.89.044910",
    journal = "Phys. Rev. C",
    volume = "89",
    number = "4",
    pages = "044910",
    year = "2014"
}

@article{Rapp:2014hha,
    author = "Rapp, Ralf and van Hees, Hendrik",
    title = "{Thermal Dileptons as Fireball Thermometer and Chronometer}",
    eprint = "1411.4612",
    archivePrefix = "arXiv",
    primaryClass = "hep-ph",
    doi = "10.1016/j.physletb.2015.12.065",
    journal = "Phys. Lett. B",
    volume = "753",
    pages = "586--590",
    year = "2016"
}

@article{Churchill:2023zkk,
    author = "Churchill, Jessica and Du, Lipei and Gale, Charles and Jackson, Greg and Jeon, Sangyong",
    title = "{Virtual Photons Shed Light on the Early Temperature of Dense QCD Matter}",
    eprint = "2311.06951",
    archivePrefix = "arXiv",
    primaryClass = "nucl-th",
    doi = "10.1103/PhysRevLett.132.172301",
    journal = "Phys. Rev. Lett.",
    volume = "132",
    number = "17",
    pages = "172301",
    year = "2024"
}

@article{STAR:2021mii,
    author = "Abdallah, Mohamed and others",
    collaboration = "STAR",
    title = "{Search for the chiral magnetic effect with isobar collisions at $\sqrt {s_{NN}}$=200 GeV by the STAR Collaboration at the BNL Relativistic Heavy Ion Collider}",
    eprint = "2109.00131",
    archivePrefix = "arXiv",
    primaryClass = "nucl-ex",
    doi = "10.1103/PhysRevC.105.014901",
    journal = "Phys. Rev. C",
    volume = "105",
    number = "1",
    pages = "014901",
    year = "2022"
}

@misc{Luo:QM2025,
  author        = {Jiaxuan Luo},
  collaboration = {STAR Collaboration},
  title         = {Thermal dielectron measurements with the STAR experiment},
  howpublished  = {Presentation at Quark Matter 2025},
  address       = {Frankfurt, Germany},
  year          = {2025},
  month         = apr,
  note          = {Quark Matter 2025, April 6--12, 2025},
  url           = {https://indico.cern.ch/event/1334113/contributions/6369529/attachments/3048099/5386383/Qm25_thermal_jxluo_v9.pdf}
}

@article{Jackson:2019mop,
    author = "Jackson, G.",
    title = "{Two-loop thermal spectral functions with general kinematics}",
    eprint = "1910.07552",
    archivePrefix = "arXiv",
    primaryClass = "hep-ph",
    doi = "10.1103/PhysRevD.100.116019",
    journal = "Phys. Rev. D",
    volume = "100",
    number = "11",
    pages = "116019",
    year = "2019"
}

@article{Ghiglieri:2014kma,
    author = "Ghiglieri, Jacopo and Moore, Guy D.",
    title = "{Low Mass Thermal Dilepton Production at NLO in a Weakly Coupled Quark-Gluon Plasma}",
    eprint = "1410.4203",
    archivePrefix = "arXiv",
    primaryClass = "hep-ph",
    doi = "10.1007/JHEP12(2014)029",
    journal = "JHEP",
    volume = "12",
    pages = "029",
    year = "2014"
}

@article{McLerran:1984ay,
    author = "McLerran, Larry D. and Toimela, T.",
    title = "{Photon and Dilepton Emission from the Quark - Gluon Plasma: Some General Considerations}",
    reportNumber = "FERMILAB-PUB-84-084-T",
    doi = "10.1103/PhysRevD.31.545",
    journal = "Phys. Rev. D",
    volume = "31",
    pages = "545",
    year = "1985"
}

@article{NA60:2005qgv,
    author = "Floris, M. and others",
    editor = "Fini, R. A. and Ghidini, B. and Lenti, V. and Virgili, T.",
    collaboration = "NA60",
    title = "{Charged particle multiplicity measurements in NA60}",
    doi = "10.1088/1742-6596/5/1/005",
    journal = "J. Phys. Conf. Ser.",
    volume = "5",
    pages = "55--63",
    year = "2005"
}

@article{Rapp:2002fc,
    author = "Rapp, Ralf",
    title = "{Hadro chemistry and evolution of (anti--) baryon densities at RHIC}",
    eprint = "hep-ph/0204131",
    archivePrefix = "arXiv",
    reportNumber = "SUNY-NTG-02-24",
    doi = "10.1103/PhysRevC.66.017901",
    journal = "Phys. Rev. C",
    volume = "66",
    pages = "017901",
    year = "2002"
}

@article{Sjostrand:2014zea,
    author = {Sj{\"o}strand, Torbj{\"o}rn and Ask, Stefan and Christiansen, Jesper R. and Corke, Richard and Desai, Nishita and Ilten, Philip and Mrenna, Stephen and Prestel, Stefan and Rasmussen, Christine O. and Skands, Peter Z.},
    title = "{An introduction to PYTHIA 8.2}",
    eprint = "1410.3012",
    archivePrefix = "arXiv",
    primaryClass = "hep-ph",
    reportNumber = "LU-TP-14-36, MCNET-14-22, CERN-PH-TH-2014-190, FERMILAB-PUB-14-316-CD, DESY-14-178, SLAC-PUB-16122",
    doi = "10.1016/j.cpc.2015.01.024",
    journal = "Comput. Phys. Commun.",
    volume = "191",
    pages = "159--177",
    year = "2015"
}

@article{ALEPH:1998rgl,
    author = "Barate, R. and others",
    collaboration = "ALEPH",
    title = "{Measurement of the spectral functions of axial - vector hadronic tau decays and determination of alpha(S)(M**2(tau))}",
    reportNumber = "CERN-EP-98-012, CERN-EP-98-12",
    doi = "10.1007/s100520050217",
    journal = "Eur. Phys. J. C",
    volume = "4",
    pages = "409--431",
    year = "1998"
}

@article{Churchill:2023vpt,
    author = "Churchill, Jessica and Du, Lipei and Gale, Charles and Jackson, Greg and Jeon, Sangyong",
    title = "{Dilepton production at next-to-leading order and intermediate invariant-mass observables}",
    eprint = "2311.06675",
    archivePrefix = "arXiv",
    primaryClass = "nucl-th",
    doi = "10.1103/PhysRevC.109.044915",
    journal = "Phys. Rev. C",
    volume = "109",
    number = "4",
    pages = "044915",
    year = "2024"
}

@article{Shen:2017bsr,
    author = {Shen, Chun and Schenke, Bj{\"o}rn},
    title = "{Dynamical initial state model for relativistic heavy-ion collisions}",
    eprint = "1710.00881",
    archivePrefix = "arXiv",
    primaryClass = "nucl-th",
    doi = "10.1103/PhysRevC.97.024907",
    journal = "Phys. Rev. C",
    volume = "97",
    number = "2",
    pages = "024907",
    year = "2018"
}

@article{Shen:2022oyg,
    author = {Shen, Chun and Schenke, Bj{\"o}rn},
    title = "{Longitudinal dynamics and particle production in relativistic nuclear collisions}",
    eprint = "2203.04685",
    archivePrefix = "arXiv",
    primaryClass = "nucl-th",
    doi = "10.1103/PhysRevC.105.064905",
    journal = "Phys. Rev. C",
    volume = "105",
    number = "6",
    pages = "064905",
    year = "2022"
}

@article{SMASH:2016zqf,
    author = "Weil, J. and others",
    collaboration = "SMASH",
    title = "{Particle production and equilibrium properties within a new hadron transport approach for heavy-ion collisions}",
    eprint = "1606.06642",
    archivePrefix = "arXiv",
    primaryClass = "nucl-th",
    doi = "10.1103/PhysRevC.94.054905",
    journal = "Phys. Rev. C",
    volume = "94",
    number = "5",
    pages = "054905",
    year = "2016"
}

@article{Shen:2023awv,
    author = {Shen, Chun and Schenke, Bj{\"o}rn and Zhao, Wenbin},
    title = "{Viscosities of the Baryon-Rich Quark-Gluon Plasma from Beam Energy Scan Data}",
    eprint = "2310.10787",
    archivePrefix = "arXiv",
    primaryClass = "nucl-th",
    doi = "10.1103/PhysRevLett.132.072301",
    journal = "Phys. Rev. Lett.",
    volume = "132",
    number = "7",
    pages = "072301",
    year = "2024"
}

@article{Jahan:2025cbp,
    author = "Jahan, Syed Afrid and Roch, Hendrik and Shen, Chun",
    title = "{Bayesian model selection and uncertainty propagation for beam energy scan heavy-ion collisions}",
    eprint = "2507.11394",
    archivePrefix = "arXiv",
    primaryClass = "nucl-th",
    doi = "10.1103/y962-1lyg",
    journal = "Phys. Rev. C",
    volume = "113",
    number = "2",
    pages = "024919",
    year = "2026"
}

@article{Jahan:2024wpj,
    author = "Jahan, Syed Afrid and Roch, Hendrik and Shen, Chun",
    title = "{Bayesian analysis of (3+1)D relativistic nuclear dynamics with the RHIC beam energy scan data}",
    eprint = "2408.00537",
    archivePrefix = "arXiv",
    primaryClass = "nucl-th",
    doi = "10.1103/PhysRevC.110.054905",
    journal = "Phys. Rev. C",
    volume = "110",
    number = "5",
    pages = "054905",
    year = "2024"
}

@article{Schenke:2010nt,
    author = "Schenke, Bjoern and Jeon, Sangyong and Gale, Charles",
    title = "{(3+1)D hydrodynamic simulation of relativistic heavy-ion collisions}",
    eprint = "1004.1408",
    archivePrefix = "arXiv",
    primaryClass = "hep-ph",
    doi = "10.1103/PhysRevC.82.014903",
    journal = "Phys. Rev. C",
    volume = "82",
    pages = "014903",
    year = "2010"
}

@article{Schenke:2010rr,
    author = "Schenke, Bjorn and Jeon, Sangyong and Gale, Charles",
    title = "{Elliptic and triangular flow in event-by-event (3+1)D viscous hydrodynamics}",
    eprint = "1009.3244",
    archivePrefix = "arXiv",
    primaryClass = "hep-ph",
    doi = "10.1103/PhysRevLett.106.042301",
    journal = "Phys. Rev. Lett.",
    volume = "106",
    pages = "042301",
    year = "2011"
}

@article{Denicol:2018wdp,
    author = {Denicol, Gabriel S. and Gale, Charles and Jeon, Sangyong and Monnai, Akihiko and Schenke, Bj{\"o}rn and Shen, Chun},
    title = "{Net baryon diffusion in fluid dynamic simulations of relativistic heavy-ion collisions}",
    eprint = "1804.10557",
    archivePrefix = "arXiv",
    primaryClass = "nucl-th",
    reportNumber = "KEK-TH-2075",
    doi = "10.1103/PhysRevC.98.034916",
    journal = "Phys. Rev. C",
    volume = "98",
    number = "3",
    pages = "034916",
    year = "2018"
}

@article{Shen:2017fnn,
    author = {Shen, Chun and Schenke, Bj{\"o}rn},
    title = "{Initial state and hydrodynamic modeling of heavy-ion collisions at RHIC BES energies}",
    eprint = "1711.10544",
    archivePrefix = "arXiv",
    primaryClass = "nucl-th",
    doi = "10.22323/1.311.0006",
    journal = "PoS",
    volume = "CPOD2017",
    pages = "006",
    year = "2018"
}

@article{Shen:2018pty,
    author = {Shen, Chun and Schenke, Bj{\"o}rn},
    editor = "Antinori, Federico and Dainese, Andrea and Giubellino, Paolo and Greco, Vincenzo and Lombardo, Maria Paola and Scomparin, Enrico",
    title = "{Dynamical initialization and hydrodynamic modeling of relativistic heavy-ion collisions}",
    eprint = "1807.05141",
    archivePrefix = "arXiv",
    primaryClass = "nucl-th",
    doi = "10.1016/j.nuclphysa.2018.08.007",
    journal = "Nucl. Phys. A",
    volume = "982",
    pages = "411--414",
    year = "2019"
}

@article{Monnai:2019hkn,
    author = {Monnai, Akihiko and Schenke, Bj{\"o}rn and Shen, Chun},
    title = "{Equation of state at finite densities for QCD matter in nuclear collisions}",
    eprint = "1902.05095",
    archivePrefix = "arXiv",
    primaryClass = "nucl-th",
    reportNumber = "KEK-TH-2106",
    doi = "10.1103/PhysRevC.100.024907",
    journal = "Phys. Rev. C",
    volume = "100",
    number = "2",
    pages = "024907",
    year = "2019"
}

@article{Denicol:2012cn,
    author = "Denicol, G. S. and Niemi, H. and Molnar, E. and Rischke, D. H.",
    title = "{Derivation of transient relativistic fluid dynamics from the Boltzmann equation}",
    eprint = "1202.4551",
    archivePrefix = "arXiv",
    primaryClass = "nucl-th",
    doi = "10.1103/PhysRevD.85.114047",
    journal = "Phys. Rev. D",
    volume = "85",
    pages = "114047",
    year = "2012",
    note = "[Erratum: Phys.Rev.D 91, 039902 (2015)]"
}

@article{Cooper:1974mv,
    author = "Cooper, Fred and Frye, Graham",
    title = "{Comment on the Single Particle Distribution in the Hydrodynamic and Statistical Thermodynamic Models of Multiparticle Production}",
    reportNumber = "Print-74-0742 (YESHIVA)",
    doi = "10.1103/PhysRevD.10.186",
    journal = "Phys. Rev. D",
    volume = "10",
    pages = "186",
    year = "1974"
}

@article{Seck:2017zjr,
    author = "Seck, F. and Galatyuk, T. and Rapp, R. and Stroth, J.",
    editor = "Destefanis, Marco and Galatyuk, Tetyana and Nicmorus, Diana and Ratti, Claudia and Tolos, Laura",
    title = "{Thermal dileptons as QCD matter probes at SIS}",
    eprint = "1710.06256",
    archivePrefix = "arXiv",
    primaryClass = "nucl-th",
    doi = "10.1088/1742-6596/1024/1/012011",
    journal = "J. Phys. Conf. Ser.",
    volume = "1024",
    number = "1",
    pages = "012011",
    year = "2018"
}

@article{Gale:1990pn,
    author = "Gale, Charles and Kapusta, Joseph I.",
    title = "{Vector dominance model at finite temperature}",
    reportNumber = "MCGILL-90-57",
    doi = "10.1016/0550-3213(91)90459-B",
    journal = "Nucl. Phys. B",
    volume = "357",
    pages = "65--89",
    year = "1991"
}

@article{Li:1998ma,
    author = "Li, Guo-Qiang and Gale, Charles",
    title = "{Intermediate mass dilepton spectra and the role of secondary hadronic processes in heavy ion collisions}",
    eprint = "nucl-th/9807005",
    archivePrefix = "arXiv",
    doi = "10.1103/PhysRevC.58.2914",
    journal = "Phys. Rev. C",
    volume = "58",
    pages = "2914--2927",
    year = "1998"
}

@article{Weldon:1990iw,
    author = "Weldon, H. A.",
    title = "{Reformulation of finite temperature dilepton production}",
    doi = "10.1103/PhysRevD.42.2384",
    journal = "Phys. Rev. D",
    volume = "42",
    pages = "2384--2387",
    year = "1990"
}

@article{STAR:2024bpc,
    author = "Aboona, B. E. and others",
    collaboration = "STAR",
    title = "{Temperature measurement of Quark-Gluon plasma at different stages}",
    eprint = "2402.01998",
    archivePrefix = "arXiv",
    primaryClass = "nucl-ex",
    doi = "10.1038/s41467-025-63216-5",
    journal = "Nature Commun.",
    volume = "16",
    number = "1",
    pages = "9098",
    year = "2025"
}

@article{Rapp:1997fs,
    author = "Rapp, R. and Chanfray, G. and Wambach, J.",
    title = "{Rho meson propagation and dilepton enhancement in hot hadronic matter}",
    eprint = "hep-ph/9702210",
    archivePrefix = "arXiv",
    reportNumber = "SUNY-NTG-97-04",
    doi = "10.1016/S0375-9474(97)00137-1",
    journal = "Nucl. Phys. A",
    volume = "617",
    pages = "472--495",
    year = "1997"
}

@article{Rapp:1999us,
    author = "Rapp, Ralf and Wambach, Jochen",
    title = "{Low mass dileptons at the CERN SPS: Evidence for chiral restoration?}",
    eprint = "hep-ph/9907502",
    archivePrefix = "arXiv",
    reportNumber = "SUNY-NTG-99-25",
    doi = "10.1007/s100500050364",
    journal = "Eur. Phys. J. A",
    volume = "6",
    pages = "415--420",
    year = "1999"
}

@article{ALICE:2023jef,
    author = "Acharya, Shreyasi and others",
    collaboration = "ALICE",
    title = "{Dielectron production in central Pb-Pb collisions at sNN=5.02TeV}",
    eprint = "2308.16704",
    archivePrefix = "arXiv",
    primaryClass = "nucl-ex",
    reportNumber = "CERN-EP-2023-194",
    doi = "10.1103/xl6m-vbqk",
    journal = "Phys. Rev. C",
    volume = "112",
    number = "5",
    pages = "054906",
    year = "2025"
}

@article{Wu:2025iix,
    author = {Wu, Xiang-Yu and Gale, Charles and Jeon, Sangyong and Paquet, Jean-Fran{\c{c}}ois and Schenke, Bj{\"o}rn and Shen, Chun},
    title = "{Electromagnetic radiation from baryon-rich matter in heavy-ion collisions}",
    eprint = "2511.08773",
    archivePrefix = "arXiv",
    primaryClass = "nucl-th",
    doi = "10.1103/nh24-jn4j",
    journal = "Phys. Rev. C",
    volume = "114",
    number = "3",
    pages = "034902",
    year = "2026"
}

@article{ALICE:2018ael,
    author = "Acharya, Shreyasi and others",
    collaboration = "ALICE",
    title = "{Measurement of dielectron production in central Pb-Pb collisions at $\sqrt{{\textit{s}}_{\mathrm{NN}}}$ = 2.76 TeV}",
    eprint = "1807.00923",
    archivePrefix = "arXiv",
    primaryClass = "nucl-ex",
    reportNumber = "CERN-EP-2018-181",
    doi = "10.1103/PhysRevC.99.024002",
    journal = "Phys. Rev. C",
    volume = "99",
    number = "2",
    pages = "024002",
    year = "2019"
}

@article{Rapp:1999zw,
    author = "Rapp, Ralf and Shuryak, Edward V.",
    title = "{Thermal dilepton radiation at intermediate masses at the CERN - SPS}",
    eprint = "hep-ph/9909348",
    archivePrefix = "arXiv",
    reportNumber = "SUNY-NTG-99-30",
    doi = "10.1016/S0370-2693(99)01367-2",
    journal = "Phys. Lett. B",
    volume = "473",
    pages = "13--19",
    year = "2000"
}

@article{Geurts:2022xmk,
    author = "Geurts, Frank and Tripolt, Ralf-Arno",
    title = "{Electromagnetic probes: Theory and experiment}",
    eprint = "2210.01622",
    archivePrefix = "arXiv",
    primaryClass = "hep-ph",
    doi = "10.1016/j.ppnp.2022.104004",
    journal = "Prog. Part. Nucl. Phys.",
    volume = "128",
    pages = "104004",
    year = "2023"
}

@article{Du:2024wjm,
    author = "Du, Lipei and Sorensen, Agnieszka and Stephanov, Mikhail",
    title = "{The QCD phase diagram and Beam Energy Scan physics: A theory overview}",
    eprint = "2402.10183",
    archivePrefix = "arXiv",
    primaryClass = "nucl-th",
    reportNumber = "INT-PUB-24-017",
    doi = "10.1142/9789811294679_0007",
    journal = "Int. J. Mod. Phys. E",
    volume = "33",
    number = "07",
    pages = "2430008",
    year = "2024"
}

@article{Jeon:2015dfa,
    author = "Jeon, Sangyong and Heinz, Ulrich",
    title = "{Introduction to Hydrodynamics}",
    eprint = "1503.03931",
    archivePrefix = "arXiv",
    primaryClass = "hep-ph",
    doi = "10.1142/S0218301315300106",
    journal = "Int. J. Mod. Phys. E",
    volume = "24",
    number = "10",
    pages = "1530010",
    year = "2015"
}

@article{Heinz:2013th,
    author = "Heinz, Ulrich and Snellings, Raimond",
    title = "{Collective flow and viscosity in relativistic heavy-ion collisions}",
    eprint = "1301.2826",
    archivePrefix = "arXiv",
    primaryClass = "nucl-th",
    doi = "10.1146/annurev-nucl-102212-170540",
    journal = "Ann. Rev. Nucl. Part. Sci.",
    volume = "63",
    pages = "123--151",
    year = "2013"
}

@book{Kapusta:2023eix,
    author = "Kapusta, Joseph I. and Gale, Charles",
    title = "{Finite-Temperature Field Theory : Principles and Applications, 2nd edition}",
    doi = "10.1017/9781009401968",
    isbn = "978-1-009-40196-8, 978-1-009-40195-1, 978-1-009-40198-2",
    publisher = "Cambridge University Press",
    year = "2007"
}

@article{STAR:2015zal,
    author = "Adamczyk, L. and others",
    collaboration = "STAR",
    title = "{Energy dependence of acceptance-corrected dielectron excess mass spectrum at mid-rapidity in Au$+$Au collisions at $\sqrt{s_{NN}} =$ 19.6  and 200 GeV}",
    eprint = "1501.05341",
    archivePrefix = "arXiv",
    primaryClass = "hep-ex",
    doi = "10.1016/j.physletb.2015.08.044",
    journal = "Phys. Lett. B",
    volume = "750",
    pages = "64--71",
    year = "2015"
}

@article{STAR:2023wta,
    author = "Abdulhamid, M. I. and others",
    collaboration = "STAR",
    title = "{Measurements of dielectron production in Au+Au collisions at sNN=27, 39, and 62.4 GeV from the STAR experiment}",
    doi = "10.1103/PhysRevC.107.L061901",
    journal = "Phys. Rev. C",
    volume = "107",
    number = "6",
    pages = "L061901",
    year = "2023"
}

@article{Gotz:2021dco,
    author = {G{\"o}tz, Niklas and Sch{\"a}fer, Anna and Garcia-Montero, Oscar and Paquet, Jean-Fran{\c{c}}ois and Elfner, Hannah and Gale, Charles},
    collaboration = "SMASH",
    title = "{Out-of-equilibrium photon production in the late stages of relativistic heavy-ion collisions}",
    eprint = "2111.13603",
    archivePrefix = "arXiv",
    primaryClass = "hep-ph",
    doi = "10.1103/PhysRevC.105.044910",
    journal = "Phys. Rev. C",
    volume = "105",
    number = "4",
    pages = "044910",
    year = "2022",
    note = "[Erratum: Phys.Rev.C 109, 049901 (2024)]"
}

@article{Rapp:2009yu,
    author = "Rapp, R. and Wambach, J. and van Hees, H.",
    editor = "Stock, R.",
    title = "{The Chiral Restoration Transition of QCD and Low Mass Dileptons}",
    eprint = "0901.3289",
    archivePrefix = "arXiv",
    primaryClass = "hep-ph",
    doi = "10.1007/978-3-642-01539-7_6",
    journal = "Landolt-Bornstein",
    volume = "23",
    pages = "134",
    year = "2010"
}

@inproceedings{Rapp:2002mm,
    author = "Rapp, Ralf",
    title = "{Thermal lepton production in heavy ion collisions}",
    booktitle = "{18th Winter Workshop on Nuclear Dynamics}",
    eprint = "nucl-th/0204003",
    archivePrefix = "arXiv",
    reportNumber = "SUNY-NTG-02-13",
    month = "4",
    year = "2002"
}

@article{NA60:2008ctj,
    author = "Arnaldi, R. and others",
    collaboration = "NA60",
    title = "{NA60 results on thermal dimuons}",
    eprint = "0812.3053",
    archivePrefix = "arXiv",
    primaryClass = "nucl-ex",
    doi = "10.1140/epjc/s10052-009-0878-5",
    journal = "Eur. Phys. J. C",
    volume = "61",
    pages = "711--720",
    year = "2009"
}

@article{Goes-Hirayama:2024aqz,
    author = "G{\'o}es-Hirayama, Renan and Elfner, Hannah",
    title = "{Distinguishing the sources of dielectron anisotropic flow at low beam energies}",
    eprint = "2408.16603",
    archivePrefix = "arXiv",
    primaryClass = "nucl-th",
    doi = "10.1103/PhysRevC.110.064903",
    journal = "Phys. Rev. C",
    volume = "110",
    number = "6",
    pages = "064903",
    year = "2024"
}

@article{vanHees:2006ng,
    author = "van Hees, Hendrik and Rapp, Ralf",
    title = "{Comprehensive interpretation of thermal dileptons at the SPS}",
    eprint = "hep-ph/0603084",
    archivePrefix = "arXiv",
    doi = "10.1103/PhysRevLett.97.102301",
    journal = "Phys. Rev. Lett.",
    volume = "97",
    pages = "102301",
    year = "2006"
}

@article{Rapp:2013nxa,
    author = "Rapp, Ralf",
    title = "{Dilepton Spectroscopy of QCD Matter at Collider Energies}",
    eprint = "1304.2309",
    archivePrefix = "arXiv",
    primaryClass = "hep-ph",
    doi = "10.1155/2013/148253",
    journal = "Adv. High Energy Phys.",
    volume = "2013",
    pages = "148253",
    year = "2013"
}

@article{Vujanovic:2013jpa,
    author = "Vujanovic, Gojko and Young, Clint and Schenke, Bjoern and Rapp, Ralf and Jeon, Sangyong and Gale, Charles",
    title = "{Dilepton emission in high-energy heavy-ion collisions with viscous hydrodynamics}",
    eprint = "1312.0676",
    archivePrefix = "arXiv",
    primaryClass = "nucl-th",
    doi = "10.1103/PhysRevC.89.034904",
    journal = "Phys. Rev. C",
    volume = "89",
    number = "3",
    pages = "034904",
    year = "2014"
}

@article{Vujanovic:2016anq,
    author = "Vujanovic, Gojko and Paquet, Jean-Fran{\c{c}}ois and Denicol, Gabriel S. and Luzum, Matthew and Jeon, Sangyong and Gale, Charles",
    title = "{Electromagnetic radiation as a probe of the initial state and of viscous dynamics in relativistic nuclear collisions}",
    eprint = "1602.01455",
    archivePrefix = "arXiv",
    primaryClass = "nucl-th",
    doi = "10.1103/PhysRevC.94.014904",
    journal = "Phys. Rev. C",
    volume = "94",
    number = "1",
    pages = "014904",
    year = "2016"
}

@article{Vujanovic:2017psb,
    author = "Vujanovic, Gojko and Denicol, Gabriel S. and Luzum, Matthew and Jeon, Sangyong and Gale, Charles",
    title = "{Investigating the temperature dependence of the specific shear viscosity of QCD matter with dilepton radiation}",
    eprint = "1702.02941",
    archivePrefix = "arXiv",
    primaryClass = "nucl-th",
    doi = "10.1103/PhysRevC.98.014902",
    journal = "Phys. Rev. C",
    volume = "98",
    number = "1",
    pages = "014902",
    year = "2018"
}

@article{Vujanovic:2019yih,
    author = "Vujanovic, Gojko and Paquet, Jean-Fran{\c{c}}ois and Shen, Chun and Denicol, Gabriel S. and Jeon, Sangyong and Gale, Charles and Heinz, Ulrich",
    title = "{Exploring the influence of bulk viscosity of QCD on dilepton tomography}",
    eprint = "1903.05078",
    archivePrefix = "arXiv",
    primaryClass = "nucl-th",
    doi = "10.1103/PhysRevC.101.044904",
    journal = "Phys. Rev. C",
    volume = "101",
    pages = "044904",
    year = "2020"
}

@article{Braaten:1990wp,
    author = "Braaten, Eric and Pisarski, Robert D. and Yuan, Tzu-Chiang",
    title = "{Production of Soft Dileptons in the Quark - Gluon Plasma}",
    reportNumber = "BNL-43941, NUHEP-TH-90-1",
    doi = "10.1103/PhysRevLett.64.2242",
    journal = "Phys. Rev. Lett.",
    volume = "64",
    pages = "2242",
    year = "1990"
}

@misc{Bao:2026HP,
  author        = {Bao, Xianwen},
  collaboration = {STAR},
  title         = {Direct Photon Yield and Thermal Dielectron Flow in Isobar Collisions at STAR},
  howpublished  = {Oral presentation at the 13th International Conference on Hard and Electromagnetic Probes of High-Energy Nuclear Collisions (Hard Probes 2026)},
  year          = {2026},
  month         = jun,
  address       = {Nashville, Tennessee, USA},
  url           = {https://indico.cern.ch/event/1428985/contributions/6999201/},
  note          = {Presented on June 24, 2026}
}

\end{document}